\documentclass[12pt]{article}
\usepackage{pstricks,pst-node,pst-coil}
\usepackage{amsfonts}
\usepackage{sectsty}
\newcommand\ZZ{{\mathbb Z}}
\newcommand\CC{{\mathbb C}}
\newcommand\NN{{\cal N}}

\newcommand\sqcdv{$\rm SQCD_5$}
\newcommand\suq{\relax\ifmmode [SU(n_c)]^N \else $[SU(n_c)]^N$\fi }
\newcommand\kcs{k_{\rm cs}}
\newcommand\tr{\mathop\mathrm{tr}\nolimits}
\newcommand\Tr{\mathop\mathrm{Tr}\nolimits}
\newcommand\Det{\mathop\mathrm{Det}\nolimits}

\newcommand\diag{\mathop\mathrm{diag}\nolimits}
\newcommand\rank{\mathop\mathrm{rank}\nolimits}
\newcommand\Res{\mathop\mathrm{Res}\limits}
\newcommand\eqcr{\buildrel {\rm cr}\over=}
\newcommand\eqdef{\buildrel {\rm def}\over=} 
\newcommand\BB{{\cal B}}
\newcommand\MM{{\cal M}}
\newcommand\AB{\widetilde{\cal B}}
\newcommand\W{{\cal W}}
\newcommand\WW{\widetilde{\cal W}}
\newcommand\Wsum{{\mathbb W}}
\newcommand\aq{\smash{\widetilde Q}\vphantom{Q}}
\newcommand\link{\Omega}
\newcommand\ev{\omega}
\newcommand\rt{\wp}
\newcommand\pp{{\cal H}}
\newcommand\pph{\widehat\pp}
\newcommand\dv{V^{Nn_c}_{\rm global}}
\newcommand\be{\begin{equation}}
\newcommand\ee{\end{equation}}
\newcommand\bea{\begin{eqnarray}}
\newcommand\eea{\end{eqnarray}}
\newcommand\distimes{$\discretionary{\hbox{${}\times{}\!$}}{}{}$}
\newcommand\vev[1]{\left\langle #1\right\rangle}

\newcommand\coeff[2]{{\textstyle{#1\over #2}}}
\newcommand\half{\coeff{1}{2}}

\newcommand\eq[1]{eq.~(\ref{#1})}
\newcommand\eqalign[1]{%
	\vcenter{%
		\normalbaselines \advance\baselineskip 5pt
		\advance\lineskip 5pt \tabskip=0pt
		\halign{%
			&\hfil $\displaystyle{##{}}$&
			$\displaystyle{{}##}$\hfil\cr
			#1\crcr
			}%
		}%
	}
\newcommand\eqrange[2]{eqs.~(\ref{#1}--\reftail{#2})}
\newcommand\dotline{\par\hbox to \hsize{\dotfill}\par}
\newcommand\crossout[1]{%
    \setbox0=\hbox{$\displaystyle{#1}$}%
    \psline[linecolor=red](0,\ht0)(\wd0,-\dp0)%
    \psline[linecolor=red](0,-\dp0)(\wd0,\ht0)%
    \box0 }
\makeatletter
\def\befored@t#1.#2.#3;{#1}
\def\afterd@t#1.#2.#3;{#2}
\def\refhead#1{\edef\next{\ref{#1}}\expandafter\befored@t\next..;}
\def\reftail#1{\edef\next{\ref{#1}}\expandafter\afterd@t\next..;}
\def\lsim{\mathrel{\mathpalette\@versim<}}
\def\gsim{\mathrel{\mathpalette\@versim>}}
\def\@versim#1#2{\vcenter{\offinterlineskip
        \ialign{$\m@th#1\hfil##\hfil$\crcr#2\crcr\sim\crcr } }}
\newcommand\becomes[1]{\mathchoice{\becomes@\scriptstyle{#1}}
   {\becomes@\scriptstyle{#1}} {\becomes@\scriptscriptstyle{#1}}
   {\becomes@\scriptscriptstyle{#1}}}
\def\becomes@#1#2{\mathrel{\setbox0=\hbox{$\m@th #1{\,#2\,}$}%
        \mathop{\hbox to \wd0 {\rightarrowfill}}\limits_{#2}}}
\makeatother
\psset{unit=1mm,linecolor=black,linewidth=1pt}
\def\normalbaselines{%
        \normalbaselineskip=20pt plus 0.2pt minus 0.1pt
	\baselineskip=\normalbaselineskip
        \lineskip=2pt plus 0.1pt minus 0.1pt
        \lineskiplimit=2pt
	}
\def\fixformat{%
        \normalbaselines
        \abovedisplayskip=15pt plus 5pt minus 3pt
        \belowdisplayskip=\abovedisplayskip
        \parskip=6pt plus 2pt minus 1pt
	\skip\footins=20pt plus 10pt minus 5pt
	\predisplaypenalty=5000
	\postdisplaypenalty=500
	\interlinepenalty=50
	\interdisplaylinepenalty=10000
	\flushbottom
        }
\sectionfont{\huge\rm\blue\nohang\centering}
\subsectionfont{\large\sc\nohang\raggedright}
\makeatletter
\def\@seccntformat#1{\@ifundefined{#1@cntformat}%
	{\csname the#1\endcsname\quad}
	{\csname #1@cntformat\endcsname}
	}
\def\section@cntformat{\thesection.\enspace}
\@addtoreset{equation}{section}
\newif\iffntmark \fntmarktrue
\renewcommand\@makefntext[1]{\noindent
	\iffntmark\llap{\@thefnmark\enspace}\fi
	#1\unskip }
\newif\ift@c \t@cfalse
\def\br@@k{\relax\ift@c\else\unskip\break\fi}
\def\brk{\protect\br@@k}
\let\t@c=\tableofcontents
\renewcommand\tableofcontents{\t@ctrue\t@c\t@cfalse}
\makeatother
\renewcommand{\theequation}{\thesection.\arabic{equation}}

\begin{document}
\fixformat
%
%
\begin{titlepage}

\rightline{\vbox{%
    \baselineskip=15pt \tabskip=0pt
    \halign{%
	#\unskip\hfil\cr
	UTTG--03--04\cr
	TAUP--2772--04\cr
	June 16, 2004\cr
	hep-th/0406122\cr
	}%
    }}
\vskip 2pc plus 1fil minus 0.5pc

\centerline{\large\bf Chiral Rings of Deconstructive \suq\
Quivers$\strut^\star$}
\vskip 2pc plus 1fil minus 0.5pc
\centerline{\large Edoardo Di Napoli$\strut^{\rm (1a)}$,
	Vadim S.\ Kaplunovsky$\strut^{\rm (1b)}$ and
	Jacob Sonnenschein$\strut^{\rm (2)}$}
\vskip 1pc minus 0.5pc
\tabskip=0pt plus 3in
\halign to \hsize{%
    \it #\hfil\cr
    \noalign{\vskip 1pc}
    \llap{\rm (1)\enspace}\relax
	Theory Group, Physics Department,\cr
    University of Texas at Austin,\cr
    Austin, TX~78712, USA\cr
    \ \rm (a) \tt edodin@physics.utexas.edu\cr
    \ \rm (b) \tt vadim@physics.utexas.edu\cr
    \noalign{\vskip 0.5pc}
    \llap{\rm (2)\enspace}\relax
	School of Physics and Astronomy,\cr
    The Raymond and Beverly Sackler Faculty of Natural Sciences,\cr 
    Tel Aviv University,\cr
    Tel-Aviv 69978, Israel\cr
    \ \tt cobi@post.tau.ac.il\cr
    }
\tabskip=0pt
\vskip 2pc plus 1fil minus 0.5pc

\begin{abstract}
Dimensional deconstruction of 5D SQCD with general $n_c$, $n_f$ and
$k_{\rm CS}$ gives rise to 4D $\NN=1$ gauge theories with large quivers
of $SU(n_c)$ gauge factors.
We construct the chiral rings of such \suq~theories, off-shell and on-shell.
Our results are broadly similar to the chiral rings of single $U(n_c)$ theories
with both adjoint and fundamental matter, but there are also some noteworthy
differences such as nonlocal meson-like operators where the quark and antiquark
fields belong to different nodes of the quiver.
And because our gauge groups are $SU(n_c)$ rather than $U(n_c)$,
our chiral rings also contain a whole zoo of baryonic and antibaryonic
operators.
\end{abstract}
\vskip 2pc plus 1fil minus 0.5pc

\begingroup
    \catcode`\@=11
    \let\@thefnmark=\relax
    \@footnotetext{%
	\nobreak
	\par\noindent \hangafter=1 \hangindent=\parindent
	{\large $\star$}\enspace
	Article based on research supported by the US National Science
	Foundation (grant PHY--0071512), by the Robert~A.\ Welsh foundation,
	by the Israel Science Foundation, and by the German--Israeli Foundation
	for Scientific Research (GIF).
	}
\endgroup

\end{titlepage}
\newpage
\pagenumbering{arabic}

\setcounter{page}{2}
\tableofcontents
%
%
\section{Introduction}
Regardless of their applicability to the weak scale phenomenology,
supersymmetric gauge theories are important from the abstract QFT
point of view because they allow for exact calculation of some
non-perturbative data.
In a 4D, $\NN=1$ theory, the exactly-computable non-perturbative data
form a mathematical structure known as {\it\blue the chiral ring}.
Physically, this ring contains invariant combinations of the
scalar VEVs, gaugino condensates and abelian gauge couplings;
usually, these data are sufficient to completely determine the phase
structure of the theory and its moduli spaces, if any.

Two recent discoveries excited much interest in
the chiral rings of gauge theories
with adjoint and fundamental matter fields:
First, Dijkgraaf and Vafa \cite{DV1,DV2,DV} found that
the gaugino condensates and the abelian gauge couplings of
4D, $\NN=1$ gauge theories correspond to perturbative amplitudes of
matrix models without any spacetime or SUSY at all.
Second, Cachazo, Douglas, Seiberg and Witten \cite{CDSW,CSW1,Seiberg,CSW2}
evaluated the entire on-shell chiral ring of an $U(N_c)$ theory with adjoint 
and fundamental matter using generalized Konishi anomaly equations
\cite{konishione,konishitwo}.
In the process, they verified the Dijkgraaf--Vafa conjecture by showing
that the loop equation of the matrix model is identical to the
anomaly equation for a particular resolvent $R(X)$
summarizing the gaugino condensates.
Both approaches --- the matrix models and the anomaly equations ---
are readily extended to the {\it quiver} theories with multiple gauge
group factors ``connected'' by the bi-fundamental matter fields.
Thus far, most work on this subject concerned the non-chiral quivers
where the bi-fundamental fields come in 
$({\bf\bar n},{\bf m})+({\bf n},{\bf\bar m})$ conjugate pairs;
This is an natural limitation of the matrix correspondence%
\footnote{%
    An un-constrained complex $n\times m$ matrix corresponds to
    a whole $\NN=2$ hypermultiplet in the bi-fundamental
    $({\bf\bar n},{\bf m})$ representation of the
    $SU(n)\times SU(m)$, or in $\NN=1$ terms to a conjugate pair
    $({\bf\bar n},{\bf m})+({\bf n},{\bf\bar m})$ of chiral multiplets.
    A chiral bi-fundamental $({\bf\bar n},{\bf m})$ multiplet
    \underline{not} accompanied by its $({\bf n},{\bf\bar m})$ conjugate
    corresponds to a matrix subject to non-linear constraints, which
    makes for a much more complicated matrix model.
    In particular, the chiral \suq\ quiver theory presented in this
    article corresponds to a model of $N$ unitary $SU(n_c)$ matrices.
    This matrix model --- and its implications for the gaugino
    condensates --- will be presented in a separate article
    \cite{dNKmatrix}.
    }
but the anomaly-equations technology 
has no particular difficulties with 4D chirality
\cite{Amati, BINOR, Chiral1,Chiral2}.

In this article we analyze the inherently chiral $\widehat A_n$ quivers
of one-way arrows
$$
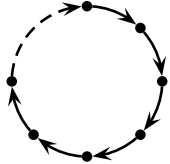
\psset{unit=1cm,linewidth=1pt,arrowscale=1.5}
\pspicture[](-1,-1)(+1,+1)
\pscircle*(0,+1){0.07}
\pscircle*(+0.71,+0.71){0.07}
\pscircle*(+1,0){0.07}
\pscircle*(+0.71,-0.71){0.07}
\pscircle*(0,-1){0.07}
\pscircle*(-0.71,-0.71){0.07}
\pscircle*(-1,0){0.07}
\psarc{<-}(0,0){1}{49}{86}
\psarc{<-}(0,0){1}{4}{41}
\psarc{<-}(0,0){1}{319}{356}
\psarc{<-}(0,0){1}{274}{311}
\psarc{<-}(0,0){1}{229}{266}
\psarc{<-}(0,0){1}{184}{221}
\psarc[linestyle=dashed]{<-}(0,0){1}{94}{176}
\endpspicture
$$
which means that all the bi-fundamental fields are chiral
$({\bf n}_i^{},{\bf\bar n}_{i+1}^{})$ multiplets \underline{not}
accompanied by their $({\bf\bar n}_i^{},{\bf n}_{i+1}^{})$
conjugates.
Also, our quivers do not have any adjoint matter fields
--- although the cyclic product of all the bi-fundamental fields
does act as some kind of a collective adjoint multiplet.
Finally, the individual gauge groups corresponding to our quivers'
nodes are of the $SU(n_c)$ rather than $U(n_c)$ type, and this
adds all kinds of {\sl baryonic} generators to the chiral ring
of the theory.

Specifically, our chiral quiver theories follow from the
{\it\blue dimensional deconstruction} of the 5D SQCD
\cite{ACG,HPW,ACG2,Csaki1,Csaki2,IK1,dNK},
hence the name {\it\blue deconstructive quivers}.
In general, dimensional deconstruction \cite{ACG,HPW,Halpern} relates simple
gauge theories in spaces of higher dimension to more complicated theories
in fewer dimensions of space:  The extra dimensions of space are
`deconstructed' into quiver diagrams of the `theory space'.
For example, starting in $4+1$ dimensions, we deconstruct the
extra space dimension by discretizing the $x^4$ coordinate
into a lattice of small
but finite spacing~$a$ and then interpreting the result as a 4D gauge
theory with a large {\sl quiver} of gauge groups.
In order to have a finite number of 4D fields, the $x^4$ is also
compactified to a large circle of length $2\pi R=Na$
(hence $N$ lattice points), but eventually
one may take the $N\to\infty$ limit  and  recover
the uncompactified 5D physics.
In this limit, the lattice spacing~$a$ remains finite and serves as
UV regulator which breaks part of the 5D Lorentz symmetry as well as
4 out of 8 supercharges but preserves the (latticized) 5D gauge symmetry
of the theory.

The deconstruction of \sqcdv\ will be discussed in detail
in a separate paper~\cite{dNK} (see also \cite{IK1} for the quarkless case).
{}From the 4D point of view, the result is an $\NN=1$ supersymmetric
gauge theory with a quiver diagram
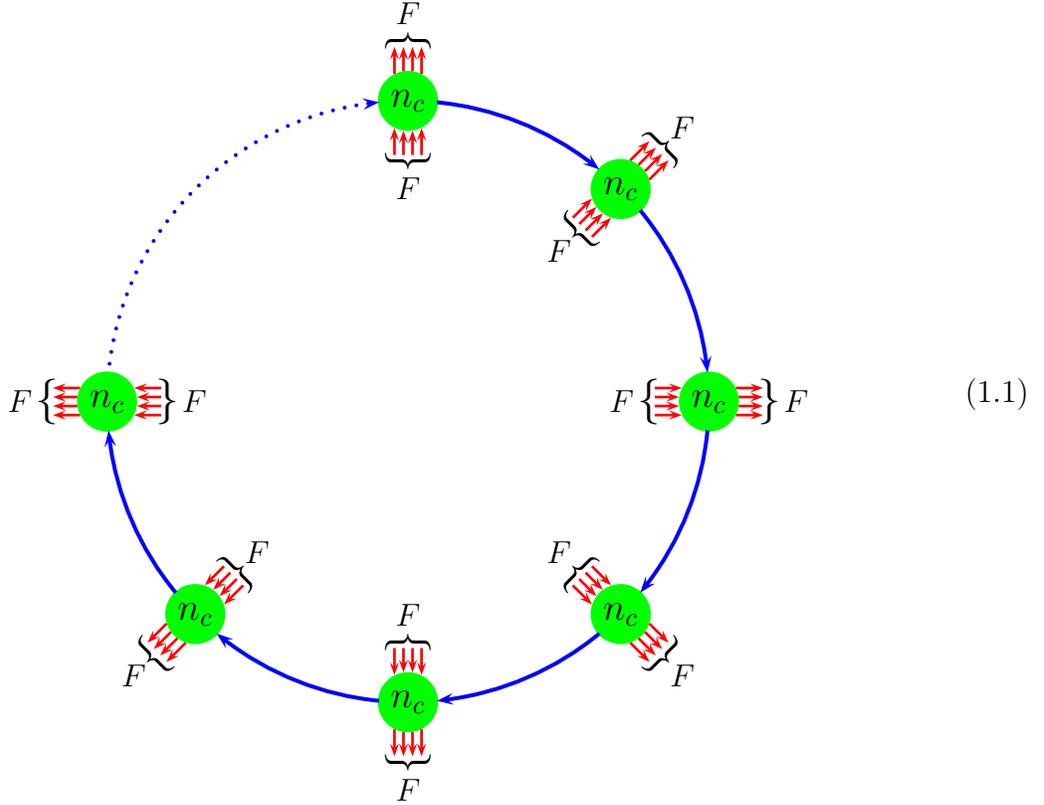
\be
\psset{unit=4mm,linecolor=black}
\def\site{%
    \pscircle*[linecolor=green]{1}\relax
    \rput{*0}(0,0){\large $n_c$}\relax
    \psline{->}(-1.8,+0.45)(-0.9,+0.45)\relax
    \psline{->}(-1.8,-0.45)(-0.9,-0.45)\relax
    \psline{->}(-1.8,-0.15)(-0.99,-0.15)\relax
    \psline{->}(-1.8,+0.15)(-0.99,+0.15)\relax
    \psline{<-}(+1.8,+0.45)(+0.9,+0.45)\relax
    \psline{<-}(+1.8,-0.45)(+0.9,-0.45)\relax
    \psline{<-}(+1.8,-0.15)(+0.99,-0.15)\relax
    \psline{<-}(+1.8,+0.15)(+0.99,+0.15)\relax
    \rput(-2,0){\Large\{}\relax
    \rput(+2,0){\Large\}}\relax
    \rput{*0}(-2.9,0){$F$}\relax
    \rput{*0}(+2.9,0){$F$}\relax
    }
\begin{pspicture}[](-14,-14)(+14,+14)
    \psset{linewidth=1pt,linecolor=red}
    \rput{0}(+10,0){\site}
    \rput{45}(+7.07,+7.07){\site}
    \rput{315}(+7.07,-7.07){\site}
    \rput{90}(0,+10){\site}
    \rput{270}(0,-10){\site}
    \rput{225}(-7.07,-7.07){\site}
    \rput{180}(-10,0){\site}
    \psset{linewidth=1.5pt,linecolor=blue}
    \psarc{<-}{10}{50.5}{84.5}
    \psarc{<-}{10}{5.5}{39.5}
    \psarc{<-}{10}{320.5}{354.5}
    \psarc{<-}{10}{275.5}{309.5}
    \psarc{<-}{10}{230.5}{264.5}
    \psarc{<-}{10}{185.5}{219.5}
    \psarc[linestyle=dotted]{<-}{10}{95.5}{174.5}
\end{pspicture}
\label{quiver}
\ee
Each  green circle of this diagram corresponds to
a simple $SU(n_c)_\ell$ factor of the net 4D gauge group
\be
G_{\rm 4D}\ =\  \prod_{\ell=1}^N [SU(n_c)]_{\,\ell}
\ee
while the red and blue arrows denote the chiral superfields:
\be
\psset{unit=0.6mm,linecolor=red,linewidth=1pt}
\eqalign{
  \vcenter{\hbox to 9mm{%
	\psline{->}(0,6)(10,6)
	\psline{->}(0,4)(10,4)
	\psline{->}(0,2)(10,2)
	\psline{->}(0,0)(10,0)
	\hfil\Large\rm\}%
      	}}\,
   \mbox{quarks}\ Q_{\ell,f}\ &
  =\ ( {\bf\square}_{\,\ell} ),\cr
  \vcenter{\hbox to 9mm{%
	\psline{<-}(0,6)(10,6)
	\psline{<-}(0,4)(10,4)
	\psline{<-}(0,2)(10,2)
	\psline{<-}(0,0)(10,0)
	\hfil\Large\rm\}%
      	}}\,
    \mbox{antiquarks}\ \aq^f_\ell\ &
  =\ ( \overline{\bf\square}_{\,\ell} ),\cr
    \vcenter{\hbox to 25mm{%
	\psline[linecolor=blue,linewidth=1.5pt]{->}(0,0)(40,0)
	}}\,
    \mbox{bifundamental~link~fields}\ \link^{}_\ell\ &
  =\ (\square_{\,\ell+1}, \overline\square_{\,\ell}),\cr
}\label{Fields}
\ee
where $f=1,2,\ldots,F$ and $\ell=1,2,\ldots,N$ is understood modulo $N$.
{}From the 4D point of view, $N$ is a fixed parameter of the
quiver theory; in our analysis we shall assume $N$ to be large but finite.

Similar to many other deconstructed theories, the quiver~(\ref{quiver})
can be obtained by orbifolding a simple 4D gauge theory with higher SUSY,
namely $\NN=2$ SQCD with $F$ flavors and $N\times n_c$ colors:
The $\ZZ_N$ twist removes the extra supercharges and reduces the gauge
symmetry from $SU(N\times n_c)$ down to
\be
S([U(n_c)]^N)\ =\ [SU(n_c)]^N\times [U(1)]^{N-1}.
\label{orbigroup}
\ee
However, the abelian photons of the orbifold theory suffer from
triangular anomalies and therefore must be removed from the
effective low-energy theory.
In string theory such removal is usually accomplished via the
Green--Schwarz terms \cite{GreenSchwarz},
but at the field theory level we simply discard the abelian factors
of the orbifolded symmetry~(\ref{orbigroup}) and interpret the nodes
(green circles)
of the quiver diagram~(\ref{quiver}) as purely non-abelian
$SU(n_c)_\ell^{}$ factors.

In this article we study chiral rings of deconstructed \sqcdv\ theories
with generic numbers of colors and flavors, Chern--Simons levels, and
sizes of the compact fifth dimension.
In 4D terms, this means  \suq\ theories with quiver diagrams
of the general form~(\ref{quiver}), but with most general numbers
$n_c$, $F$, $N$, as well as generic quark masses; for the sake of
4D chirality we assume $N\ge3$ quiver nodes.
We shall see that the chiral rings of such theories resemble
the rings of refs.\ \cite{CDSW,CSW1,Seiberg,CSW2}
but also have two novel features:
First, our rings have meson-like generators involving the quark and
the antiquark fields belonging to different nodes of the quiver.
Such operators are non-local from the 5D point of view, but in 4D
terms they are legitimate generators of the chiral ring.
Second, in the absence of abelian gauge fields, all kinds of baryonic
and antibaryonic operators are gauge-invariant and thus also belong
to the chiral ring; in fact, there is a whole zoo of such operators.

The main part of this paper (after this introduction)
is organized as follows:
In section \S2 we study the deconstructive \suq\ quiver theories
at the classical level.
First, in \S2.1 we spell out the superpotentials and explain how
the 4D quiver theories deconstruct the 5D~SQCD.
Next, in \S2.2 we survey the classical vacua of the quiver theories and find
that they form the same Coulomb, mesonic, and baryonic moduli spaces as
the 5D theories compactified on a large circle.
And then in \S2.3 we break the correspondence by deforming the 4D
superpotentials in order to trigger the gaugino condensation at the
quantum level.
This deformation is similar to the adjoint field's superpotential
in the single--$U(n_c)$ theory and has similar consequences for the
classical vacua of a quiver theory:
The Coulomb moduli space  breaks up into a large discrete set of
isolated vacua.

In section \S3 we study the fully-quantum \suq\ quiver theories and
derive their chiral rings, or rather the non-baryonic sub-rings.
In \S3.1 we construct the off-shell chiral rings (un-constrained by the
anomalous equations of motion).
Similar to the single--$U(n_c)$ theory of \cite{CDSW,CSW1,Seiberg,CSW2},
the non-baryonic generators of the quivers' off-shell rings combine
into several resolvents where the cyclic-ordered product
$\link_N\link_{N-1}\cdots\link_2\link_1$ of the bi-fundamental
link fields plays the role of the adjoint field $\Phi$.
The main difference from the single--$U(n_c)$ theory is a much richer
set of mesonic generators and hence resolvents:
In a quiver theory, the quark and the antiquark fields of a meson-like
chiral operator may belong to different quiver nodes as long as they
are connected by a chain of link fields which maintain the gauge invariance.
From the 5D point of view such operators are non-local and they create /
annihilate un-bound $\bar qq$ pairs, but in 4D they are local, chiral
and gauge invariant and thus do belong to the chiral ring.

In \S3.2 we turn to the on-shell chiral rings:
We calculate the generalized Konishi anomalies for suitable variations
of the quark, antiquark and link fields of the quiver theories and
derive the anomalous equations of motions for all the resolvents.
We solve the equations in terms of a few polynomials and find that
all the on-shell resolvents are single-valued on the same hyperelliptic
Riemann surface $\bf\Sigma$ defined by the quadratic equation~(\ref{Rreseq})
for the gaugino-bilinear resolvent.
Physically, $\bf\Sigma$ is the Seiberg--Witten curve \cite{SWcurve}
of the theory, and in \S3.3 we use analytic considerations to show
that it indeed looks like the SW curve of the \sqcdv\ compactified
on a circle \cite{HananyOz,ArgyresPlesserShapere}.
We also find that this curve is completely determined at the level of
{\sl one diagonal instanton}, that is one instanton of the
$SU(n_c)_{\rm diag}\equiv\diag[SU(n_c)^N]$ or equivalently one instanton
in each and every $SU(n_c)_\ell$ factor of the total 4D gauge group.
Finally, we study the vacua of the quantum quiver theories and show that
in the weak coupling limit we have exactly the same Coulomb, Higgs,
pseudo-confining, {\it etc.}, vacua as expected in a semisclassical theory,
but in the strong coupling regime all vacua with similar numbers of
massless photons are interchangeable by the monodromies in the parameter
space of the theory.

In section \S4 we complete the chiral rings by adding all kinds of
baryonic, antibaryonic and other generators with non-trivial
$[U(1)_B]^N$ quantum numbers.
In \S4.1 we warm up by studying baryon-like generators in a theory
with a single $SU(n_c)$ gauge group, $n_f$ quarks and antiquarks and
an adjoint field $\Phi$:
Because the gauge group is $SU(n_c)$ rather than $U(n_c)$, the chiral
baryon operators are gauge invariant, and so are the $\Phi$--baryons
comprised of $n_c$ quarks plus any number of adjoint fields $\Phi$.
Off-shell, such operators exist for any $n_f\ge 1$, and we summarize
them in baryonic resolvents.
On-shell however, the $\Phi$--baryons follow from the ordinary baryons,
hence no baryonic VEVs whatsoever for $n_f<n_c$, and even for $n_f\ge n_c$
baryonic VEVs exist only for the classical-like branches of the moduli
space.

In \S4.2 we analyze the baryonic generators of the \suq\ quiver theory.
Off-shell, we find a whole zoo of baryon-like generators comprised of
$n_c$ quarks belonging to different quiver nodes and connected to
each other by chains of link fields --- and each chain may wrap a few
times around the whole quiver to emulate the $\Phi$ fields of $\Phi$--baryons.
Again, there is a big pile of independent baryon-like operators for
any $n_f\ge 1$, but only off-shell.
On-shell, we tame the zoo by solving the equations of motions for the baryonic
resolvents and showing that all baryon-like VEVs follow from those
of ordinary baryonic operators
(all quarks at the same node and no link fields).
Consequently, all baryonic VEVs of the quantum theory follow the classical
rules:  They require $n_f\ge n_c$ as well as overdetermined Coulomb moduli
of the baryonic branch.
However, the precise constraint on the quark masses due to baryonic branch's
existence is subject to quantum corrections at the one-diagonal-instanton
level.

In \S4.3 we calculate quantum corrections to the determinants
of link chains,
\be
\det\Bigl(\link_{\ell_2}^{}\cdots\link_{\ell_1}^{}\Bigr)\
=\ \det(\link_{\ell_2}^{})\times\cdots\times\det(\link_{\ell_2}^{})\
+\ \rm corrections.
\ee
We find that for chains of less than $N$ links the corrections come from
instantons in the individual $SU(n_c)_\ell$ gauge groups rather than
the diagonal instantons, but the determinant $\det(\link_N\cdots\link_1)$
of the whole quiver is subject to separate individual-instanton
and diagonal-instanton corrections.
We evaluate the corrections and summarize their effect on the Coulomb
moduli space of a deconstructive quiver theory.
A particularly technical part of our analysis is removed to the Appendix
of this paper.

Finally, in section \S5 we discuss the open questions related to the
present research.

 
%
%
\section{Deconstructive Quiver Theories \brk and their Classical Vacua}
Deconstruction of \sqcdv\ with general numbers of colors and flavors
will be explained in much detail in the companion paper~\cite{dNK}.
In this section, we summarize the salient features of the deconstructed
theory from the 4D point of view.
In the immediately following \S2.1 we write down the superpotential of
the 4D theory and briefly explain how deconstruction works
for the most symmetric vacuum of the 5D theory.
Of course, the \suq\ quiver theory has many other classical vacua,
and we describe them in \S2.2.
Finally, in \S2.3 we deform the quiver's superpotential in order to
trigger some kind of a gaugino condensation and describe the effects of this
deformation on the quiver's vacua.
(The gaugino condensates themselves will be discussed in the later section~\S3).

%
\subsection{Deconstruction Summary}
The 4D gauge theory of the deconstructed \sqcdv\ is
\be
G_{\rm 4D}\ =\ \prod_{\ell=1}^N [SU(n_c)]_{\,\ell}
\label{G4SU}
\ee
with equal gauge couplings $g_\ell\equiv g$ for all the factors
to assure discrete translation invariance in the $x^4$ direction.
The chiral superfields comprise the quarks, the antiquarks,
and the bilinear link fields specified in the table~(\ref{Fields}),
and also singlets $s_\ell$ (one for each $\ell=1,2,\ldots,N$).
The superpotential has two distinct parts,
$W=W_{\rm hop}+W_\Sigma$, where
\be
W_{\rm hop}\
=\ \gamma\sum_{\ell=1}^N\sum_{f=1}^{F}\Bigl(
	\aq^f_{\ell+1} \link^{}_\ell Q_{\ell,f}\
	-\ \mu_f\, \aq^f_\ell Q_{\ell,f}
	\Bigr)
\label{Whop}
\ee
facilitates the propagation of the (anti)~quark fields
in the $x^4$ direction, while
\be
W_\Sigma\ =\ \beta\sum_{\ell=1}^N
s_\ell\,\left(\det \link_\ell\,-\,v^{n_c}\right)
\label{Wsigma}
\ee
sets up an $SL(n_c,\CC)$ (AKA complexified $SU(n_c)$) linear sigma model
at each link of the latticized 5D theory.%
\footnote{%
	Although the $\beta$ coupling is non-renormalizable for $n_c>2$,
	all the resulting divergences can be regularized via
	higher-derivative Lagrangian terms for the singlet fields~$s_\ell$
	without disturbing the chiral ring of the theory.
	}
Disregarding the massive ``radial'' mode, we have
\be
\link_\ell(x)\ =\ v\times
\mathop{\hbox{\Large\rm exp}}\limits_{\rm Path\atop ordered}\left(
	 \int\limits_{a\ell}^{a(\ell+1)}\!\!\!dx^4
	 \Bigl(iA_4(x)+\phi(x)\Bigr)\right)\
+\ \mbox{fermionic terms}
\label{Dmap}
\ee
where $A_\mu(x)$ and $\phi(x)$ are the 5D vector fields and their scalar
superpartners.
The simplest 5D vacuum with $\phi\equiv A^4\equiv 0$ corresponds
to the 4D field configuration
\be
\vev{\link_\ell}\ \equiv\ v\times\hbox{\Large 1}_{n_c\times n_c}
\label{SymVacSU}
\ee
which Higgses the 4D gauge symmetry \suq{} down to its diagonal subgroup
$SU(n_c)_{\rm diag}=\diag\left(\prod_\ell SU(n_c)_\ell\right)$.
The rest of the 4D vectors acquire masses
\be
M^2(k)\ =\ 4g^2|v|^2\,\sin^2\frac{\pi k}{N}\
=\ \frac{4}{a^2}\,\sin^2\frac{aP_4}{2}
\label{VectorMasses}
\ee
where the second equality follows from identifying the lattice spacing $a$ as
\be
a\ =\ \frac{1}{g|v|}
\label{Vspacing}
\ee
and the lattice momentum $P_4$ as
\be
P_4\ =\ \frac{2\pi k}{Na}\,,\qquad k=1,2,\ldots\ \mbox{modulo}\ N.
\ee
In the  large $N$ limit, the bottom end of
the spectrum~(\ref{VectorMasses}) becomes a Kaluza--Klein tower
\be
M^2\ \approx\ P_4^2\ =\ \left(\frac{2\pi k}{Na}\right)^2
\ee
of a massless relativistic 5D vector field
(compactified on a circle of length $2\pi R=Na$);
this is the momentum-space view of the dimensional deconstruction.

Similarly, the low-energy end of the 4D quark spectrum comprises
Kaluza--Klein towers
for relativistic 5D hypermultiplets of masses $m_f\ll(1/a)$.
Indeed, for a quark flavor of 4D mass $\gamma\mu_f$,
the mass matrix for the $Q_{\ell,f}$ and $\aq^f_\ell$ fields
($\ell=1,2,\ldots,N$) has
eigenvalues
\be
M^2(f,k)\ =\ |\gamma|^2\,\left| v e^{2\pi ik/N}\,-\,\mu_f \right|^2,
\ee
and {\blue for $|\mu_f|\approx|v|$},
the bottom end of this spectrum becomes (in the large $N$ limit)
\be
M^2\ \approx\ m_{\rm 5D}^2\ +\ P_4^2
\label{QuarkMasses}
\ee
where
\bea
m_{\rm 5D} &=& |\gamma|\times\Bigl(|\mu_f|\,-\,|v|\Bigr)\
	\ll\ \frac{1}{a}\,, \label{QuarkMass}\\
P_4 &=& \frac{2\pi k}{Na}\ +\ \mbox{constant (Wilson line)}, \\
\mbox{and}\ a &=& \frac{1}{|\gamma v|}\,.
\label{Qspacing}
\eea
The 4D quarks with $|\mu_f|\not\approx|v|$
do not have low-mass $M^2\ll(1/a)^2$ modes and  do not deconstruct
any 5D particles.
For $\mu_f\gg v$, the 4D quark decouples above the deconstruction
threshold $(1/a)$ and has no low-energy effect whatsoever, but
quarks with $\mu_f\ll v$ decouple at the threshold itself and
modify  the Chern--Simons interactions  of the deconstructed~\sqcdv.
Since the Chern--Simons level~$\kcs$ affects the moduli space geometry
and even the phase structure of the 5D theory,
it must be deconstructed correctly.
Thus, the deconstructive quiver should have
\be
F\ =\ n_f\ +\ \Delta F\ \ge\ n_f
\label{Nflavors}
\ee
4D flavors, where $n_f$ is the number flavors in 5D,
\be
\Delta F\ =\ n_c\ -\ \frac{n_f}{2}\ -\ \kcs\,,\qquad
0\,\le\,\Delta F\,\le\,2n_c\,-\,n_f\,,
\label{DeltaDef}
\ee
and
\be
\mu_f\ =\ v\times\exp(a m_f^{\rm 5D})\ \sim v\quad
\mbox{for}\ f\le n_f\,\quad
\mbox{but}\ \mu_f\ \ll\ v\quad\mbox{for}\ f>n_f\,.
\ee
However, from the holomorphic, purely-4D point of view,
there is no qualitative difference between $\mu_f\sim v$
and $\mu_f\ll v$, but there is a difference between $\mu_f\neq 0$
and between $\mu_f=0$.
Hence, for the purposes of this article, we shall assume
\be
\mbox{generic}\ \mu_f\ \neq\ 0\quad\mbox{for}\ f=1,2,\ldots,n_f\quad
\mbox{but}\ \mu_f\ =\ 0\quad\mbox{for}\ f=(n_f+1),\ldots,F\,.
\label{masses4D}
\ee

Note that consistency between eqs.~(\ref{Vspacing}) and~(\ref{Qspacing})
requires equal gauge and Yukawa couplings, $g=|\gamma|$.
In a quantum theory, this means equality of the renormalized physical
couplings,
\be
g^{\rm phys}\ =\ |\gamma|^{\rm phys},
\label{Ggamma}
\ee
or in non-perturbative terms, in the very low energy limit $E\ll (1/Na)$
the effective theory
(the diagonal $SU(n_c)$  with an adjoint field $\Phi$ and several
quark flavors)
should be $\NN=2$  supersymmetric.
Without this condition, the deconstructed theory would have quarks
and gluons with different effective speed of light in the $x^4$ direction.
This is a common problem in lattice theories with some continuous dimensions
({\it eg.} Hamiltonian lattice theories with continuous time but
discrete space),
and the common solution is fine-tuning of the lattice parameters.
For the deconstructed \sqcdv, the fine tuning involves the K\"ahler parameters
(such as coefficients of the quarks', antiquarks' and links'
 kinetic-energy Lagrangian terms)
and does not affect any of the holomorphic properties of the quiver
such as its chiral ring.
Consequently, {\sl in the present article} we may disregard \eq{Ggamma}
and treat the holomorphic $\gamma$ and the gauge coupling $g$ (or rather
its dimensional transmutant $\Lambda$) as free parameters of
the quiver theory.

We conclude this section by acknowledging that there are many ways
to skin a cat or to deconstruct \sqcdv\ with given $n_c$, $n_f$ and $\kcs$.
For example, one can pack several 5D quark flavors into a single 4D
flavor with a complicated dispersion relation $M_{\rm 4D}(P_4)$
by generalizing the hopping superpotential~(\ref{Whop}) to
allow the quark to hop over several lattice spacing at once.
Indeed,
\be
W_{\rm hop}\ =\ \sum_{q=0}^p\Gamma_p\sum_{\ell=1}^N
\aq_{\ell+q}\link_{\ell+q-1}\cdots\link_{\ell} Q_{\ell}
\label{LongHop}
\ee
endows a single 4D flavor with $p$ light modes when the polynomial
\be
H_p(x)\ =\,\sum_{q=0}^p \Gamma_p x^p
\ee
has all $p$ of its roots $\mu_1,\ldots,\mu_p$ located close to the circle
$|x|=|v|$.
We found however that such $p$-fold quarks are pretty much
equivalent to $p$ ordinary flavors with masses $\mu_1,\ldots,\mu_p$.
Consequently, we shall henceforth stick to the $p=1$ model~(\ref{Whop})
because of its twin virtues of relative simplicity and renormalizability.

%
\subsection{The Classical Vacua.}
As explained in \cite{dNK}, the classical moduli space of the
\suq{} quiver theory has exactly the same
Coulomb, Higgs and mixed branches as the un-deconstructed~\sqcdv.
The Coulomb branch is distinguished by zero classical values of the
quark and antiquark scalars while the link fields have non-zero VEVs
subject to D/F term constraints
\be
\forall\ell:\quad
\vev{\link_\ell}^\dagger\vev{\link_\ell}\
-\ \vev{\link_{\ell-1}}\vev{\link_{\ell-1}}^\dagger\
\propto\ \hbox{\Large 1}_{n_c\times n_c}\,,\quad
\Det\vev{\link_\ell}\ =\ v^{n_c} .
\ee
Consequently, all the $\vev{\link_\ell}$ matrices are equal and diagonal
modulo an $\ell$-dependent gauge transform:
\be
\forall\ell:\ \vev{\link_\ell}\
=\ \diag\left(\ev_1,\ev_2,\ldots,\ev_{n_c}\right)
\label{CoulombVEV}
\ee
for some complex moduli
$(\ev_1,\ev_2,\ldots,\ev_{n_c})$ satisfying $\prod_j\ev_j=v^{n_c}$.
Note that each $\ev_j$ is gauge-equivalent to $\ev_j\times\root N\of{1}$,
hence the $(\ev_1^N,\ev_2^N,\ldots,\ev_{n_c}^N)$ makes a better coordinate
system for the moduli space, although it's still redundant with respect
to permutations of the $\ev_j^N$.
Generically, all the $\ev_j^N$ are distinct and the 4D gauge symmetry is 
broken all the way down to the Cartan $(U(1))^{n_c-1}$ subgroup of the
$SU(n_c)_{\rm diag}=\diag\Bigl[\prod_\ell SU(n_c)_\ell\Bigr]$,
but a non-abelian subgroup  $SU(k)\subset SU(n_c)_{\rm diag}$ survives
un-Higgsed when $k$ of the $\ev_j^N$ happen to coincide.

According to the deconstruction map~(\ref{Dmap}),
\be
\ev_j\ =\ v\times\exp\bigl(a(\phi_j+iA^4_j)\bigr)
\ee
where $\phi_j$ are the real 5D moduli scalars and $A^4_j\times Na$
are the  Wilson lines of the diagonal gauge fields
around the deconstructed dimension.
Of course, the deconstruction works only for
$\phi_j\ll(1/a)\ \Longrightarrow\ \ev_j\sim v$, but
this restriction does not affect the 4D theory as such.

Unlike the Coulomb branch which exists for any quark masses,
the Higgs and the mixed branches
require coincidences between the $\mu_f$ or rather
among the non-zero $\mu_f^N$.
For example, for $\mu_1^N=\mu_2^N\neq 0$ there is a mixed mesonic branch
where one of the $\ev_j^N$ is frozen at the same value.
Indeed, let
\be
\ev_1\ =\ e^{2\pi i k_1/N}\,\mu_1\ =\ e^{2\pi i k_2/N}\,\mu_2
\label{MesonicOrigin}
\ee
for some integer $(k_1,k_2)$; then the quark mass matrix due to
the superpotential~(\ref{Whop}) allows for the squark VEVs
\be
\vev{Q^j_{\ell,f}}\ =\ e^{2\pi i k_f\ell/N} Q_f,\qquad
\vev{\aq^f_{\ell,j}}\ =\ e^{-2\pi i k_f\ell/N} \aq^f\quad
\mbox{for}\ j=1\ \&\ f=1,2\ \mbox{only},
\label{SquarkVEVs}
\ee
subject to F-term and D-term constraints
\bea
Q_1\,\aq^1\ +\ Q_2\,\aq^2 &=& 0, \label{FtermMesonic} \\
(|Q_1|^2\,-\,|\aq^1|^2)\ +\ (|Q_2|^2\,-\,|\aq^2|^2) &=& 0.
\label{DtermMesonic}
\eea
These VEVs Higgs the $(SU(n_c))^N$ symmetry down to
$(SU(n_c-1))^N$,
which is further broken by the link VEVs $\vev{\link_\ell}$
down to a subgroup of the $SU(n_c-1)_{\rm diag}=
 \diag\Bigl[\prod_\ell SU(n_c-1)_\ell\Bigr]$.
For generic values of the un-frozen Coulomb moduli
$(\ev_2^N,\ldots,\ev_{n_c}^N)$,
the surviving gauge symmetry is $U(1)^{n_c-2}$, but coincidences
among these moduli allow for un-Higgsing of a non-abelian
$SU(k)\subset SU(n_c-1)_{\rm diag}$.

The mesonic  branch of the quiver deconstructs the mesonic
 branch of the \sqcdv\ where $\phi_1=m_1^{\rm 5D}=m_2^{\rm 5D}$.
Although the deconstruction requires $\phi_j\ll(1/a)$
and hence $m_{1,2}^{\rm 5D}\ll(1/a)$,
this restriction does not affect the 4D theory as such.
In 4D, a coincidence $\mu_1^N=\mu_2^N$ gives rise to a
a mesonic branch regardless of whether $\mu_{1,2}$ is larger,
smaller, or similar to $v$, {\it as long as} $\mu_{1,2}^N\neq 0$.
On the other hand, having  two or more exactly massless
4D flavors ($i.\,e.,\ \Delta F\ge 2$)
does not lead to a mesonic  branch of the \suq{}
quiver because the link eigenvalues~$\ev_j$ cannot vanish.
(Note the constraint $\det(\link_\ell)=v^{n_c}\neq 0$.)

Further coincidences among the $\mu_f^N$ allow for multi-mesonic mixed
branches with more squark VEVs, more frozen Coulomb moduli $\ev_j^N$
({\it eg.}, $\ev_1^N=\mu_1^N=\mu_2^N,\ \ev_2^N=\mu_3^N=\mu_4^N$),
and a lower rank of the un-Higgsed gauge symmetry.
Such multi-mesonic  branches work similarly to the single-meson mixed
branch, so we need not discuss them any further.
Instead, let us consider the purely-Higgs baryonic branch which exists for
$n_f\ge n_c$ when $n_c$ of the 5D quark masses add to zero,
or in 4D terms, when the product of $n_c$ of the $\mu_f^N$
happens to equal to the $(v^{n_c})^N$.
Indeed, for $\mu^N_1\times\mu^N_2\times\cdots\times\mu^N_{n^{}_c}=v^{Nn_c}$
we may freeze {\sl all} of the Coulomb moduli at
\be
\ev_j\ =\ e^{2\pi ik_j/N}\,\mu_j\quad\forall j=1,2,\ldots,n_c\,,
\quad k_j\in\ZZ,
\ee
which gives zero modes to all quark colors $j=1,\ldots,n_c$ for
$f=j$ and allows non-zero VEVs
\be
\vev{Q^j_{\ell,f}}\ =\ \delta^j_f e^{2\pi i k_j\ell/N} Q^j,\qquad
\vev{\aq^f_{\ell,j}}\ =\ \delta_j^f\, e^{-2\pi i k_j\ell/N}\,\aq_j
\ee
subject to the D term constraint
\be
\mbox{same}\ (|Q^j|^2\,-\,|\aq_j|^2)\ \forall j
\label{DtermBaryonic}
\ee
and the F-term constraint
\be
\frac{\partial W}{\partial\link^j_{\ell,j}}\
=\ \gamma e^{-2\pi ik_j/N}\,Q^j\aq_j\
+\ \beta s_\ell\times\frac{v^{n_c}}{\ev_j}\
=\ 0.
\label{FtermBaryonic}
\ee
The simplest solutions to these constraints are either $\aq_j\equiv0$, same
$Q^j\equiv Q\ \forall j$ (baryonic VEVs only) or {\it vice verse} $Q^j\equiv0$,
same $\aq_j\equiv\aq\ \forall j$ (antibaryonic VEVs only),
but thanks to the singlet fields $s_\ell$ enforcing
the $\det\link_\ell=v^{n_c}$ constraints, there are other solutions
where both baryonic and antibaryonic VEVs are present at the same time.
In the deconstruction limit $\phi_j=m_j^{\rm 5D}\ll(1/a)\ \Longrightarrow\
\ev_j\approx v$ (up to a phase) in \eq{FtermBaryonic},
we have $Q^j\equiv Q,\ \aq_j\equiv\aq$ and hence
\be
\vev{Q^j_{\ell,f}}\ =\ \delta^j_f e^{2\pi i k_j\ell/N}\times Q,\qquad
\vev{\aq^f_{\ell,j}}\ =\ \delta_j^f e^{-2\pi i k_j\ell/N}\times\aq
\ee
for some arbitrary pair $(Q,\aq)$ of complex moduli which deconstruct
the baryonic hyper-modulus of the \sqcdv.
Outside the deconstruction limit, the 4D baryonic branch exists anyway,
albeit with more complicated anti/squark VEVs.
In any case, there are two complex moduli and
the \suq{} gauge symmetry is completely Higgsed down.

This completes our survey of the classical moduli space of the
\suq{} quiver.
The bottom line is, all the classical vacua of this quiver are
deconstructive, $i.\,e.$ correspond to the \sqcdv's vacua according
to the deconstruction map~(\ref{Dmap}) and the zero modes of the
massless 5D gauge bosons match all the massless 4D vector fields.

%
\subsection{Deforming the Superpotential.}
In the extreme infrared limit $E\ll(1/Na)$, the $\NN=1$ quiver
theory reduces to the $\NN=2$ SQCD$_4$ with several flavors, and
thanks to this extra supersymmetry, the gauginos do not form
bilinear condensates.
According to Cachazo, Douglas, Seiberg and Witten
\cite{CDSW,Seiberg,CSW1,CSW2},
the gaugino condensates play a key role in the chiral ring
of the $U(n_c)$ theory with an adjoint chiral field~$\Phi$.
Indeed, the best way for understanding the $\NN=2$ SQCD$_4$
involves deforming the theory to $\NN=1$ (via superpotential
$\tr \W(\Phi)$ for the adjoint field) in order to turn on
the gaugino condensation, although eventually,
{\em after} solving the anomaly equations of the chiral
ring of the deformed theory (including the chiral gaugino 
condensates) one may turn off the deformation and return to
$\NN=2$ SUSY. 

In the quiver theory, the role of the adjoint field $\Phi$
is played by the quiver-ordered product
$(\link_1\link_2\cdots\link_N)$ of the link fields.
Hence, to study the chiral ring of the quiver, we need to
temporarily deform the superpotential according to
\be
W\ =\  W_{\rm hop}\ +\ W_\Sigma\ \longrightarrow\
W\ =\ \ W_{\rm hop}\ +\ W_\Sigma\ +\ W_{\rm def}
\label{Wdeformation}
\ee
where
\be
W_{\rm def}\ =\ \tr\W(\link_N\link_{N-1}\cdots\link_2\link_1)\
\eqdef\,\sum_{k=1}^d\frac{\nu_k}{k}\,
\tr\left(\left(\link_N\link_{N-1}\cdots\link_2\link_1\right)^k\right).
\label{Wdef}
\ee
Although this deformation does not make any sense from the
5D point of view --- indeed, it is utterly non-local in the
$x^4$ direction --- as well as grossly non-renormalizable in 4D,
it does lead to non-zero gaugino condensates which will help us
later in~\S3.
But before we study such non-perturbative effects, we need
to know the effect of the deformation~(\ref{Wdeformation})
on the classical vacua of the theory.

The general effect is similar to the deformed $\NN=2$ SQCD$_4$.
The Coulomb branch collapses to a discrete set of isolated vacua
where each Coulomb modulus $\ev_j^N$ takes one of $d$ possible values
$(\rt_1,\rt_2,\ldots,\rt_d)$, namely the roots of the polynomial
\be
\WW(X)\ =\,\sum_{k=1}^d \nu_k X^k\ +\ \beta \vev{s}v^{n_c}
\label{polydef}
\ee
where $\vev{s}\equiv\vev{s_\ell}$ is the common expectation value
of the singlet fields $s_\ell$ which adjusts itself to assure
$\prod_j\ev_j=v^{n_c}$.
Indeed, given the Coulomb VEVs (\ref{CoulombVEV}),
\be
\frac{\partial W}{\partial\ev_j}\ =\ \frac{N}{\ev_j}\,\WW(\ev_j^N)\
\Longrightarrow\ \forall j:\ \WW(\ev_j^N)\ =\ 0.
\ee
Note that each root $\rt_i$ of the $\WW$ polynomial may capture
several moduli $\ev_j^N$, or just one, or even none at all.
The individual Coulomb vacua of the quiver are distinguished by
the `occupation numbers' $n_i=\#\{j:\ev_j^N=\rt_i\}=0,1,2,\ldots$
for $i=1,\ldots d$; altogether, $\sum_i n_i=n_c$.
For any given set of the $n_i$,
the surviving gauge symmetry of the \suq\ quiver theory is
\be
G_{\rm unbroken}\ =\
S\left[\prod_{i} U(n_i)\right]\ \subset\ SU(n_c)_{\rm diag}
\label{SurvivingSymmetrySU}
\ee
where each $n_i\ge 2$ gives rise to a nonabelian factor $SU(n_i)$.
In the quantum theory, such nonabelian factors develop mass gaps
due to pseudo-confinement\footnote{%
	Following the terminology of ref.~\cite{CSW2}
	we call an $SU(n_i)$ gauge factor pseudo-confining
	rather than confining because the overall quiver theory
	contains fields with anti/fundamental quantum numbers
	(with respect to the $SU(n_i)$) which prevent the
	complete confinement.
	In this terminology, the ordinary QCD with finite-mass
	quarks is also pseudo-confining rather than confining.
	}
and the ultimate low-energy limit of the theory has only the
abelian photons.
The net number of such surviving photons is
\be
n_{\rm Abel}\ =\ \#\{i:n_i\neq 0\}\ -\ 1\ \le\ n_c\,-\,1.
\label{PhotonNumberSU}
\ee

Besides the purely--Coulomb vacua, the deformed quiver also has
discrete mesonic vacua where the squark VEVs are determined by the
deformation~(\ref{Wdef}).
In such vacua, some of the Coulomb moduli $\ev_j^N$
are frozen at non-zero, non-degenerate  $\mu_f^N$
while the rest  follow the roots $\rt_i$ of the
deformation polynomial~(\ref{polydef}).
For example, let $\ev_1^N=\mu_1^N\neq 0$ while
the $(\ev_2^N,\ldots,\ev_{n_c}^N)$ are captured by the $\rt_i$.
Then at this point, the D/F term constraints {\em require}
non-zero squark and antisquark VEVs for $j=f=1$ only:
\be
\vev{Q^1_{\ell,1}}\ =\ e^{2\pi ik\ell/N}Q,\quad
\vev{\aq^1_{\ell,1}}\ =\ e^{-2\pi ik\ell/N}\aq,\quad
|Q|^2\,=\,|\aq|^2,\quad
Q\aq\ =\ -\frac{\WW(\mu_1^N)}{\gamma\mu_1}\ \neq 0.
\label{HiggsVac}
\ee
In this vacuum $\sum_i n_i=n_c-1<n_c$, which reduces the unbroken
gauge symmetry according to \eq{SurvivingSymmetrySU}.
Likewise, we may have fixed squark VEVs for several distinct
$(j,f)$ pairs when the corresponding moduli $\ev_j^N$
are trapped at the $\mu_f^N\neq 0$ instead of the roots $\rt_i$;
this results in even lower $\sum_i n_i\le n_c-2$ and hence further
Higgsing down of the gauge symmetry.

In the quantum theory, the Higgs mechanism is complementary to
pseudo-confinement and the two types of vacua are
continuously connected in the overall parameter/moduli space
of the theory.
The way this duality works in the deformed $\NN=2$ SQCD$_4$
is explained in detail in \cite{CSW1,CSW2}, and the same arguments
apply to the deconstructive quiver theories under discussion.
The bottom line is, all the Higgs and the pseudo-confining Coulomb
vacua (of the same theory) which have the same
{\em abelian rank}~$n_{\rm Abel}$
are continuously connected to each other in the quantum quiver theory.
The purely-abelian Coulomb vacua with no non-abelian factors at all
form a separate class because they have higher abelian rank
than any Higgs or pseudo-confining vacuum.

Eventually, we shall un-deform the quiver theory by taking the
limit $\nu_k\to 0$ in the deformation superpotential~(\ref{Wdef}).
We should be careful to maintain finite roots $\rt_i$ with $n_i>0$
and to allow them to move all over the complex plane
(subject to the constraint $\prod_i \rt_i^{n_i}\ =\ v^{Nn_c}$ in the \suq\ case)
while the overall scale of the polynomial $\WW(X)$ diminishes away.
In this limit, the purely-Coulomb vacua
where each root $\rt_i$ captures a single modulus $\ev_i^N$
span the whole Coulomb moduli space of the un-deformed quiver
while each individual vacuum state adiabatically recovers
its un-deformed properties.
At the same time, the Higgs and the pseudo-confining vacua
asymptote to the Coulomb vacua we already have while losing
their distinguishing features.
For example, the Higgs vacuum~(\ref{HiggsVac}) loses the squark VEVs
in the $\WW(X)\to 0$ limit and becomes indistinguishable from a
Coulomb vacuum which simply happens to have $\ev_1^N=\mu_1^N$
and hence a massless quark mode with $j=f=1$.
Likewise, the pseudo-confining vacuum with $\ev_1^N=\ev_2^N=\rt_1$
becomes indistinguishable from the ordinary Coulomb vacuum with
$\rt_1\approx \rt_2$ when the $SU(2)$ sector loses its mass gap
in the un-deformed limit of the quiver.

Therefore, as far as the un-deformed deconstructive quiver theory
is concerned, the pseudo-confining and the isolated-Higgs vacua
are artifacts of the deformation and we should focus on the
abelian Coulomb vacua with $n_i=1$ (or 0) only.
Nevertheless, the very existence of the pseudo-confining and
isolated-Higgs vacua of the deformed theory affects its chiral ring,
and so we will take them into consideration in~\S3.

We conclude this section by discussing the mesonic and baryonic branches
of the {\sl deformed} quiver theories, assuming the
quark masses allow their existence in the first place.
The mesonic branches are of mixed Coulomb+Higgs type, and the two kinds
of moduli sub-spaces are affected in two different ways:
The Coulomb subspace of a mesonic branch (the $\ev_j^N$ which are
not frozen by the squark VEVs) becomes discretized similarly
to the main Coulomb branch, but the Higgs subspace remains continuous,
although its complex structure {\sl may} be deformed.
For example, the mesonic Higgs branch~(\ref{SquarkVEVs}) which exists
for $ev_1^N=\mu_1^N=\mu_2^N\neq 0$ has its
F-term constraint~(\ref{FtermMesonic})
for the squark and antisquark VEVs deformed to
\be
Q_1\aq^1\ +\ Q_2\aq^2\
=\ -\frac{\WW(\mu_1^N)}{\gamma\,\mu_1}\,,
\label{FmesonicDeformed}
\ee
but despite this deformation, we still have continuously variable anti/squark
VEVs governed by two independent complex Higgs moduli.
On the other hand, the un-frozen Coulomb moduli $(\ev_2^N,\ldots,\ev_{n_c}^N)$
of the deformed quiver are no longer continuously variable but restricted
to the discrete set of the $\rt_i$ roots.

The multi-mesonic  Coulomb+Higgs branches suffer similar effects:
The Coulomb moduli  become discretized but the Higgs moduli remain
continuously variable, although the complex structure of the mesonic
moduli space suffers a deformation.
Likewise, a baryonic Higgs branch of the \suq\ quiver
survives as a continuous moduli
space with two complex moduli, but its complex structure is deformed
as the F-term constraint~(\ref{FtermBaryonic}) becomes
\be
\gamma e^{2\pi ik_j/N}\,Q^j\aq_j\ +\ \frac{\beta\vev{s} v^{n_c}}{\ev_j}\
+\,\sum_k\nu_k\,\ev_j^{kN-1}\ =\ 0\quad \forall j.
\label{FbaryonicDeformed}
\ee

This completes our summary of the classical quiver theories.
The quantum theories and their chiral rings
will be addressed in the following sections \S\S3--4.

 
%
%
\section{The Non--Baryonic Chiral Ring}
The main subject of this paper is the quantum chiral ring
of the \suq\ quiver theory from the purely $D=4$, $\NN=1$ point of view.
Thus, we consider the size $N$ of the quiver as a fixed,
finite parameter of the theory and allow the superpotential
deformation (\ref{Wdef}) despite its non-locality in the $x^4$ direction.
Using the techniques of Cachazo, Douglas, Seiberg and Witten
\cite{CDSW,Seiberg,CSW1,CSW2},
we package the chiral ring's generators into several
resolvent functions of an auxiliary complex variable $X$,
and then derive and solve the anomaly equations for these resolvents.

The new aspects of the present work (compared to Cachazo {\it et al.})
are due to a more complicated object of study:  A whole quiver of
$N$ gauge groups instead of just one, and each gauge group is
$SU(n_c)$ rather than $U(n_c)$.
In this section we focus on the quiver issues and study the
non-baryonic generators of the chiral ring.
The baryons and other generators allowed by the $SU(n_c)$ rather than
$U(n_c)$ symmetries will be addressed in the following section~\S4.

Let us start with a brief review of the chiral ring basics.
Most generally, the chiral ring of a 4D $\NN=1$
gauge theory is the $\overline{Q}$~cohomology in the algebra of
local {\sl gauge-invariant} operators ${\cal O}(x)$ of the theory.
That is, we consider chiral operators
$[\overline{Q}^{\dot\alpha},{\cal O}\}=0$
and identify them modulo $\overline{Q}$ commutators,
\be
{\cal O}_1\ \eqcr\ {\cal O}_2\ \Longleftrightarrow\
{\cal O}_1\, -\, {\cal O}_2\
=\ \left[\overline{Q}^{\dot\alpha},{\cal O}'\right\}
\ee
where the operator ${\cal O}'(x)$ is also local and gauge invariant.
In the superfield formalism we may use the $\overline D^{\dot\alpha}$
super-derivative instead of the $\overline Q^{\dot\alpha}$ supercharges,
thus chiral operators ${\cal O}(z)$ satisfy $\overline D^{\dot\alpha}\,
{\cal O}=0$ and in the chiral ring
\be
{\cal O}_1\ \eqcr\ {\cal O}_2\ \Longleftrightarrow\ {\cal O}_1\,-\,{\cal O}_2\
=\ \overline D^{\dot\alpha}\Bigl(\mbox{gauge-invariant}\ {\cal O}'\Bigr).
\label{ChiralRingDef}
\ee
In the chiral ring the spacetime location of an operator is irrelevant,
\be
\partial_{\alpha\dot\alpha}{\cal O}\
=\ \coeff{1}{2i}\,\overline D_{\dot\alpha}(D_\alpha {\cal O})\
\eqcr\ 0\ \Longrightarrow\
\forall x_1,x_2:\ {\cal O}(x_1)\ \eqcr\ {\cal O}(x_2),
\ee
and therefore all operator products are also position independent,
\be
{\cal O}_1(x_1)\,{\cal O}_2(x_2)\,\cdots {\cal O}_n(x_n)\
\eqcr\ \mbox{same}\ {\cal O}_1 {\cal O}_2\cdots {\cal O}_n\quad
\forall\, x_1,x_2,\ldots,x_n\,.
\label{OperatorProducts}
\ee
This position independence distinguishes the chiral ring from the
more general operator algebra of the quantum theory and makes
it exactly solvable.
It also makes it a bona-fide ring, which simplifies the analysis:
Once we construct all the independent generators from the fundamental
fields of the theory, the operator products (\ref{OperatorProducts})
follow from the ring structure without any further work.

Finally, note the distinction between the {\it off-shell} and the
{\it on-shell} chiral rings of the same theory:
In the off-shell chiral ring the equivalence relations~(\ref{ChiralRingDef})
must be operatorial identities of the quantum theory, but in the
{\it on-shell} chiral ring we use both the identities and the equations
of motion.
Classically
\be
\frac{\partial W}{\partial\phi}\
=\ \coeff{1}{4}\,\overline D^2\left(\frac{\partial K}{\partial\phi}\right)\
\eqcr\ 0
\label{ClassicalOnShell}
\ee
for any independent chiral field $\phi$, but at the quantum level
eqs.~(\ref{ClassicalOnShell}) are corrected by the generalized Konishi
anomalies, {\it cf.} \eq{GenericAnomaly} on page~\pageref{GenericAnomaly}.

%
\subsection{Generating the Chiral Ring}
In this subsection we generate ($i.\,e.$, construct independent generators of)
the off-shell chiral rings of the \suq\ quiver theories.
Or rather the {\sl almost} off-shell rings where the operatorial identities
of the theory are supplemented by the anomaly-free equations of motion
for the singlet fields $s_\ell$:
\be
\forall\ell:\quad \frac{\partial W}{\partial s_\ell}\ \eqcr\ 0\quad
\Longrightarrow\quad \det(\link_\ell)\ \eqcr\ v^{n_c} .
\label{CRdet}
\ee
We group these particular equations of motion with the operatorial
identities because the $s_\ell$ fields do not do anything interesting besides
imposing the constraints~(\ref{CRdet}) on the link fields to set up the
$SL(n_c,\CC)_\ell$ sigma models.

We begin with the chiral ring generators made from the link fields~$\link_\ell$
and nothing else.
Because of eqs.~(\ref{CRdet}) we cannot form chiral gauge invariants
from the individual link fields; instead, we have to take traces
$\tr(\link_N\link_{N-1}\cdots\link_2\link_1)^k$ of whole chains of links
wrapped several times around the quiver.
Also, thanks to $\det(\link_\ell)\neq 0$ the matrix inverses
$\link_{\ell}^{-1}$ are well-defined chiral operators;
this allows us to takes traces
$\tr(\link_1^{-1}\link_2^{-1}\cdots\link_N^{-1})^k$
of the inverse link chains
wrapped around the quiver in the opposite direction.
Conveniently, both types of traces can be summarized via a single resolvent
\be
\eqalign{
T(X)\ &=\ \tr\left(\frac{1}{X-\link_N\cdots\link_1}\right)\cr
&=\, \sum_{k=0}^\infty\frac{1}{X^{k+1}}\times
	\tr\left(\link_N\link_{N-1}\cdots\link_2\link_1\right)^k \cr
&=\ -\sum_{k=1}^\infty X^{k-1}\times
	\tr\left(\link_1^{-1}\link_2^{-1}\cdots
		\link_{N-1}^{-1}\link_N^{-1}\right)^k .\cr
}\label{Tresolvent}
\ee
Classically, this resolvent has simple poles at the Coulomb moduli
of the quiver
\be
T(X)\ = \sum_{j=1}^{n_c}\frac{1}{X\,-\,\ev_j^N}
\label{Tclassical}
\ee
and we may use contour integrals
\be
n({\cal C})\ = \oint\limits_{\cal C}\!\frac{dX}{2\pi i}\,T(X)
\label{polecount}
\ee
as gauge-invariant counts of the Coulomb moduli inside any particular
contour~$\cal C$;
for example, $n_i=n({\cal C}_i)$ for a sufficiently small contour ${\cal C}_i$
surrounding the deformation root~$\rt_i$.

In the quantum theory of the quiver, the resolvent~(\ref{Tresolvent})
behaves similarly to its $\tr\left(\frac{1}{X-\Phi}\right)$ analogue
in the deformed $\NN=2$ SQCD$_4$:
The poles at $\rt_i$ become $\root 2\of{\phantom{2}}$ branch cuts,
but the $n_i$ --- defined as $n({\cal C}_i)$ for suitable contours
${\cal C}_i$ --- remain exactly integer.
And since the monodromies in the parameter space of the deformed quiver theory
can entangle a ${\cal C}_i$ with any other cycle of the Riemann
surface of the $T(X)$, it follows that
{\em\blue all closed-contour integrals (\ref{polecount})
of the link resolvent~(\ref{Tresolvent}) must have integer values.}
Indeed, consider the differential
\be
T(X)\,dX\ =\ d\tr\log(X-\link_N\cdots\link_1)\
=\ d\log\det(X-\link_N\cdots\link_1).
\label{TdX}
\ee
Regardless of any quantum corrections to the determinant
$\det(X-\link_N\cdots\link_1)$, its logarithm will always have exactly
integer${}\times 2\pi i$ differences between its branches.
Consequently, the quantum quiver theory has exactly integer contour integrals
(\ref{polecount}) for all contours $\cal C$ which are closed
on the Riemann surface of the~$T(X)$.

The readers who find the above argument too heuristic are referred
to Cachazo {\it et~al} for a rigorous proof;
the arguments of ref.~\cite{CSW2} apply equally well to the present case
and we don't see the need of repeating them here almost verbatim.

\smallskip
\centerline{\blue\large $\star\quad\star\quad\star$}
\smallskip

Next, let us add the quark and antiquark fields to the picture
and form all kinds of chiral operators with mesonic quantum numbers.
Besides the true mesons
\be
\left[M_\ell\right]^{f'}_{\,f}\ =\ \aq^{f'}_\ell Q_{\ell,f}
\label{Mesons}
\ee
which are local in 5D as well as in 4D, there are other meson-like chiral
gauge-invariant
operators where the quark and the antiquark are located at different
quiver nodes $\ell\neq\ell'$ but are connected to each other by
a chain of link fields, {\it eg.,}
$\aq_{\ell'}^{f'}\link_{\ell'-1}\cdots\link_{\ell}Q_{\ell,f}$.
From the 5D point of view, these are {\em bi-local} operators which
create/annihilate {\em un-bound} pairs of quarks and antiquarks,
while the link chains deconstruct the un-physical Wilson strings which allow
for manifest gauge invariance of such bi-local operators:
\bea
[{\bf M}(x_2,x_1)]^{f'}_{\,f}\ &=&
\aq^{f'}(x_2)\times
    \mathop{\hbox{\Large\rm exp}}\limits_{\rm Path\atop ordered}\left(
	i\int\limits_{x_1}^{x_2}\!\!dx^\mu\,A_\mu(x)\right)
    \times Q_{f}(x_1) \nonumber \\
&\longrightarrow &
\aq_{\ell'}^{f'}\,\link_{\ell'-1}\link_{\ell'-2}\cdots
	\link_{\ell+1}\link_{\ell}\,Q_{\ell,f}
    \label{SplitPair} \\
&&\qquad\mbox{for}\ x_1^{0,1,2,3}=x_2^{0,1,2,3},\
	 x_1^4=a\ell\ \mbox{and}\ x_2^4=a\ell'. \nonumber
\eea
From the 4D point of view however, these operators are local, chiral
and gauge invariant ---
and therefore belong to the chiral ring of the quiver.%
\footnote{%
	Actually, to make the chiral operator on the second line of
	\eq{SplitPair}, the Wilson string on the top line must be
	modified to incorporate the 5D scalar field $\phi(x)$
	along with the gauge field $A_4(X)$.
	Indeed, according to the deconstruction map~(\ref{Dmap}),
	$$
	\aq_{\ell'}^{f'}\,\link_{\ell'-1}\link_{\ell'-2}\cdots
		\link_{\ell+1}\link_{\ell}\,Q_{\ell,f}\
	=\ \aq_{\ell'}^{f'}\times \mathop{\hbox{\Large\rm exp}}
		\limits_{\rm Path\atop ordered}
	\left(\int\limits_{\,a\ell}^{\,a\ell'}\!dx^4\,(\phi(x)+iA_4(x))\right)
		\times Q_{\ell,f}\,.
	$$
	Despite this correction, the Wilson string remains un-physical
	and the quark-antiquark pair remains unbound.
	}

Besides the ``split mesons'' (\ref{SplitPair}) where the link chain
runs directly from the quark to the antiquark, we may have the chain
going several times around the whole quiver, thus
$\aq_{\ell'}^{f'}\link_{\ell'-1}\cdots\link_1\distimes
	(\link_N\cdots\link_1)^k\distimes
	\link_N\cdots\link_{\ell+1}\link_\ell Q_{\ell,f}$,
or in reverse direction (via inverse links), thus
$\aq^{f'}_{\ell'}\link^{-1}_{\ell'}\link^{-1}_{\ell'+1}\cdots
\link^{-1}_{\ell-2}\link^{-1}_{\ell-1}Q_{\ell,f}$\hskip 0.5em plus 1em
or even\hskip 0.5em plus 1em
$\aq_{\ell'}^{f'}\link^{-1}_{\ell'}\link^{-1}_{\ell'+1}\cdots\link^{-1}_N
	\distimes(\link^{-1}_1\cdots\link^{-1}_N)^k\distimes
	\link^{-1}_1\cdots\link^{-1}_{\ell-1}Q_{\ell,f}$.
To summarize all these meson-like operators, we define mesonic resolvents
\be
\MM_{\ell',\ell}(X)\
=\ \aq_{\ell'}\,\frac{\link_{\ell'-1}\cdots\link_\ell}{X-\link\cdots\link}\,
Q_{\ell}
\label{Mresolvents}
\ee
where the flavor indices of the quarks and antiquarks are suppressed
for notational simplicity (or in other words, each $\MM_{\ell',\ell}(X)$
is an $F\times F$ matrix),
the quiver indices are understood modulo~$N$, and the
$\displaystyle\frac{\link_{\ell'-1}\cdots\link_\ell}{X-\link\cdots\link}$
is a short-hand for
\be
\link_{\ell'-1}\cdots\link_{\ell}\times
\frac{1}{X-\link_{\ell-1}\cdots\link_1\link_N\cdots\link_{\ell}}
=\ \frac{1}{X-\link_{\ell'-1}\cdots\link_1\link_N\cdots\link_{\ell'}}
\times\link_{\ell'-1}\cdots\link_{\ell}\,.\qquad
\ee
The resolvents~(\ref{Mresolvents}) with $\ell\le\ell'<\ell+N$
suffice to generate all the meson-like operators; 
for $\ell'=\ell+N$ there is a periodicity equation
\be
\MM_{\ell+N,\ell}(X)\ 
=\ X\times\MM_{\ell,\ell}(X)\ -\ M_\ell
\label{Mperiodicity}
\ee
where the last ($X$-independent) term on the right hand side is the
matrix of the true mesons~(\ref{Mesons}).
Classically, the resolvents (\ref{Mresolvents})
have simple poles at $X=\mu_f^N$, but only for mesonic vacua with
$\vev{Q_f},\vev{\aq_f}\neq 0$.
In the quantum theory, such poles exist for all vacua, but only the
mesonic vacua have them on the ``physical sheet'' of the quiver's
spectral curve; we shall explain this issue the following subsection~\S3.2.

\smallskip
\centerline{\blue\large $\star\quad\star\quad\star$}
\smallskip

Meanwhile,  consider the chiral gaugino superfields
$W^\alpha_\ell=\lambda^\alpha_\ell+F_\ell^{\alpha\beta}\theta_\beta+\cdots
{\in\mathop{\rm Adj}(SU(n_c)_\ell)}$
and their gauge-invariant combinations with the other chiral fields
of the quiver.
Although there is a great multitude of such combinations,
most of them turn out to be total $\overline{D}^2$
super-derivatives and thus do not belong to the chiral ring.
This follows from the appearance of the tensor sum of all  gaugino
superfields
\be
\Wsum^\alpha\ =\ \bigoplus_{\ell=1}^N W_\ell^\alpha
\ee
in the anti/commutation algebra of the gauge-covariant
spinor derivatives $\nabla^\alpha$ and $\overline\nabla^{\dot\beta}$:
\be
\left[\overline\nabla_{\dot\alpha},
	\left\{\overline\nabla_{\dot\beta},\nabla_\gamma\right\}\right]\
=\ 4\epsilon_{\dot\alpha\dot\beta}\Wsum_\gamma\qquad
\Longrightarrow\ \forall\ \mbox{chiral}\ \Phi:\quad
-\coeff18\overline\nabla^2\nabla^\alpha\Phi\
=\ \Wsum^\alpha\Phi,\qquad
\label{SuperAlgebra}
\ee
thus
\be
\eqalign{
-\coeff18\overline\nabla^2\nabla^\alpha Q_\ell^f\ &
=\ W^\alpha_\ell Q_\ell^f\,,\cr
-\coeff18\overline\nabla^2\nabla^\alpha \aq_\ell^f\ &
=\ -\aq_\ell^f W^\alpha_\ell\,,\cr
-\coeff18\overline\nabla^2\nabla^\alpha \link_\ell\ &
=\ W^\alpha_{\ell+1}\link_\ell\,-\,\link_\ell W^\alpha_\ell\,,\cr
-\coeff18\overline\nabla^2\nabla^\alpha W^\beta_\ell\
    = -\coeff18\overline\nabla^2\nabla^\beta W^\alpha_\ell\ &
=\ \left\{W^\alpha_\ell,W^\beta_\ell\right\}\,.\cr
}\qquad\label{DbarIdents}
\ee
Therefore, any gauge invariant combination of chiral superfields
which includes both gauginos and quarks or antiquarks is a total
$\overline D^2$ super-derivative --- which does not belong to the chiral ring.
For example,
\be
\aq_{\ell'}\link_{\ell'-1}\cdots\link_{\ell''}W^\alpha_{\ell''}
	\link_{\ell''-1}\cdots\link_{\ell}Q_{\ell}\
=\ -\coeff18\,\overline D^2\Bigl(
	\left(\aq_{\ell'}\link_{\ell'-1}\cdots\link_{\ell''}\right)
	\nabla^\alpha
	\left(\link_{\ell''-1}\cdots\link_{\ell}Q_{\ell}\right)
    \Bigr)\
\eqcr\ 0.\hskip 4em minus 4em
\label{NoWMesons}
\ee
Furthermore, for any gauge-covariant combination $\Xi_{\ell,\ell'}$
of gaugino and link fields which transforms as
$(\square_\ell,\overline\square_{\ell'})$, we have
\be
\!\tr\left(\Xi_{\ell,\ell'}\,W^\alpha_{\ell'}\,
	\link_{\ell'-1}\cdots\link_{\ell}\right)\
-\ \tr\left(\Xi_{\ell,\ell'}\,\link_{\ell'-1}\cdots\link_{\ell}\,
	W^\alpha_\ell\right)\
=\ -\coeff18\overline D^2\Bigl(\Xi_{\ell,\ell'}\,
	\nabla^\alpha\left(\link_{\ell'-1}\cdots\link_{\ell}\right)\Bigr)\
\eqcr\ 0.
\label{Wslide}
\ee
In particular, for any $\ell$ and $\ell'$,
\be
\tr\left( W^\alpha_\ell\,
	(\link_{\ell-1}\cdots\link_{\ell+1}\link_{\ell})^k\right)\
\eqcr\ \tr\left( W^\alpha_{\ell'}\,
	(\link_{\ell'-1} \cdots \link_{\ell'+1}\link_{\ell'})^k\right)
\ee
and likewise
\be
\tr\left( W^\alpha_\ell\,
	(\link^{-1}_{\ell}\link^{-1}_{\ell+1}\cdots\link^{-1}_{\ell-1})^k
	\right)\
\eqcr\ \tr\left( W^\alpha_{\ell'}\,
	(\link^{-1}_{\ell'}\link^{-1}_{\ell'+1}\cdots\link^{-1}_{\ell'-1})^k
	\right),
\ee
which means that all chiral ring's generators which
involve a single gaugino operator are summarized in a single
$\ell$--independent gaugino resolvent
\be
\Psi^\alpha(X)\ \eqdef\ \frac{1}{4\pi}\,
\tr\left(\frac{W^\alpha}{X-\link\cdots\link}\right)\
\equiv\ \frac{1}{4\pi}\,\tr\left(W^\alpha_\ell\times
	\frac{1}{X-\link_\ell\cdots\link_{\ell-1}}\right)\quad
\langle\!\langle\mbox{same}\ \forall\ell \rangle\!\rangle.
\label{PSIresolvent}
\ee

Physically, this resolvent
encodes the exactly massless abelian photinos of the \suq\ quiver.
Indeed, contour integrals of the $\Psi(X)$ yield traces of the
diagonal gaugino fields of the $SU(n_c)_{\rm diag}$ over a subspace
where the Coulomb moduli $\ev_j^N$ happen to lie inside the integration
contour:
\be
\oint\limits_{\cal C}\!\frac{dX}{2\pi i}\>\Psi^\alpha(X)\
=\ \frac{1}{4\pi}\,\tr\left(
	 \left.W^\alpha\right|_{\ev^N\ \rm inside\ \cal C}
	\right)
\label{GauginoContour}
\ee
where the $W^\alpha$ can be thought as belonging to the $SU(n_c)_{\rm diag}$
because any $W^\alpha_\ell$ would yield  the same generator of the
chiral ring regardless of $\ell$.
In particular, for the ${\cal C}_i$ contour surrounding
a deformation root $\rt_i$ (or in the fully quantum theory, surrounding
the branch cut near an $\rt_i$), the integral~(\ref{GauginoContour})
restricts the gaugino fields $W^\alpha$ to the $U(n_i)$ subgroup
of the unbroken gauge symmetry~(\ref{SurvivingSymmetrySU})
--- and then the trace extracts the abelian photino $W^\alpha_i$ in
the $U(1)_i$ center of the $U(n_i)$:
\be
4\pi\oint\limits_{{\cal C}_i}\!\frac{dX}{2\pi i}\>\Psi^\alpha(X)\ 
=\ \tr\left(\left.W^\alpha\right|_{U(n_i)}\right)\
=\ W_i^\alpha .
\label{PhotinoIntegrals}
\ee
Note that for the whole $SU(n_c)_{\rm diag}$, $\tr(W^\alpha)=0$ and hence
$\sum_i W_i^\alpha=0$;
in terms of the gaugino resolvent $\Psi^\alpha(X)$, it means no residue
at $X=\infty$ and $\Psi^\alpha(X)=O(1/X^2)$ rather than $O(1/X)$.

Next, consider the chiral ring generators involving two gaugino operators
$W^\alpha_{\ell_1}$ and $W^\beta_{\ell_2}$ inserted into a closed chain
$\tr(\link_N\cdots\link_1)^k$ of link operators.
Again, the specific points of insertion do not matter:
For any $\ell_1$ and $\ell_2$ and any $k_1+k_2=k-1$,
\be
\eqalign{
\tr\Bigl( W^\alpha_{\ell_1}\,
	(\link_{\ell_1-1}\cdots\link_{\ell_1})^{k_1}\, &
\link_{\ell_1-1}\cdots\link_{\ell_2}\,W^\beta_{\ell_2}\,
	(\link_{\ell_2-1}\cdots\link_{\ell_2})^{k_2}\,
	\link_{\ell_2-1}\cdots\link_{\ell_1} \Bigr) \cr
&\eqcr\ \tr\Bigl( W_{\ell_1}^\alpha W_{\ell_1}^\beta
	(\link_{\ell_1-1}\cdots\link_{\ell_1})^k\Bigr)\cr
&\eqcr\ \tr\Bigl( W_{\ell}^\alpha W_{\ell}^\beta
	(\link_{\ell-1}\cdots\link_{\ell})^k\Bigr)
	\qquad\qquad\forall\ \mbox{other}\ \ell\cr
&\eqcr\ -\half\,\epsilon^{\alpha\beta}\times
      \tr\Bigl( W_\ell^2\,(\link_{\ell-1}\cdots\link_{\ell})^k\Bigr)\cr
}\hskip 6em minus 6em \label{TwoGauginos}
\ee
where the last equality follows from the fourth \eq{DbarIdents}.
Thanks to reversibility of the link matrices $\link_\ell$ in the
\suq\ quiver theory, eqs.~(\ref{TwoGauginos}) extend to negative
$k_{1,2}$; in particular, for $k_1+k_2+1=k=0$ we have
\be
\forall\ell_1,\ell_2:\quad
\tr\left( W_{\ell_1}^2\right)\ \eqcr\ \tr\left( W_{\ell_2}^2\right)
\label{SameS}
\ee
without any link fields being involved at all
(except at the intermediate stages).%
\footnote{%
	Strictly speaking, the identities (\ref{SameS}) depend
	on the on-shell equations~(\ref{CRdet}).
	The off-shell operatorial identities of the quiver's chiral ring
	have form
	$$
	\det(\link_\ell)\times\left( \tr(W^2_\ell)\,
		-\,\tr(W^2_{\ell+1})\right)\ \eqcr\ 0
	$$
	and imply eqs.~(\ref{SameS}) if and only if
	all $\det(\link_\ell)\neq 0$.
	}
Consequently, all generators involving two gauginos are summarized
in a single $\ell$--independent ``gaugino bilinear'' resolvent
\be
R(X)\ \eqdef\ \frac{1}{32\pi^2}\,
\tr\left(\frac{W^2}{X-\link\cdots\link}\right)\
\equiv\ \frac{1}{32\pi^2}\,\tr\left(W^\alpha_\ell W_{\ell,\alpha}\times
	\frac{1}{X-\link_{\ell-1}\cdots\link_\ell}\right)\quad
\langle\!\langle\mbox{same}\ \forall\ell \rangle\!\rangle.
\label{Rresolvent}
\ee
The contour integrals of this resolvent encode the ``gaugino condensates''
of the non-abelian factors $SU(n_i)$ of the unbroken gauge
symmetry~(\ref{SurvivingSymmetrySU}):
\be
{\cal S}_i\ \equiv\,\oint\limits_{{\cal C}_i}\!\!\frac{dX}{2\pi i}\,R(X)\
=\ \frac{1}{32\pi^2}\,\tr\left(\left.W^\alpha W_\alpha\right|_{U(n_i)}\right).
\label{CondensateIntegrals}
\ee
Or rather, the ``gaugino condensation'' in the $SU(n_i)$ factor is the
leading contribution to the ${\cal S}_i$ for $n_i\ge 2$, in which case
${\cal S}_i\sim e^{-8\pi^2/n_ig^2_{\rm diag}}$
develops at the fractional instanton level~$1/n_i$.
For $n_i=1$ there is no gaugino condensation {\it per~se}, but the
${\cal S}_i$ ``condensate'' develops anyway at the one-whole-instanton level
${\cal S}_i\sim e^{-8\pi^2/g^2_{\rm diag}}$, thanks to coset
instantons in the broken $SU(n_c)_{\rm diag}/U(1)_i$.
In any case, we count instanton levels with respect to the {\em diagonal}
$SU(n_c)$ subgroup of the quiver; in terms of the whole \suq\ gauge group,
{\blue\em one diagonal instanton means one instanton in
\underline{each}~$SU(n_c)_\ell$ factor} according to
\be
\exp\left(-\frac{8\pi^2}{g^2_{\rm diag}}\right)\
=\ \prod_\ell \exp\left(-\frac{8\pi^2}{g^2_\ell}\right),
\label{InstantonCount}
\ee
or in un-deconstructed 5D terms, one Euclidean ${0^{\rm time}+1^{\rm space}}$
instanton brane wrapped around the compactified $x^4$ dimension.
However, apart from this quiver-specific instanton counting, the gaugino
bilinear resolvent~(\ref{Rresolvent}) behaves similarly to its
$\tr\left(\frac{W^\alpha W_\alpha}{X-\Phi}\right)$ analogue in the
deformed $\NN=2$ SQCD$_4$.

Finally, the quiver's chiral ring does not have any generators
involving three or more gaugino fields.
Indeed, let us insert $W^\alpha_{\ell'},W^\beta_{\ell''},\ldots,
W^\gamma_{\ell'''}$ into a closed chain $\tr(\link_N\cdots\link_1)^k$
of the link operators.
Applying \eq{Wslide} several times, we can move all the
gaugino operators to the same quiver node $\ell$, thus
$\tr\left( W_\ell^\alpha W_\ell^\beta\cdots W_\ell^\gamma\,
(\link_{\ell-1}\cdots\link_{\ell})^k\right)$, but then
the fourth \eq{DbarIdents} implies
$W_\ell^\alpha W_\ell^\beta\cdots W_\ell^\gamma
=\overline\nabla^2({\rm something})$ and therefore,
the whole shmeer${}=\overline D^2(\rm something\ else)\eqcr0$
and does not belong to the chiral ring.

{\bf Altogether},
we have constructed all the generators of the \suq\ quiver's chiral ring,
except for the baryons, the antibaryons, and their multi-local cousins
comprised of $n_c$ quarks or antiquarks located at different quiver nodes
connected by chains of link operators.
The \suq\ quiver has a whole zoo of such multi-local baryon-like chiral
operators, and we prefer to discuss them in a separate section~\S4.2.

%
\subsection{Anomalous Equations of Motion and their Solutions}
Thus far, we generated the (almost) {\em off-shell} chiral ring
of the \suq\ quiver.
In this section, we focus on the {\em on-shell} chiral ring
in which the resolvents $T(X)$, $\MM_{\ell',\ell}(X)$,
$\Psi^\alpha(X)$ and $R(X)$ satisfy the equations of motion
of the quantum quiver theory.
Generically, such equations follow from infinitesimal
gauge-covariant field-dependent variations of the fundamental
chiral fields of the theory
\be
\Phi\ \to\ \Phi\ +\ \delta\Phi(\Phi,\ \mbox{other chiral operators}).
\ee
Classically, $\delta W^{\rm tree}\equiv
\frac{\partial W^{\rm tree}}{\partial\Phi}\times\delta\Phi\eqcr 0$,
but in the quantum theory generalized Konishi anomalies change this
equation to
\be
\frac{\partial W^{\rm tree}}{\partial\Phi}\times\delta\Phi\
\eqcr\ \frac{1}{32\pi^2}\,\Tr\left(
	\frac{\Wsum^\alpha\Wsum_\alpha\,\partial\delta\Phi}{\partial\Phi}
	\right)
\label{GenericAnomaly}
\ee
where the trace on the right hand side is taken with respect to all
indices of the field~$\Phi$, color and flavor.
As a specific example in the quiver context, let $\Phi$ be the quark
field $Q_{\ell'}$ and consider the variation
\be
\delta Q_{\ell'}\ =\
\frac{\link_{\ell'-1}\link_{\ell'-2}\cdots\link_{\ell}}{X-\link\cdots\link}
\times Q_\ell \times \varepsilon
\label{QvarM}
\ee
where $\varepsilon$ is an infinitesimal $F\times F$ matrix
in the flavor space.
For this variation,
\be
\eqalign{
\delta W^{\rm tree}\ =\ \delta W^{\rm hop}\ &
=\ \gamma\left[
	\tr\left(\aq_{\ell'+1}\link_{\ell'}\times\delta Q_{\ell'}\right)\
	- \tr\left(\mu\times\aq_{\ell'}\times\delta Q_{\ell'}\right)\right]\cr
&=\ \gamma\left[\tr\left(\aq_{\ell'+1}\times
	\frac{\link_{\ell'}\link_{\ell'-1}\cdots\link_{\ell}}
	{X-\link\cdots\link}\times Q_\ell\times\varepsilon\right)
	\right.\cr
&\qquad\left.{}-\ \tr\left(\mu\times\aq_{\ell'}\times
	\frac{\link_{\ell'-1}\link_{\ell'-2}\cdots\link_{\ell}}
	{X-\link\cdots\link}\times Q_\ell\times\varepsilon\right)
	\right] \cr
&=\ \gamma\,\tr\Bigl(
	\left(\MM_{\ell'+1,\ell}(X)\,-\,\mu\times\MM_{\ell',\ell}(X)\right)
	\times\varepsilon\Bigr)\cr
}\label{MesonicDW}
\ee
while the Konishi anomaly exists only for $\ell'=\ell$
(otherwise $\delta Q_{\ell'}$ does not depend on the $Q_{\ell'}$ itself)
and amounts to
\be
\frac{1}{32\pi^2}\,
\Tr\left( W^\alpha_{\ell'}W_{\ell',\alpha}\,
	\frac{\delta_{\ell',\ell}}{X-\link_{\ell'-1}\cdots\link_{\ell'}}\
	\otimes\ \varepsilon\right)\
\eqcr\ \delta_{\ell',\ell}\,R(X)\times\tr(\varepsilon).
\label{KonishiQuark}
\ee
Substituting these formul\ae\ into the generic \eq{GenericAnomaly}, we arrive
at the anomalous equations of motion for the on-shell mesonic resolvents:
In $F\times F$ matrix notations,
\be
\MM_{\ell'+1,\ell}(X)\ -\ \mu\times\MM_{\ell',\ell}(X)\
\eqcr\ \delta_{\ell',\ell}\,\gamma^{-1}\,R(X)
\times\hbox{\large 1}_{F\times F}.
\label{MResEqQ}
\ee
Consequently,
\be
\MM_{\ell',\ell}(X)\ \eqcr\ \mu^{\ell'-\ell}\times\MM_{\ell,\ell}(X)\
+\ \gamma^{-1}\mu^{\ell'-\ell-1}\times R(X)\qquad
\mbox{for}\ \ell<\ell'\le\ell+N, \nonumber
\ee
and hence in light of the periodicity equation~(\ref{Mperiodicity}),
\be
\mbox{\bf on shell:}\quad
\left\{\eqalign{
	\MM_{\ell'=\ell}(X)\ &
	=\ \frac{1}{X-\mu^N}\times\left(\gamma^{-1}\mu^{N-1}\,
			 R(X)\ +\ M_\ell\right),\cr
	\MM_{\ell'>\ell}(X)\ &
	=\ \frac{\mu^{\ell'-\ell-1}}{X-\mu^N}\times
		\left(\gamma^{-1}\,XR(X)\ +\ \mu\times M_\ell\right).\cr
	}\right.
\label{MResQ}
\ee
Thus, we have solved for {\em all} of the on-shell mesonic resolvents
in terms of the ordinary mesons $M_\ell$ and the gaugino bilinear
resolvent~$R(X)$.

Likewise, starting with the infinitesimal antiquark variation
\be
\delta \aq_{\ell}\ =\ \varepsilon\times\aq_{\ell'}\times
\frac{\link_{\ell'-1}\link_{\ell'-2}\cdots\link_{\ell}}{X-\link\cdots\link}
\label{AQvarM}
\ee
we also arrive at anomalous equations of motion for the on-shell
mesonic resolvents, but this time we have
\be
\MM_{\ell',\ell-1}(X)\ -\ \MM_{\ell',\ell}(X)\times\mu\
\eqcr\ \delta_{\ell',\ell}\,\gamma^{-1}\,R(X)
\times\hbox{\large 1}_{F\times F}
\label{MResEqA}
\ee
and consequently
\be
\mbox{\bf on shell:}\quad
\left\{\eqalign{
	\MM_{\ell'=\ell}(X)\ &
	=\ \left(\gamma^{-1}\mu^{N-1}\, R(X)\ +\ M_{\ell'}\right)
		\times\frac{1}{X-\mu^N}\,,\cr
	\MM_{\ell'>\ell}(X)\ &
	=\ \left(\gamma^{-1}\,XR(X)\ +\ M_{\ell'}\times\mu\right)
		\times\frac{\mu^{\ell'-\ell-1}}{X-\mu^N}\,.\cr
	}\right.
\label{MResA}
\ee
Note that \eqrange{MResEqA}{MResA} are just as valid as
\eqrange{MResEqQ}{MResQ}, and to assure mutual consistency
of the two equation systems, all meson matrices $M_\ell$ --- and
hence all the mesonic resolvent matrices $\MM_{\ell',\ell}(X)$ ---
must commute with the $\mu^N$ matrix and therefore must be block-diagonal
in its eigenbasis.
Furthermore, if $\mu_{f'}=\mu_f$ whenever $\mu^N_{f'}=\mu^N_f$, then
all the $M_\ell$ matrices must be equal to each other, $M_\ell\equiv M$;
otherwise $M_\ell=\mu^\ell\times M\times \mu^{-\ell}$.
In terms of the matrix elements, we have
\be
[M_\ell]^{f'}_{\,f}\
=\ M^{f'}_{\,f}\times\left(\frac{\mu_{f'}}{\mu_f}\right)^\ell\quad
\mbox{for}\ \mu_{f'}^N\,=\,\mu_f^N\,\neq\,0\ \mbox{only;}\qquad
\mbox{otherwise,}\ [M_\ell]^{f'}_{\,f}\ =\ 0.\qquad
\label{MesonRelation}
\ee
In particular, the matrix elements with $\mu_{f'}=\mu_f=0$ must vanish because
the on-shell mesonic resolvents of the \suq\ quiver
{\em cannot} have poles at $X=0$.
Indeed,
\be
\mbox{for}\ X\to 0,\qquad
\MM_{\ell',\ell}(X)\ \to\ \aq_{\ell'}\,\link_{\ell'}^{-1}
\link_{\ell'+1}^{-1}\cdots\link_{\ell-1}^{-1}\,Q_\ell\ \neq\ \infty
\ee
because on shell $\det\link_\ell\neq 0$ and the inverse links are well-defined
chiral operators.

Next, let us vary a link field $\link_\ell$ according to
\be
\delta \link_\ell\ =\ \varepsilon\,\frac{\link_\ell}{X-\link\cdots\link}
\label{linkvarT}
\ee
where $\varepsilon$ is now an infinitesimal number rather than a matrix.
For this variation,
\be
d\delta\link_\ell\
=\ \frac{\varepsilon}{X-\link_\ell\cdots\link_{\ell+1}}\times
d\link_\ell\times\frac{X}{X-\link_{\ell-1}\cdots\link_\ell}
\ee
while
\be
\Wsum^\alpha\Wsum_\alpha\,d\delta\link_\ell\
=\ W^\alpha_{\ell+1}W_{\ell+1,\alpha}\,d\delta\link_\ell\
-\ 2W^\alpha_{\ell+1}\,d\delta\link_\ell\,W_{\ell,\alpha}\
+\ d\delta\link_\ell\,W^\alpha_\ell W_{\ell,\alpha}\,,
\ee
hence the Konishi anomaly comes up to
\be
\eqalign{
\frac{1}{32\pi^2}\, &
\Tr\left(\frac{\Wsum^\alpha\Wsum_\alpha\,d\delta\link_\ell}{d\link_\ell}
	\right)\cr
\noalign{\vskip 10pt}
&=\ \frac{\varepsilon X}{32\pi^2}\,\Tr\left(\eqalign{
	\frac{W^\alpha_{\ell+1}W_{\ell+1,\alpha}}{X-\link\cdots\link}
		\otimes\frac{1}{X-\link\cdots\link}\ &
	-\ 2\,\frac{W^\alpha_{\ell+1}}{X-\link\cdots\link}\otimes
		\frac{W_{\ell,\alpha}}{X-\link\cdots\link}\cr
	&+\ \frac{1}{X-\link\cdots\link}\otimes
		\frac{W^\alpha_{\ell}W_{\ell,\alpha}}{X-\link\cdots\link}\cr
    }\right) \cr
\noalign{\vskip 10pt}
&=\ \varepsilon X \Bigl( R(X)\times T(X)\
	-\ \Psi^\alpha(X)\times\Psi_\alpha(X)\ +\ T(X)\times R(X)\Bigr).\cr
}\qquad\label{KonishiLink}
\ee
At the same time, the classical superpotential varies according to
\bea
\delta W_{\rm hop} &=&
\varepsilon\gamma\,\tr\left(\aq_{\ell+1}\,
	\frac{\link_\ell}{X-\link\cdots\link}\,Q_\ell\right)\
    \equiv\ \varepsilon\gamma\,\tr\Bigl(\MM_{\ell+1,\ell}(X)\Bigr), \\
\delta W_\Sigma \vrule width 0pt height 15pt depth 10pt &=&
\varepsilon\beta s_\ell\,
	\tr\left(\frac{\det\link_\ell}{X-\link\cdots\link}\right)\
    =\ \varepsilon\beta s_\ell\,
    	\Bigl(\det\link_\ell\,=\,v^{n_c}\Bigr)\times T(X),\qquad\\
\delta W_{\rm def} &=&
\varepsilon\sum_{k=1}^d\nu_k\,\tr\left(
	\frac{(\link\cdots\link)^k}{X-\link\cdots\link}\right)\
    =\ \varepsilon\Bigl[ X\W'(X)\,T(X)\Bigr]_-\,.
\eea
where following the notations of \cite{CDSW}, the $[X\W'(X)T(X)]_-$
stands for the negative-power part of the $X\W'(X)T(X)$
with respect to the power series expansion around $X=\infty$.
Thus, we arrive at an anomalous equation of motion
\be
\left[\WW(X)\,T(X)\right]_-\ +\ \gamma\,\tr\Bigl(\MM_{\ell+1,\ell}(X)\Bigr)\
\eqcr\ 2X\,R(X)\,T(X)\ -\ X\,\Psi^\alpha(X)\,\Psi_\alpha(X)
\label{Treseq}
\ee
where $\WW(X)=X\W'(X)+\beta v^{n_c}s_\ell$ according to \eq{polydef} ---
and all singlets are equal on shell, $s_\ell\equiv s$.
Solving \eq{Treseq} for the on-shell link resolvent, we have
\be
T(X)\
=\ \frac{t(X)\,-\,\gamma\tr\MM_{\ell+1,\ell}(X)\,-\,X\Psi^2(X)}
	{\WW(X)\,-\,2XR(X)}
\label{Tres}
\ee
for some polynomial $t(X)=[\WW(X)\,T(X)]_+$ of $\rm degree\le(d-1)$;
we shall derive a more specific formula later in this section.

The anomalous equation of motion for the gaugino resolvent $\Psi^\alpha(X)$
also follows from varying a link field $\link_\ell$, but this time we have
\be
\delta \link_\ell\
=\ \frac{\link_\ell}{X-\link\cdots\link}\times
\frac{W^\alpha_\ell\,\varepsilon_\alpha}{4\pi}
\label{linkvarPsi}
\ee
where $\varepsilon_\alpha$ is an infinitesimal spinor.
Consequently, $\delta W_{\rm hop}\eqcr 0$ (because of anti/quark and gaugino
operators present in the same expression) and
\be
\delta W_{\rm tree}\ \eqcr\ \delta W_{\rm def}\ +\ \delta W_\Sigma\
=\ \left[\WW(X)\Psi^\alpha(X)\varepsilon_\alpha\right]_-
\ee
while the Konishi anomaly is
\be
\eqalign{
\frac{1}{32\pi^2}\, &
\Tr\left(\frac{\Wsum^\alpha\Wsum_\alpha\,d\delta\link_\ell}{d\link_\ell}
	\right)\cr
\noalign{\vskip 10pt}
&=\ \frac{X}{128\pi^3}\,\Tr\left(\eqalign{
	\frac{W^\beta_{\ell+1}W_{\ell+1,\beta}}{X-\link\cdots\link}\otimes
		\frac{W_\ell^\alpha\varepsilon_\alpha}
			{X-\link\cdots\link}\ &
	-\ 2\,\frac{W^\beta_{\ell+1}}{X-\link\cdots\link}\otimes
		\frac{W_\ell^\alpha\varepsilon_\alpha W_{\ell,\beta}}
			{X-\link\cdots\link}\cr
	&+\ \frac{1}{X-\link\cdots\link}\otimes
		\frac{W_\ell^\alpha\varepsilon_\alpha
			W^\beta_{\ell}W_{\ell,\beta}}{X-\link\cdots\link}\cr
    }\right) \cr
\noalign{\vskip 10pt}
&\eqcr\ X \Bigl( R(X)\times \Psi^\alpha(X)\varepsilon_\alpha\
	+\ \Psi^\beta(X)\times R(X)\,\varepsilon_\beta\
	+\ T(X)\times 0\Bigr),\cr
}\qquad\label{KonishiLinkW}
\ee
and hence
\be
\left[\WW(X)\Psi^\alpha(X)\right]_-\
\eqcr\ 2XR(X)\Psi^\alpha(X).
\label{PSIreseq}
\ee
This anomalous equation of motion has a particularly simple solution for
the on-shell gaugino resolvent:
\be
\Psi^\alpha(X)\ =\ \frac{\zeta^\alpha(X)}{\WW(X)\,-\,2XR(X)}
\label{PSIres}
\ee
where $\zeta^\alpha(X)=\left[\WW(X)\Psi^\alpha(X)\right]_+$ is a spinor-valued
polynomial of $X$ of $\rm degree\le(d-2)$.
(Note $\Psi^\alpha(X)=O(1/X^2)$ for $X\to\infty$ because
$\tr(W^\alpha_\ell)\equiv 0$.)

Finally, consider yet another link variation
\be
\delta\link_\ell\
=\ \frac{\link_\ell}{X-\link\cdots\link}\times
\frac{W^\alpha_\ell W_{\ell,\alpha}\times\varepsilon}{32\pi^2}
\label{linkvarR}
\ee
where $\varepsilon$ is once again an infinitesimal c-number.
This time, the Konishi anomaly comes up to
\be
\frac{1}{32\pi^2}\Tr\left(\frac{\Wsum^\alpha\Wsum_\alpha\,d\delta\link_\ell}
			{d\link_\ell}\right)\
\eqcr\ \varepsilon X\Bigl( R(X)\times R(X)\ +\ 0\ +\ 0\Bigr)
\label{KonishiLinkWW}
\ee
while
\be
\delta W_{\rm tree}\ \eqcr\ \left[\WW(X)R(X)\varepsilon\right]_-\,,
\ee
and therefore, the on-shell gaugino bilinear resolvent $R(X)$ satisfies
the quadratic equation
\be
X\,\left[R(X)\right]^2\ =\ \left[\WW(X)R(X)\right]_-\
\equiv\ \WW(X)\times R(X)\ -\ F(X)
\label{Rreseq}
\ee
where $F(X)=\left[\WW(X)R(X)\right]_+$ is yet another polynomial of $X$
of $\rm degree\le(d-1)$.
Consequently,
\be
R(X)\ =\ \frac{\WW(X)\,\mp\,\sqrt{\WW^2(X)\,-\,4XF(X)}}{2X}
\label{Rformula}
\ee
and according to eqs.~(\ref{MResQ}), (\ref{MResA}), (\ref{Tres})
and (\ref{PSIres}),
{\em\blue \underline{all} the on-shell resolvents of the quiver theory
--- $\MM_{\ell',\ell}(X)$, $T(X)$, $\Psi^\alpha(X)$ and $R(X)$ ---
are meromorphic functions of the coordinates $X$ and $Y$
of the hyperelliptic Riemann surface $\bf\Sigma$ of}
\be
Y^2\ =\ \WW^2(X)\ -\ 4XF(X).
\label{SpectralCurve}
\ee
In particular,
\be
\Psi^\alpha(X,Y)\ =\ \frac{\zeta^\alpha(X)}{Y}
\label{PsiFormula}
\ee
which means that $\blue\bf\Sigma$ is the
{\em\blue Seiberg--Witten spectral curve}
encoding the abelian gauge couplings of the quantum quiver theory
modulo the $Sp(n_{\rm Abel},\ZZ)$ electromagnetic duality group.

The sign choice in \eq{PsiFormula} corresponds to
\be
R(X,Y)\ =\ \frac{\WW(X)\,-\,Y}{2X}\,.
\label{RFormula}
\ee
In the $X\to\infty$ limit $Y\approx\pm\WW(X)$ depending on the sheet
of the Riemann surface $\bf\Sigma$;
on the $Y\approx+\WW(X)$ sheet, the gaugino bilinear resolvent behaves
physically as $R\approx\frac{F(X)}{\WW(X)}=O(1/X)$ while on the other
sheet we have un-physical divergence $R\approx\frac{\WW(X)}{X}=O(X^{d-1})$.
Likewise, for $X\to 0$ $R$ is regular on the first sheet but has an unphysical
pole on the second sheet,
\be
R(X,Y)\ \becomes{X\to 0}\
\cases{ \mathrm{finite} & on the $Y\approx+\WW(X)$ sheet,\cr
	\frac{\beta v^{n_c}s}{X} & on the $Y\approx-\WW(X)$ sheet,\cr }
\label{Rpole}
\ee
so following Cachazo {\it et~al} we shall refer to the two
sheets of $\bf\Sigma$  ``the physical sheet'' and ``the unphysical sheet''.
As in \cite{CSW1,CSW2}, the distinction between the two sheets is clear over
most of the $X$ plane in the weakly coupled regime of the quiver but
becomes blurred in the strongly coupled regime
(except for $X\to\infty$ or $X\to 0$).

\smallskip
\centerline{\blue\large $\star\quad\star\quad\star$}
\smallskip

By this point, we have solved the anomalous equations of motion for
the whole on-shell chiral ring of the quiver in terms of
the three polynomials $t(X)$, $\zeta^\alpha(X)$ and $F(X)$
and one $X$-independent meson matrix $M^{f'}_{\,f}$.
However, there are additional constraints on these parameters
following from a yet another anomalous equation of motion
due to quark-dependent variation of the link field
\be
\delta\Omega_\ell\
=\ \frac{Q_{\ell+1}\varepsilon\aq_\ell}{X-\link\cdots\link}
\label{linkvarM}
\ee
where $\varepsilon$ is an infinitesimal $F\times F$ matrix
in the flavor space.
This time, the Konishi anomaly is
\bea
\frac{1}{32\pi^2}\,\Tr\left(
	\frac{\Wsum^\alpha\Wsum_\alpha\,d\delta\link_\ell}{d\link_\ell}
	\right) &
\eqcr&\frac{1}{32\pi^2}\,\Tr\left(0\ +\ 0\ +\
	\frac{Q_{\ell+1}\varepsilon\aq_\ell\link_{\ell-1}\cdots\link_{\ell+1}}
		{X-\link\cdots\link}
	\otimes\frac{W^\alpha_\ell W_{\ell,\alpha}}{X-\link\cdots\link}
	\right) \hskip 0pt plus 5em \nonumber\\
\noalign{\vskip 10pt}
&=&\tr\Bigl(\varepsilon\,\MM_{\ell+N,\ell+1}(X)\Bigr)
	\times R(X)
\label{KonishiLinkQA}
\eea
while the tree-level superpotential varies according to
\bea
\delta W_{\rm hop} &=&
\gamma\tr\Bigl(\varepsilon\,\MM_{\ell,\ell}(X)\times M_{\ell+1}\Bigr),\\
\noalign{\penalty500\vskip 10pt}
\delta W_{\rm def} &=&
\Bigl[ W'(X)\,\tr\left(\varepsilon\,\MM_{\ell+N,\ell+1}(X)\right)
	\Bigr]_-,\\
\noalign{\penalty500\vskip 10pt}
\delta W_\Sigma &=&
\beta v^{n_c}s\,\tr\left(\varepsilon\,\MM_{\ell,\ell+1}(X)\right)\\
\noalign{\penalty9999}
&=& \frac{\beta v^{n_c}s}{X}\,\tr\left(
	\varepsilon\times\Bigl(
	    \MM_{\ell+N,\ell+1}(X)\,-\,\aq_\ell\link_\ell^{-1}Q_{\ell+1}
	    \Bigr)\right) .\nonumber
\eea
This gives us an anomalous equation
\be
R(X,Y)\times\MM_{\ell+N,\ell+1}(X,Y)\
\eqcr\ \gamma\MM_{\ell,\ell}(X,Y)\times M_{\ell+1}\
+\ \frac{\WW(X)}{X}\times \MM_{\ell+N,\ell+1}(X,Y)\
-\ \frac{C(X)}{X}
\label{MixedEq}
\ee
where $C(X)$ is yet another degree${}\le d-1$ polynomial of $X$.
When combined with the on-shell equations (\ref{MResQ}),
(\ref{MResA}) and (\ref{Rformula}) for the mesonic and gaugino-bilinear
resolvents, \eq{MixedEq} yields a quadratic equation for the meson
matrix $M$, namely
\be
(\gamma\mu M)^2\ +\ \frac{\mu^N}{X}\,\WW(X)\times (\gamma\mu M)\
+\ \mu^N F(X)\
=\ \gamma\mu^2\,\frac{X-\mu^N}{X}\,C(X) .
\label{MquadraticX}
\ee
In the eigenbasis of the quark mass matrix $\mu$ (or rather of the $\mu^N$)
we may sequentially substitute $X=\mu_f^N$ and apply all the resulting
equations at once since the $M$ matrix does not depend on $X$.
Consequently, the right hand side of \eq{MquadraticX} vanishes regardless
of the $C(X)$ polynomial, while the left hand side yields a matrix
equation:
\be
(\gamma\mu M)^2\ +\ \WW(\mu^N)\times (\gamma\mu M)\
+\ \mu^N F(\mu^N)\ =\ 0.
\label{Mquadratic}
\ee

Now consider a quark flavor $f$ with a non-degenerate $\mu_f^N$.
According to \eq{Mquadratic}, the on-shell value of
the meson operator $M^f_{\,f}$ satisfies a quadratic equation which has
two solutions
\be
M^f_{\,f}\ =\ \frac{-\WW(\mu_f^N)\mp
\sqrt{\WW^2(\mu_f^N)-4\mu_f^N F(\mu_f^N)}}{2\gamma\mu_f}\
=\ -\left.\frac{XR(X,\mp Y_{\rm phys})}{\gamma\mu_f}\right|_{X=\mu^N_f}
\label{Mvalues}
\ee
corresponding to two different vacua of the quiver.
The physical identities of these vacua become apparent in the
weakly coupled regime of the theory where $XF(X)\ll\WW^2(X)$
over most of the complex $X$ plane\footnote{%
	We shall see later in this section that
	$F(X)=O(\Lambda^{N(2n_c-F)})$.}:
For the upper-sign solution of \eq{Mvalues} we have
\be
M^f_{\,f}\ \equiv\ \aq^f Q_f\ \approx\ -\frac{\WW(\mu_f^N)}{\gamma\mu_f}
\label{HiggsVacQuantum}
\ee
precisely as in \eq{HiggsVac} --- which strongly suggest that this is the
discrete mesonic vacuum with one of the Coulomb moduli
frozen at $\ev_j^N=\mu_f^N$ by the squark VEVs \eq{HiggsVacQuantum}.

To confirm the frozen Coulomb modulus in the fully-quantum language of
the chiral ring we turn to the link resolvent $T(X,Y)$ and check its
analytic structure near $X=\mu^N$: According to \eq{polecount},
a frozen Coulomb modulus will manifest itself via a simple pole of
residue exactly $+1$; more generally, $k$ Coulomb moduli frozen at the
same value yield a pole of residue $k$.
According to \eq{Tres}, the poles of $T(X,Y)$ at finite $X$ follow
from the poles of the mesonic resolvent $\MM_{\ell+1,\ell}(X,Y)$
at $X=\mu_f^N$ and have residues
\bea
\Res_{X=\mu_f^N}\Bigl[T(X,Y)\Bigr] &
=& -\frac{\gamma}{Y}\times\Res_{X=\mu_f^N}
	\Bigl[\tr \MM_{\ell+1,\ell}(X)\Bigr]. \\
\langle\!\langle\mbox{by \eq{MResQ}}\rangle\!\rangle && \nonumber\\
&=& -\left.\frac{\gamma\mu_f^{}M_f^f\,+\,XR(X,Y)}{Y}\right|_{X=\mu_f^N}
	\label{Mpoles}\\
\langle\!\langle\mbox{by \eq{Mvalues}}\rangle\!\rangle && \nonumber\\
&=&\cases{+1 & on the physical sheet of $\Sigma$,\cr
	0 & on the unphysical sheet.\cr} \nonumber
\eea
Note that in terms of the spectral curve $\bf\Sigma$, $X=\mu^N$ describes
two distinct points but only one of them carries a pole of the link resolvent.
For the upper-sign solution (\ref{Mvalues}) the pole is on the physical sheet,
but it moves to the unphysical sheet for the lower-sign solution where
\be
\Res_{X=\mu_f^N}\Bigl[T(X,Y)\Bigr]\
=\,\cases{ 0 & on the physical sheet of the $\bf\Sigma$,\cr
	+1 & on the unphysical sheet.\cr }
\label{Tpoles}
\ee
Only the physical sheet of $\bf\Sigma$ is visible at the classical and
perturbative levels of the string theory, so the upper-sign solution
(\ref{Mvalues}) indeed has a Coulomb modulus frozen at $X=\mu_f^N$ but
the lower-sign solution does not have any frozen moduli.
Classically, this goes along with the zero meson VEV, but the quantum
corrections generate $M^f_{\,f}\neq 0$; however, in the weak coupling regime
\be
M^f_{\,f}\ \equiv\ \aq^f Q_f\
\approx\ -\frac{\mu_f^{N-1}F(\mu_f^N)}{\gamma\WW(\mu_f^N)}\
\propto\ \Lambda^{N(2n_c-F)}\ \longrightarrow\ 0.
\label{WeakMeson}
\ee

Now consider a degenerate pair of quark flavors, say $\mu_1^N=\mu_2^N\neq 0$.
In this case, the corresponding $2\times2$ block $M_2$ of the meson matrix $M$
satisfies the quadratic equation (\ref{Mquadratic}) as a matrix,
so each of the two block's eigenvalue may independently choose either
root~(\ref{Mvalues}).
This gives us two discrete solutions plus continuous family:
\par\begin{enumerate}
\item Both eigenvalues pick the bigger root (the upper sign in \eq{Mvalues}).
    This is a discrete mesonic vacuum where two Coulomb moduli are frozen
    at the same value $\ev_1^N=\ev_2^N=\mu_{1,2}^N$ by two quark flavors. 
\item One eigenvalue picks the bigger root and the other picks the smaller
    root --- which gives us a continuous family of solutions parameterizing
    the spontaneous breakdown $SU(2)\to U(1)$ of the flavor symmetry.
    This is the continuous mesonic branch of the quiver where two degenerate
    flavors freeze one Coulomb modulus at $\ev_1^N=\mu_{1,2}^N$.
    Classically, the $M_2$ block of the meson matrix satisfies
    $\rank(M_2)=1$ as well as the trace condition~(\ref{FmesonicDeformed});
    in the quantum theory, the $M_2$ has two non-zero eigenvalues
    but one of them is much smaller than the other (in the weak coupling
    regime) while the trace condition remains unchanged.
\item Both eigenvalues pick the smaller root (the lower sign in \eq{Mvalues}).
    This is a discrete Coulomb vacuum where none of the $\ev_j^N$ are
    frozen at $X=\mu_{1,2}^N$, and in the weakly coupled regime,
    the whole $M_2$ meson block becomes small according to \eq{WeakMeson}.
\end{enumerate}
\par\noindent
To confirm our identification of the above vacua we calculate the
poles of the link resolvent $T(X,Y)$ at $X=\mu_{1,2}^N$ and find
\bea
\Res_{X=\mu_{1,2}^N}\Bigl[T(X,Y)\Bigr] &
=& -\left.\frac{\gamma\mu_{1,2}^{}\tr(M_2)\,+\,2XR(X,Y)}{Y}\right|_{X=\mu_f^N}
	\label{Mpoles2}\\
&=& \cases{ k & on the physical sheet,\cr
	2-k & on the unphysical sheet,\cr }
\eea
where $k=0,1,2$ is the number of eigenvalues of $M_2$ equal to the
bigger root of \eq{Mquadratic}.
Again, the pole count {\sl on the physical sheet} gives us the
correct number of frozen Coulomb vacua for each of the three solutions.

Likewise, for $m>2$ degenerate flavors we have two discrete solutions
plus $m-1$ continuous families distinguished by the number $k=0,1,2.\ldots,m$
of meson eigenvalues which pick the bigger root of the quadratic equation.
The residue calculation yields
\be
\Res_{X=\mu_{1,2}^N}\Bigl[T(X,Y)\Bigr]\
=\,\cases{ k & on the physical sheet,\cr
	m-k & on the unphysical sheet,\cr }
\label{TpolesGen}
\ee
which confirms the physical meaning of $k$ as the number of Coulomb moduli
frozen at $X=\mu_n$.
Classically, up to $m$ moduli can be frozen at this point, and this is
exactly what we see in the quantum theory as well.

Note that all poles of the link resolvent $T(X,Y)$ on both sheets
of the spectral curve $\bf\Sigma$ have non-negative integer residues.
({\it cf.}\ \eqrange{Tpoles}{TpolesGen}).
For the poles on the physical sheet this follows from the physical
meaning of the poles as frozen Coulomb moduli, the residue being
the number of such moduli frozen at the same point.
The poles on the unphysical sheet do not have a clear physical meaning
--- indeed the whole unphysical sheet of $\bf\Sigma$ does not exist
at the perturbative level of the quiver theory ---
but their residues are subject to the same rules as the physical poles
by reasons of analytic continuation:
As one wanders around the parameter space of the quiver theory --- and in
particular changes the roots $\rt_i$ of the deformation $\WW(X)$ ---
the point $X=\mu_f^N$ may cross the branch cut of the Riemann surface
$\bf\Sigma$ and move the pole from the unphysical sheet to the physical sheet
or vice verse.
Thus any pole is physical {\sl somewhere} in the parameter space
of the theory --- and that's why in \eq{TpolesGen} the residue $m-k$
on the unphysical sheet has the same spectrum of values as the
physical sheet's residue~$k$, namely $m-k=0,1,2,\ldots,m$.

Actually, for the \suq\ quiver theory with some exactly
massless quark flavors ($i.\,e.$, for $\Delta F>0$),
the link resolvent $T(X,Y)$ has a pole
stuck to the unphysical sheet at $X=0$.
Indeed, because of the unphysical pole~(\ref{Rpole}) of the $R(X,Y)$
at $(X=0,Y_{\rm unphys})$,
the massless block of the mesonic resolvent $\MM_{\ell+1,\ell}$
also have a pole of residue $(+\beta v^{n_c}s=-Y)\gamma^{-1}\delta_f^{f'}$
at this point ({\it cf.}\ eqs.~(\ref{MResQ}) and (\ref{MResA})),
and consequently
\be
T(X,Y)\becomes{X\to 0}\
\cases{ \mathrm{finite} & on the physical sheet,\cr
	\frac{\Delta F}{X} & on the unphysical sheet.\cr }
\label{BadPole}
\ee
Note that despite its unphysical nature, this
pole has a non-negative integer residue anyway.
Again, this is related to our ability to move the poles all over the
spectral curve~$\Sigma$ --- including the physical sheet ---
by changing the parameters of the theory.
Indeed, having $\Delta F$ exactly massless quark flavors is simply
a choice of parameters $\mu_{n_f^{}+1},\ldots,\mu_{n_F^{}+\Delta F}$,
and once we change these parameters to $\mu'\neq 0$, the
pole~(\ref{BadPole}) moves away from zero to $X=\mu^{\prime N}$, which
may end up on the physical sheet when $\mu^{\prime N}$ crosses a branch
cut of the spectral curve.

\subsection{Analytic Considerations}
Having learned what we could from the anomaly equations of the quiver,
let us now consider the analytic properties of the link resolvent $T(X,Y)$.
As in \cite{CSW1,CSW2}, the differential $T(X,Y)dX$ is meromorphic, and
has exactly integer periods (in units of $2\pi i$) for all closed contours
on the Riemann surface~$\bf\Sigma$, including the little contours around
the poles at $X=\mu_f^N$ as well as big contours around the branch cuts
of $\bf \Sigma$ over the $X$ plane.
Consequently,
\be
T(X,Y)\,dX\ =\ \frac{d\,\Xi}{\Xi}
\label{Tformula}
\ee
for some meromorphic function $\Xi(X,Y)$, and since all poles of $T$ at
finite $X$ have positive residues, it follows that $\Xi(X,Y)$
has zeros rather than poles at $X\neq\infty$.
Furthermore, the product $\Xi(X,+Y)\times\Xi(X,-Y)$ is a single-valued
function of $X$, has a simple zero at each $X=\mu_f^N$ (regardless of the
corresponding pole of $T(X,Y)$ being on the physical or the unphysical sheet),
and has no essential singularity at $X=\infty$, which immediately
implies polynomial behavior
\be
\Xi(X,+Y)\times\Xi(X,-Y)\ =\ \alpha B(X)\
\equiv\ \alpha\prod_{f=1}^F(X-\mu_f^N)\
=\ \alpha\,X^{\Delta F}\prod_{f=1}^{n_f}(X-\mu_f^N)
\label{XiProd}
\ee
for some constant ($i.\,e.\ X$--independent) $\alpha$.
Also, for $X\to\infty$ on the physical sheet $T(X,Y)\approx n_cX^{-1}$,
which translates to $\Xi(X,Y)\propto X^{n_c}$ --- and therefore
on the unphysical sheet, $\Xi(X,Y)\propto X^{F-n_c}$.
For $F\le 2n_c$ this means that the sum $\Xi(X,+Y)+\Xi(X,-Y)$
--- which is also a single-valued holomorphic function of $X$ ---
grows like $X^{n_c}$ at $X\to\infty$, hence
\be
\Xi(X,+Y)\,+\,\Xi(X,-Y)\ =\ P(X)\
\equiv\ \prod_{j=1}^{n_c}(X-\varpi_j)
\label{XiSum}
\ee
is another polynomial of $X$ of degree $n_c$
with some roots $\varpi_1,\ldots,\varpi_{n_c}$
 --- and therefore
\be
\Xi(X,Y)\ =\ \half\left( P(X)\pm\sqrt{P^2(X)\,-\,4\alpha B(X)}\right).
\label{Xiformula}
\ee
Finally, to assure that this function is single-valued on $\bf\Sigma$,
the ratio
$$
\frac{\sqrt{P^2(X)-4\alpha B(X)}}{Y}\
=\ \sqrt{\frac{P^2(X)-4\alpha B(X)}{\WW^2(X)-4XF(X)}}
$$
must be a rational function of $X$
--- which means polynomial factorization
\bea
\WW^2(X)\ -\ 4XF(X) &=& H(X)\times K^2(X), \label{Yfactor}\\
P^2(X)\ -\ 4\alpha B(X) &=& H(X)\times G^2(X), \label{Tfactor}
\eea
where $H(X)$, $K(X)$ and $G(X)$ are three more polynomials of $X$
of respective degrees $2h$, $(d-h)$ and $(n_c-h)$
where $d=\mathop{\rm deg}(\WW)$ and $h=1,2,\ldots,\min(d,n_c)$
depending on a particular vacuum of the quantum quiver theory.
Specifically, a vacuum which supports $n^{\rm free}_{\rm photon}$
free massless photon abelian gauge fields should have
$h=n^{\rm free}_{\rm photon}+1$
and the $H(X)$ polynomial should have $2h$ distinct simple roots.
Indeed, according to eqs.~(\ref{Yfactor}) and (\ref{PsiFormula}),
these are conditions for the gaugino resolvent $\Psi(X)$ to have
precisely $h$ branch cuts and therefore $h-1=n^{\rm free}_{\rm photon}$
independent zero modes corresponding to $h-1$ species of free massless photons.%
\footnote{%
	In terms of the contour integrals $\oint_{\cal C}\Psi^\alpha(X)dX$,
	contours around branch cuts yield free photons while contours
	around poles (if any) yield photons interacting with massless
	quarks and antiquarks (or monopoles and antimonopoles, dyons
	and antidyons, {\it etc., etc.}
	In the deformed quiver theory, such massless quarks, {\it etc.},
	generally develop non-zero VEVs, so only the free photons
	remain exactly massless.
	In terms of the resolvent $\Psi^\alpha(X)$ it means no poles
	but only branch cuts, hence in \eq{PsiFormula} the spinor-valued
	polynomial $\zeta^\alpha(X)$ must factorize as 
	$\zeta^\alpha(X)=K(X)\times\hat\zeta^\alpha(X)\ \Longrightarrow\
	\Psi^\alpha(X)=\pm\frac{\hat\zeta^\alpha(X)}{\sqrt{H(X)}}$
	where $\hat\zeta^\alpha(X)$ has rank $h-2$ and hence
	$h-1=n^{\rm free}_{\rm photon}$ independent spinor-valued
	coefficients $\Longrightarrow$ zero modes.
	}

But regardless of a particular solution, \eqrange{Yfactor}{Tfactor}
allow us to rewrite the Seiberg--Witten curve of the quiver
in terms of the parameters of the link resolvent $T(X)$ rather than
gaugino condensate resolvent.
Specifically, for $\hat Y=\frac{G(X)}{2K(X)}Y+\frac{\WW(X)}{2}$ we have
\be\blue
{\bf\Sigma}\ =\ \{(X,\hat Y):\
	\hat Y^2\ -\ P(X)\times \hat Y\ +\ \alpha B(X)\ =\ 0\}
\black,
\label{SeibergWitten}
\ee
which looks exactly like the SW curve \cite{HananyOz,ArgyresPlesserShapere}
of the $\NN=2$ SQCD$_4$ with $n_c$ colors and $F$ flavors.
This similarity is no accident but a reflection of the ultra-low
energy limit of the quiver theory:
For finite $N$, energies $E\ll 1/(Na)$ are governed by the dimensional
reduction of the deconstructed \sqcdv\ to 4D --- which is nothing but
the $\NN=2$ SQCD$_4$ whose gauge symmetry is the diagonal $SU(n_c)$
subgroup of the \suq\ quiver.
In this effective $\NN=2$ SQCD$_4$, the $\alpha$ parameter in
\eq{SeibergWitten} is generated at the one-instanton level of the
diagonal $SU(n_c)$;
in terms of the whole quiver, this means one instanton in each and
every $SU(n_c)_\ell$ factor according to \eq{InstantonCount}.
Indeed, it is easy to see that were we to turn off quantum effects
in a single  $SU(n_c)_\ell$ factor (by turning off $\Lambda_\ell\to 0$),
then the low-energy limit of the whole quiver theory would become that
classical $SU(n_c)_\ell$ gauge theory with $F$ quark flavors and
an adjoint chiral field $\Phi_\ell$
made from $\link_{\ell-1}\cdots\link_1\link_N\cdots\link_\ell$,
and without the IR-visible quantum effects we would have $\alpha=0$.
Hence, in the fully-quantum quiver
\be
\alpha\ \sim\ \exp\left( -\frac{8\pi^2}{g_{\rm diag}^2}\right)\
\sim\ \prod_\ell\Lambda_\ell^{2n_c-F}\
=\ \left(\Lambda^{2n_c-F}\right)^N
\label{Alevel}
\ee
where the last equality assumes
$\Lambda_\ell\equiv\Lambda$, as required by the
discretized translational invariance of the deconstructed \sqcdv.

To verify that $\alpha\neq 0$ is indeed generated at the
one-diagonal-instanton level --- and also to determine
the pre-exponential factors in \eq{Alevel} --- we use the anomalous
$U(1)_A\times U(1)_\link\times U(1)_R$ symmetry of the quiver which
transforms both the chiral fields and the couplings according to
the following charge table:
\be
\arrayrulewidth=0.5pt
\doublerulesep=1pt
\def\vsr{\vrule width 0pt height 15 pt depth 5pt}
\begin{tabular}{||c||c|c|c||@{\vsr}}
\hline\hline
FIELDS \&		&	\multicolumn{3}{c||@{\vsr}}{CHARGES} \\
\cline{2-4}
PARAMETERS		&	$q_A$	&	$q_\link$ &	$q_R$ \\
\hline\hline
$Q_\ell$		&	1	&	0	&	0 \\
\hline
$\aq_\ell$		&	1	&	0	&	0 \\
\hline
$\link_\ell$		&	0	&	1	&	0 \\
\hline
$W^\alpha_\ell$		&	0	&	0	&	1 \\
\hline\hline
$\mu_f$			&	0	&	1	&	0 \\
\hline
$\gamma$		&	-2	&	-1	&	2 \\
\hline
$\Lambda_\ell^{2n_c-F}$	&	$2F$	&	$2n_c$	&	$-2F$ \\
\hline\hline
$X$			&	0	&	$N$	&	0 \\
\hline
$\WW(X)$		&	0	&	0	&	2 \\
\hline
$F(X)$			&	0	&	$-N$	&	4 \\
\hline
$\alpha$		&	0	&  $(2n_c-F)N$	&	0 \\
\hline\hline
\end{tabular}
\label{Charges}
\ee
In light of these charges, the only allowed formula for $\alpha$
of the type $\alpha=\Lambda^{\rm power}\times{}$something else is
\be
\alpha\
=\ \left(\Lambda^{2n_c-F}\gamma^F\right)^N\times
\mbox{a numeric constant},
\label{Aformula1}
\ee
so the one-diagonal-instanton level of \eq{Alevel} is indeed correct.

Finally, the numeric factor in \eq{Aformula1} follows from the
decoupling of ultra-heavy quark flavors from the low-energy chiral
ring of the quiver.
Indeed, suppose one of the $n_f$ massive quark flavors of the quiver theory
becomes ultra-heavy, {\it eg.}\ $\gamma\mu_1\to\infty$.
In this limit, the chiral ring equations should approximate those
of the theory without the ultra-heavy flavor;
in particular, in \eq{XiProd} we should have
\be
\forall\ \mbox{fixed finite}\ X:\
\left.\alpha B(X)\right|_{n_f^{}}\ \longrightarrow\
\left.\alpha' B'(X)\right|_{n'_f=n^{}_f-1}\
\Longrightarrow\ \alpha'\ =\ \alpha\times(-\mu_1^N).
\ee
At the same time, for each $SU(n_c)_\ell$ gauge factor of the quiver
we decouple one ultra-heavy quark flavor of mass $(-\gamma\mu_a^{})\to\infty$
out of $n_c+F$ effective flavors altogether, hence
\be
\forall\ell:\ (\Lambda'_\ell)^{2n_c-F+1}\
=\ (\Lambda_\ell^{})^{2n_c-F}\times(-\gamma\mu_a^{}).
\label{Lambdaformula}
\ee
Consequently, mutual consistency of \eqrange{Aformula1}{Lambdaformula}
completely determines
\be
\alpha\ =\ (-1)^F\left((-\gamma)^F\Lambda^{2n_c-F}\right)^N.
\label{Aformula}
\ee

\smallskip
\centerline{\blue\large $\star\quad\star\quad\star$}
\smallskip

For the $\NN=2$ SQCD$_4$, roots of the $P(X)$ polynomial
in \eq{SeibergWitten} are the Coulomb moduli of the theory;
classically, $P=\det(X-\Phi)$ and the roots are the eigenvalues
of the adjoint field.
The same holds true for the \suq\ quiver theory:
In the weak coupling limit $\alpha\to 0$,
\eq{Xiformula} yields $\Xi(X,Y)\approx P(X)$
{\it on the physical sheet of the spectral curve $\bf\Sigma$}
--- and therefore
\be
\left.T(X)\right|_{\rm physical\atop sheet}\
\approx\,\sum_{j=1}^{n_c}\frac{1}{X-\varpi_j}\,,
\label{Tlimit}
\ee
exactly as in the classical formula (\ref{Tclassical}),
with the roots $\varpi_j$ of the $P(X)$ polynomial playing
the role of the classical Coulomb moduli $\ev_j^N$.
In other words, classically $\varpi_j\equiv\ev_j^N$ and
$P(X)=\det(X-\link_N\cdots\link_1)$.
In the quantum theory, we need to redefine the Coulomb moduli
because the link eigenvalues $\ev_j$ are no longer chiral
gauge-invariant operators;
in light of \eq{Tlimit}, we {\it define} the Coulomb moduli
of the quantum quiver as the roots of the $P(X)$,
\be
\ev_j^N\ \eqdef\ \varpi_j\ \ i.\,e.\quad
P(X)\ \equiv\ \prod_{j=1}^{n_c}(X-\ev_j^N).
\label{CoulombModuli}
\ee
For $F<n_c$, this is equivalent to identifying $P(X)$ with
the $\det(X-\link_N\cdots\link_1)$ operator in the quantum theory,
but for $F\ge n_c$ the relation is more complicated;
we shall return to this issue in section~\S4.3.

\par\vskip 0pt plus 20pt\goodbreak\begingroup
\postdisplaypenalty=10000
For a deformed quiver with $\WW(X)\not\equiv0$, the Coulomb
moduli~(\ref{CoulombModuli})
are constrained by the need to satisfy both \eqrange{Yfactor}{Tfactor}
at the same time as well as
$$
\refstepcounter{equation}
\prod_{j=1}^{n_c}\ev_j^N\
\equiv\ (-1)^{n_c}P(X=0)\ =\ V^{Nn_c}\
=\ \Bigl[v^{n_c}\,+\,\mbox{quantum corrections}\Bigr]^N
\label{DetCon}\eqno (\theequation)\rlap{\strut\footnotemark}
$$
\footnotetext{%
	According to \eq{Pzero} in \S4.3, $V^{Nn_c}=V_1^{Nn_c}+V_2^{Nn_c}$,
	where $V_{1,2}^{n_c}$ are the two roots of \eq{V1V2}.
	}%
which translates the determinant constraint (\ref{CRdet})
into the language of the Coulomb moduli (\ref{CoulombModuli}).
Consequently, the continuous moduli space of the un-deformed
deconstructive quiver breaks into a discrete set of solutions
describing different vacua of the deformed quiver.
For example, let us consider the $[SU(2)]^N$ quiver with one quark
flavor and degree $d=2$ deformation superpotential
$\W(X)=\nu_2(\half X^2+AX)$.
For this quiver, we have one $h=2$ solution and five $h=1$
solutions.
\endgroup
The $h=2$ solution
\bea
P(X) &=& X^2\ +\ AX\ +\ V^{2N}, \nonumber\\
B(X) &=& X\ -\ \mu^N, \nonumber\\
G(X) &\equiv& 1,\quad K(X)\ \equiv\ \nu_2,\nonumber\\
\beta v^{n_c}s &=& \nu_2\sqrt{V^{4N}+4\alpha\mu^N}\,,
	\label{OneSolution}\\
& \hbox{\rput{90}(0,0){$\Longleftarrow$}} &\nonumber\\
\WW(X) &=& \nu_2\left( X^2\ +\ AX\ +\ \sqrt{V^{4N}+4\alpha\mu^N}\right ),
	\nonumber \\
F(X) &=& \frac{\nu_2^2}{2}\left(
	(\sqrt{V^{4N}+4\alpha\mu^N}-V^{2N})\times(X+A)\
	+\ 2\alpha\right),\nonumber
\eea
describes the unique discrete Coulomb vacuum with unbroken
$U(1)\subset SU(2)_{\rm diag}$.
Classically, this calls for $n_1=n_2=1$ `occupation numbers'
for the two roots $\rt_{1,2}$ of the $\WW(X)$ polynomial --- and indeed,
in the weak coupling limit $\alpha\to 0$ we have $\WW(X)\approx\nu_2 P(X)$
and hence $\ev_1^N\approx\rt_1$ and $\ev_2^N\approx\rt_2$.
Also, in this limit
\be
F(X)\ \propto\ \alpha
\Longrightarrow\ \mbox{gaugino condensates}\
{\cal S}_{1,2}\ \propto\ \alpha
\ee
as expected for a purely abelian Coulomb vacuum where gaugino bilinears
are generated by the coset instantons of the $SU(2)_{\rm diag}\big/ U(1)$.
Outside the weak coupling limit, the Coulomb moduli $\ev^N_{1,2}$
move away from the roots $\rt_{1,2}$ of the $\WW(X)$ polynomial
and the occupation numbers $n_{1,2}$ become ill-defined.
Nevertheless, me may identify the $h=2$ solution with the Coulomb vacuum
of the quiver simply because $h=2$ implies a Seiberg--Witten curve
of genus $g=1$ and hence a single exactly massless photon of the
unbroken $U(1)\subset SU(2)_{\rm diag}$.

The $h=1$ solutions describe mesonic and pseudo-confining vacua
without any massless photons whatsoever.
The general form of these solutions is given by
\bea
P(X) &=& X^2\ -\ (c+2d)X\ +\ V^{2N}, \nonumber\\
B(X) &=& X\ -\ \mu^N, \nonumber\\
G(X) &=& X\ -\ (c+d), \nonumber\\
H(X) &=& (X-d)^2\ +\ 4e, \nonumber\\
K(X) &=& \nu_2(X+A+d), \label{FiveSolutions}\\
\beta v^{n_c}s &=& -\nu_2(A+d)\sqrt{d^2+4e}\,, \nonumber\\
& \hbox{\rput{90}(0,0){$\Longleftarrow$}} &\nonumber\\
\WW(X) &=& \nu_2\left[X^2\ +\ AX\ -\ (A+d)\sqrt{d^2+4e}\right], \nonumber\\
F(X) &=& -\nu_2^2e\left[
	\frac{2(A+d)}{d+\sqrt{d^2+4e}}\,(X+A)\
	+\ X\ +\ 2(A+d)
	\right],\nonumber
\eea
where the parameters $c,d,e$ satisfy a quintic
equation system
\be
c\ +\ d\ -\ \frac{e}{c}\ =\ \mu^N,\qquad
ce\ =\ \alpha,\qquad
2e\ +\ cd\ +\ d^2\ =\ V^{2N},
\label{QuinticSystem}
\ee
hence five solutions.
In the weak coupling limit $\alpha\to 0$, four solutions
\bea
d\ \approx\ \pm V^N,&\qquad&
c\ \approx\ \pm'\sqrt{\alpha\over \pm V^N-\mu^N}\,,\qquad
    e\ \approx\ \pm'\sqrt{\alpha(\pm V^N-\mu^N)}\,,\nonumber\\
& \hbox{\rput{90}(0,0){$\Longleftarrow$}} &\nonumber\\
P(X) &\approx& (X\mp V^N)^2\ \mp'\ 4\sqrt{\alpha(\pm V^N-\mu^N)},
	\label{ConfiningSolutions}\\
\WW(X) &\approx& \nu_2(X\mp V^N)(X+A\pm V^N), \nonumber
\eea
describe two pairs of pseudo-confining vacua
($i.\,e.$, two classical vacua solutions denoted by the un-primed $\pm$ signs,
each giving rise to two quantum vacuum states denoted by the primed
signs~$\pm'\,$),
while the fifth solution
\bea
c\ \approx\ \frac{\mu^{2N}-V^{2N}}{\mu^N}\,,&\quad&
d\ \approx\ \frac{V^{2N}}{\mu^N}\,,\qquad
    e\ \approx\ \frac{\alpha\mu^N}{\mu^{2N}-V^{2N}}\,,\nonumber\\
& \hbox{\rput{90}(0,0){$\Longleftarrow$}} &\nonumber\\
P(X) &\approx& \Bigl(X-{V^{2N}\over\mu^N}\Bigr)\times\Bigl(X-\mu^N\Bigr) ,
	\label{MesonicSolution}\\
\WW(X) &\approx& \nu_2\Bigl(X-{V^{2N}\over\mu^N}\Bigr)
	\times\Bigl(X+A+{V^{2N}\over\mu^N}\Bigr) \nonumber
\eea
describes a discrete mesonic vacuum of the quiver.
Indeed, the solutions~(\ref{ConfiningSolutions}) has both Coulomb moduli /
roots of $P(X)$ captured by the same deformation root,
$\varpi_1\approx\varpi_2\approx\rt_1\ \Longrightarrow{}$ classically
unbroken pseudo-confining $SU(2)_{\rm diag}$,
but the (\ref{MesonicSolution}) solution has only one Coulomb modulus
$\varpi_1\approx\rt_1$ captured by the deformation root while the other
modulus is frozen at the quark mass, $\varpi_2\approx\mu^N$
regardless of the deformation superpotential.
Also, for the fifth solution~(\ref{MesonicSolution}), $T(X,Y)$ has a
pole at $X=\mu^N$ on the physical sheet --- and hence large mesonic VEV
$\vev{M}$ according to \eq{HiggsVacQuantum} --- while the first four
solutions ~(\ref{ConfiningSolutions}) have this pole on the un-physical sheet
--- and hence small mesonic VEVs according to \eq{WeakMeson}.

For all five $h=1$ solutions, the $R(X)$ resolvent has a
single branch cut near the first deformation root $\rt_1\approx d$
while the second deformation root $\rt_2$ remains unoccupied.
In the weak coupling limit the branch cut becomes a pole,
and indeed all five solutions have
\be
\left.R(X)\right|^{}_{\rm physical\atop sheet}\ \approx\
\frac{F(X)}{\WW(X)}\ \approx\ -\frac{\nu_2 e(2+A/d)}{X-d}
\ee
with a single pole of residue
\be
{\cal S}\ =\ -\nu_2(2+A/d)e\ =\ \cases{
	O(\sqrt{\alpha}) & for the first four solutions
		 (\ref{ConfiningSolutions}), and\cr
	O(\alpha) & for the fifth solution
		 (\ref{MesonicSolution}).\cr }
\ee
This gives us yet another way of distinguishing between the mesonic and the
pseudo-confining vacua of a weakly coupled quiver:
The solutions with ${\cal S}\propto\alpha^{1/2}$ describe the
pseudo-confining vacua with classically unbroken $SU(2)_{\rm diag}$
which produces the gaugino condensate at the half-instanton level,
while the solution with ${\cal S}\propto\alpha^1$ describes the
discrete mesonic vacuum where the $SU(2)_{\rm diag}$ is broken at
the classical level and it takes a whole (diagonal) instanton to
generate ${\cal S}\neq 0$.

For the strongly coupled quiver theory, the distinction between the
mesonic and the pseudo-confining $h=1$ solutions becomes moot.
Indeed, for large $\alpha$ all five solutions have both Coulomb moduli /
roots of $P(X)$  nowhere near deformation roots $\rt_{1,2}$ or
the quark mass $\mu^N$, so the occupation numbers are of no use.
Likewise, all five solutions have large mesonic VEVs as well as
large gaugino condensates $\cal S$ which depend on $\alpha$ in
a complicated way.
Furthermore, all five solutions are permuted into each other by
monodromies of the $(\mu^N,\alpha,V^{2N})$ parameter space
around the discriminant locus of eqs.~(\ref{QuinticSystem}),
hence it's mathematically impossible to find any global distinction
which remains valid for all parameter values.
Physically, this means that the difference between the pseudo-confining
and the mesonic vacua of the quantum quiver theory is an artifact
of the weak coupling approximation.
At strong coupling, there are simply five photon-less vacua,
without any meaningful way of telling which is mesonic and which is
pseudo-confining:
This is an example of the $\rm confinement\leftrightarrow Higgs$ duality
which shows up in all kinds of supersymmetric gauge theories.

Other quiver theories with more colors and/or flavors exhibit the
same general behavior:
In the weak coupling regime of the deformed quiver, the solutions
of \eqrange{Yfactor}{Tfactor} and (\ref{DetCon}) approximate
the classical vacua discussed in section \S2.4; in the $\alpha\to 0$
limit, the approximation becomes exact.
When the coupling becomes strong we still have exactly the same
number of solutions, but their physical identities become blurred
and only the net abelian rank $n_{\rm Abel}=h-1$ and the number
$n_{\rm mes}$ of {\it continuous} mesonic moduli remain clear.
Generally, all vacua of the same $n_{\rm Abel}$ and $n_{\rm mes}$
are connected by the monodromies of the parameter space and thus
cannot be physically told apart in the strongly coupled regime.

Finally, consider what happens when we un-deform a deconstructive
quiver by removing the deformation superpotential~(\ref{Wdef}).
Specifically, let us take to zero the leading coefficient $\nu_d$
of the polynomial $\WW(X)$ while keeping its roots $\rt_1,\ldots,\rt_d$
finite.
In this limit, \eq{Yfactor} yields $K(X)\propto\nu_d^{}$,
$F(X)\propto\nu_d^2$, hence $Y\propto\nu^{}_d$,
$R(X)\propto\nu^{}_d$ and therefore all gaugino condensates diminish
as $O(\nu_d)$.
And according to \eq{Mvalues}, all {\sl discrete} mesonic VEVs
also diminish as $O(\nu_d)$.
Thus, in the undeformed limit $\nu_d=0$, the pseudo-confining and the
discrete mesonic vacua lose their characteristic VEVs and become
indistinguishable from the Coulomb vacua which just happen to have
similar moduli $(\ev^N_1,\ldots,\ev^N_{n_c})$.

Ultimately, for $\nu_d=0$ \eq{Yfactor} becomes a trivial $0=0$
identity and the roots (\ref{CoulombModuli})
become free moduli of the continuous Coulomb moduli space.
For generic values of these moduli, the Seiberg--Witten spectral
curve~(\ref{SeibergWitten}) has maximal genus $g=n_c-1$
describing a purely-Coulomb vacuum with a maximal number
$n_{\rm Abel}=n_c-1$ of massless photons.
There are also interesting curves of reduced genus, eg.\
$g=n_c-2$ when $B(X)$ has a double zero and $P(X)$ has a simple zero
at {\it exactly} the same point ---
physically, this corresponds to a mesonic vacuum family with a frozen
Coulomb modulus at a degenerate quark mass, $\ev_1^N=\mu_1^N=\mu_2^N$.
Finally, there are reduced-genus curves at special loci in the
Coulomb moduli space where some roots of the $P^2(X)-4\alpha B(X)$
just happen to coincide.
In this case, the low-energy theory retains it full complement of
$n_{\rm Abel}=n_c-1$ photons but in addition
some electrically or magnetically charged particles become exactly massless,
and the spectral curve has a reduced rank
$g<n_{\rm Abel}$ because some  photons are no longer free in the
infrared limit.
This works exactly as in the $\NN=2$ SQCD$_4$ --- and indeed 
the very-low-energy limit of the \suq\ quiver theory
{\bf is} the $\NN=2$ SQCD$_4$ with the $SU(n_c)_{\rm diag}$ gauge
symmetry.\footnote{%
	From the deconstruction point of view, this effective
	very-low-energy theory is the
	Kaluza--Klein reduction of the \sqcdv\  on a circle of size
	$2\pi R=Na$.
	and any 4D non-perturbative effects at $E<(Na)^{-1}$
	are artifacts of the finite-size compactification.
	However, as far as a finite--$N$ quiver is concerned,
	the non-perturbative effects associated with the low-energy
	$SU(n_c)_{\rm diag}$ are just as important as any other
	non-perturbative effect in the theory.
	}

This completes our study of the \suq\ quiver's chiral ring ---
or rather its sub-rings generated by the non-baryonic operators.
The baryonic and the anti-baryonic generators are presented
in the following section~\S4.

 
%
%
%
\section{Baryons and Other Determinants}
%
%
In this section we complete the chiral ring of the \suq\ quiver theory
by adding all kinds of baryonic and antibaryonic generators.
Note that besides the ordinary baryonic operators
$B^{(\ell)}_{f_1,\ldots,f_{n_c}}=\epsilon_{j_1,\ldots,j_{n_c}}
Q^{j_1}_{\ell,f_1}\cdots Q^{j_{n_c}}_{\ell,f_{n_c}}$ comprised
of $n_c$ quarks at the same quiver node~$\ell$, there is a great multitude
of baryon-like gauge-invariant operators where the quarks sit
at different nodes $\ell_1,\ell_2,\ldots,\ell_{n_c}$
 but are connected to each other via chains of link fields.
Or rather each quark is connected by a link chain
$\link_{\ell-1}\link_{\ell-2}\cdots\link_{\ell_i}$ to a common node $\ell$
where $n_c$ quark indices of the $SU(n_c)_\ell$ are combined into a gauge
singlet,
\be\eqalign{
B_{\,f_1,f_2,\ldots, f_{n_c}}(\ell;\ell_1,\ell_2,\ldots,\ell_{n_c})\
    =\ \epsilon_{j_1,j_2,\ldots,j_{n_c}}
    \Bigl(\link_{\ell-1}\cdots \link_{\ell_1}Q_{\ell_1,f_1}\Bigr)^{j_1}&
\Bigl(\link_{\ell-1}\cdots \link_{\ell_2}Q_{\ell_2,f_2}\Bigr)^{j_2}\cdots{}\cr
{}\cdots{}&\Bigl(\link_{\ell-1}\cdots \link_{\ell_{n_c}}Q_{\ell_{n_c},f_{n_c}}
	\Bigr)^{j_{n_c}}\,.\cr
}\ee
{}From the 5D point of view, these are {\it multi-local} operators which
create/annihilate {\it un-bound} sets of $n_c$ quarks while the link
chains deconstructs the un-physical Wilson strings which allow for
manifest gauge invariance of such multi-local operators
\bea
[B(x;x_1,\ldots,x_{n_c})]_{f_1,\ldots,f_{n_c}}
&=& \epsilon_{j_1,\ldots,j_{n_c}}\,\prod_{i=1}^{n_c}
	\mathop{\hbox{\large\rm exp}}\limits_{\rm Path\atop ordered}
	\left(i\!\int_{x_i}^x\!\!dx^\mu A_\mu(x)\right)^{\!j_i}_{\,j'_i}
	Q_{\ell_i,f_i}^{j'_i} \nonumber\\
&\longrightarrow& \epsilon_{j_1,\ldots,j_{n_c}}\,\prod_{i=1}^{n_c}
	\left(\link_{\ell-1}\cdots \link_{\ell_i}Q_{\ell_i,f_i}\right)^{j_i}
		\label{SplitBaryon} \\
&&\qquad \mbox{for}\ x_i^{0,1,2,3}\equiv x^{0,1,2,3},\
	x^4=a\ell\ \mbox{and}\ x_i^4=a\ell_i\,. \nonumber
\eea
However, just like their mesonic counterparts~(\ref{SplitPair}),
from the 4D point of view these are local chiral gauge-invariant operators
and we must include them in the \suq\ quiver's chiral ring.

Note that each of the $n_c$ link chains $\link_{\ell-1}\cdots\link_{\ell_i}$
may wrap a few times around the whole quiver ($\link_{\ell-1}\cdots\link_1
(\link_N\cdots\link_1)^k\link_N\cdots\link_{\ell_i}$), or go in reverse
direction ($\link_{\ell}^{-1}\link_{\ell+1}^{-1}\cdots\link_{\ell_i+1}^{-1}$
for $\ell_i>\ell$), or both, and all these possibilities give
rise to a whole zoo of baryonic generators.
To make sure the readers do not get lost in this big zoo, we would like
to begin our presentation with a smaller menagerie of baryonic operators
in a single $SU(n_c)$ gauge theory with an adjoint field $\Phi$ as well
as quarks and antiquarks.
Once we explore this menagerie in section~\S4.1 below, we shall return
to the baryonic and antibaryonic operators of the \suq\ quiver
in section~\S4.2.
Finally, in section~\S4.3 we shall study quantum corrections to
determinants of link chains (as in $\det(\link_{\ell'}\cdots\link_{\ell})
=\det(\link_{\ell'})\cdots\det(\link_\ell)+{}$corrections).
Of particular importance is the determinant
$\det(\link_N\cdots\link_1)$ of the whole quiver:
It governs quantum corrections to $V^{n_c}=v^{n_c}+\cdots{}$
in \eq{DetCon} and affects the origin of the quivers' baryonic Higgs branch
in the Coulomb moduli space.

\subsection{$\Phi$--Baryons in the Single $SU(n_c)$ Theory}
Consider the chiral ring of the $\NN=1$ $SU(n_c)$ gauge theory
with $n_f$ chiral quark and antiquark fields $Q^f$ and $\aq_f$,
a single adjoint field $\Phi$, and the superpotential
\be
W\ =\ \tr\Bigl(\W(\Phi)\Bigr)\ +\ \aq^{f'}m_{\!f'}^{\,f}(\Phi) Q_f 
\label{DSQCDdef}
\ee
where $m_{\!f'}^{\,f}(\Phi)$ is a matrix-valued polynomial of $\Phi$
of degree $p\ge 1$.
The non-baryonic generators of this ring are \underline{\it exactly} as in
the $U(n_c)$ theory with the same chiral fields:
According to Cachazo {\it et al} \cite{CDSW,Seiberg,CSW1,CSW2},
all of these generators are encoded in just
four resolvents
\be
\eqalign{
T(X)\ =\ \tr\frac{1}{X-\Phi}\,,&
\qquad \Psi^\alpha(X)\ =\ \frac{1}{4\pi}\tr\frac{W^\alpha}{X-\Phi}\,,\cr
R(X)\ =\ \frac{1}{32\pi^2}\tr\frac{W^\alpha W_\alpha}{X-\Phi}\,,&
\qquad \MM^{f'}_{\,f}(X)\ =\ \aq_{f'}\frac{1}{X-\Phi}Q_f\,.\cr
}\ee
Un-gauging the $U(1)$ factor of the $U(n_c)$ makes all kinds of
(anti) baryonic chiral operators gauge-invariant and hence adds them to
the chiral ring of the $SU(n_c)$ theory.
Note that besides the ordinary baryons $B_{f_1,\ldots,f_{n_c}}=
\epsilon_{j_1,\ldots,j_{n_c}}Q^{j_1}_{f_1}\cdots Q^{j_{n_c}}_{f_{n_c}}$
made of $n_c$ quarks and nothing else, there are baryonic generators
comprising  $n_c$ quarks plus any number of the adjoint operators $\Phi$:
\be
B_{f_1,f_2,\ldots,f_{n_c}}(k_1,k_2,\ldots,k_{n_c})\
=\ \epsilon_{j_1,j_2,\ldots,j_{n_c}}\,
\Bigl(\Phi^{k_1}Q_{f_1}\Bigr)^{j_1}
\Bigl(\Phi^{k_2}Q_{f_2}\Bigr)^{j_2}
\Bigl(\Phi^{k_{n_c}}Q_{f_{n_c}}\Bigr)^{j_{n_c}} ;
\label{PhiBaryons}
\ee
henceforth, we shall call such operators $\Phi$--baryons.
Likewise, the anti-baryonic operators include the ordinary antibaryons as well
as $\Phi$--antibaryons
\be
\tilde B^{f_1,f_2,\ldots,f_{n_c}}(k_1,k_2,\ldots,k_{n_c})\
=\ \epsilon^{j_1,j_2,\ldots,j_{n_c}}\,
\Bigl(\aq^{f_1}\Phi^{k_1}\Bigr)_{j_1}
\Bigl(\aq^{f_2}\Phi^{k_2}\Bigr)_{j_2}
\Bigl(\aq^{f_{n_c}}\Phi^{k_{n_c}}\Bigr)_{j_{n_c}}\,.
\label{PhiABaryons}
\ee
Note that the $\Phi$--baryons and the $\Phi$--antibaryons are
antisymmetric with respect to simultaneous permutations of the flavor
indices $f_1,f_2,\cdots,f_{n_c}$ and the $\Phi$--numbers
$k_1,k_2,\ldots,k_{n_c}$, but there is no antisymmetry with respect to
the flavor indices only.
Consequently, the $\Phi$--baryons and the $\Phi$--antibaryons exist
as non-trivial generators of the (off-shell) chiral ring for any number
of flavors $n_f\ge 1$, unlike the ordinary baryons and antibaryons which exist
only when $n_f\ge n_c$.
Indeed, even for a single flavor we can build $\Phi$--baryons such as
\be
B(0,1,2,\ldots,n_c-1)\ =\ \epsilon_{j_1,j_2,\ldots,j_{n_c}}\,
(Q)^{j_1}(\Phi Q)^{j_2}(\Phi^2 Q)^{j_3}\cdots(\Phi^{n_c-1}Q)^{j_{n_c}},\quad
{\it etc.,\ etc.}
\ee
However, we shall see that  all the baryonic and antibaryonic operators
{\it vanish on shell} unless $n_f\times p\ge n_c$.
For the simplest case of (deformed) $\NN=2$ SQCD with linear quark masses
$m^{f'}_{\,f}=\delta^{f'}_f(\Phi-\mu_f)$, this means
{\it no {\blue on-shell} baryonic or antibaryonic generators unless
$n_f\ge n_c$}, and furthermore, all the on-shell $\Phi$--baryons
are completely determined by the ordinary baryons of similar flavors,
and ditto for the $\Phi$--antibaryons.

Similarly to the mesonic resolvent $\MM^{\!f'}_{\,f}(X)$ encoding
both the ordinary mesons and the $\Phi$--mesons $\aq^{f'}\Phi^k Q_f$,
we define the baryonic and antibaryonic resolvents
\bea
\BB_{f_1,f_2,\ldots,f_{n_c}}(X_1,X_2,\ldots,X_{n_c}) &=&
\epsilon_{j_1,j_2,\ldots,j_{n_c}}\,\prod_{i=1}^{n_c}\left(
	\frac{1}{X_i-\Phi}\,Q_{f_i}\right)^{j_i}
	\label{BaryonResolvent} \\
&=& \sum_{k_1,k_2,\ldots,k_{n_c}}
	\frac{B_{f_1,f_2,\ldots,f_{n_c}}(k_1,k_2,\ldots,k_{n_c})}
		{X_1^{k_1+1} X_2^{k_2+1}\cdots X_{n_c}^{k_{n_c}+1}}\,,
		 \qquad\nonumber\\
\AB^{f_1,f_2,\ldots,f_{n_c}}(X_1,X_2,\ldots,X_{n_c}) &=&
\epsilon^{j_1,j_2,\ldots,j_{n_c}}\,\prod_{i=1}^{n_c}\left(
	\aq^{f_i}\frac{1}{X_i-\Phi}\right)_{j_i}
	\label{ABaryonResolvent} \\
&=& \sum_{k_1,k_2,\ldots,k_{n_c}}
	\frac{\tilde B^{f_1,f_2,\ldots,f_{n_c}}(k_1,k_2,\ldots,k_{n_c})}
	{X_1^{k_1+1} X_2^{k_2+1}\cdots X_{n_c}^{k_{n_c}+1}}
		 \nonumber
\eea
encoding all kinds of baryonic and antibaryonic operators.
Unlike the mesonic resolvent, the (anti) baryonic resolvents depend
on $n_c$ independent complex variables $X_1,X_2,\ldots,X_{n_c}$ ---
this is necessary to encode the (anti) baryonic operators
with arbitrary \underline{independent} numbers $k_1,k_2,\ldots,k_{n_c}$
attached to each (anti) quark.
By construction, the (anti) baryonic resolvents
(\ref{BaryonResolvent}--\reftail{ABaryonResolvent})
are antisymmetric with respect
to simultaneous permutations of the flavor indices $f_1,f_2,\ldots,f_{n_c}$
and the arguments $X_1,X_2,\ldots,X_{n_c}$ but not under the separate
permutations;
this allows for the non-trivial {\it off-shell} (anti) baryonic resolvents
for any flavor number $n_f\ge 1$.

The \underline{\it on-shell} baryonic resolvents are subject to equations
of motion stemming from the quark-dependent variations of the antiquark
fields, namely
\be
\delta \aq^f_j\
=\ \left(\frac{1}{X_1-\Phi}\right)_{\!j}^{\,j_1}\times
\varepsilon^{f,f_2^{},\ldots,f_{n_c}^{}}_{}\,
\epsilon_{j_1^{},j_2^{},\ldots,j^{}_{n_c}}^{}
\left(\frac{1}{X_2-\Phi}\,Q^{}_{f_2}\right)^{j_2}\cdots
\left(\frac{1}{X_{n_c}-\Phi}\,Q^{}_{f_{n_c}}\right)^{j_{n_c}}
\label{QDAV}
\ee
where $\varepsilon^{f,f_2^{},\ldots,f_{n_c}^{}}_{}$ is an infinitesimal
tensor in the flavor space ($n_c$ indices, no particular symmetry).
This variation is anomaly free since the $\delta\aq^f_j$
does not depend on the antiquark field $\aq^f_j$ itself,
hence the equation is simply
\bea
\delta W_{\rm tree} &
=& \delta\aq^f_j\times [m_{\!f}^{\,f_1}(\Phi)]^{\!j}_{\,j'} Q^{j'}_{f_1}
	\nonumber\\
&=& \varepsilon^{f,f_2^{},\ldots,f_{n_c}^{}}_{}\,
    \epsilon_{j_1^{},j_2^{},\ldots,j^{}_{n_c}}^{}
    \left(\frac{m_{\!f'}^{\,f_1}(\Phi)}{X_1-\Phi}\,Q^{}_{f_1}\right)^{j_1}
    \left(\frac{1}{X_2-\Phi}\,Q^{}_{f_2}\right)^{j_2}\cdots
    \left(\frac{1}{X_{n_c}-\Phi}\,Q^{}_{f_{n_c}}\right)^{j_{n_c}}
	\nonumber\\
&=& \varepsilon^{f,f_2^{},\ldots,f_{n_c}^{}}_{}\times
    \Bigl[ m_{\!f'}^{\,f_1}(X_1)\,
	\BB^{}_{f_1,f_2,\ldots,f_{n_c}}(X_1,X_2,\ldots,X_{n_c})
	\Bigr]_{-\ \mbox{wrt}\ X_1}\
    \eqcr\ 0
\eea
where the subscript `$-$ wrt $X_1$' means `the negative-power part of the
power series with respect to the $X_1\to\infty$,
but only with respect to the $X_1$'.
Hence,
\be\eqalign{
m_{\!f'}^{\,f_1}(X_1)\times
    \BB_{f_1,f_2,\ldots,f_{n_c}}(X_1,X_2,\ldots,X_{n_c})\,\eqcr\,{}&
\Bigl[ m_{\!f'}^{\,f_1}(X_1)\,
    \BB_{f_1,f_2,\ldots,f_{n_c}}(X_1,X_2,\ldots,X_{n_c})
    \Bigr]_{+\ \mbox{wrt}\ X_1} \cr
{}={}\,&\mathop{\rm Polynomial}(X_1)\ \mbox{of degree}\le p-1 \cr
&\omit with coefficients depending on $X_2,\ldots,X_{n_c}\,$.
	\vrule width 0pt depth 10pt\hfil\cr
}\ee
Furthermore, the $X_2$ dependence of these coefficients is restricted
in exactly the same way as the $X_1$ dependence of the baryonic resolvent
itself, thus
\be
m_{\!f'_1}^{\,f_1}(X_1)\,m_{\!f'_2}^{\,f_2}(X_2)\times
\BB_{f_1,f_2,\ldots,f_{n_c}}(X_1,X_2,\ldots,X_{n_c})\
\eqcr\ \mathop{\rm Polynomial}(X_1,X_2)
\ee
with coefficients depending on the $X_3,\ldots,X_{n_c}$.
Iterating this argument, we arrive at
\be
m_{\!f'_1}^{\,f_1}(X_1)\,m_{\!f'_2}^{\,f_2}(X_2)\,\cdots\,
m_{\!f'_{n_c}}^{\,f_{n_c}}(X_{n_c})\times
\BB_{f_1,f_2,\ldots,f_{n_c}}(X_1,X_2,\ldots,X_{n_c})
\ee
being a polynomial of degree${}\le p-1$ with respect to each of the variables
$X_1,X_2,\ldots,X_{n_c}$;
in other words, the on-shell baryon resolvent has form
\bea
\BB_{f_1,\ldots,f_{n_c}}(X_1,\ldots,X_{n_c}) &\eqcr&
[m^{-1}(X_1)]_{\!f_1}^{\,f'_1}\,
    \cdots\,[m^{-1}(X_{n_c})]_{\!f_{n_c}}^{\,f'_{n_c}}\,\times{}
\label{BaryonFormula}\\
&&{}\times\sum_{k_1=0}^{p-1}\cdots\sum_{k_{n_c}=0}^{p-1}
    b_{f'_1,\ldots,f'_{n_c}}(k_1,\ldots,k_{n_c})\,
	X_1^{k_1}\cdots X_{n_c}^{k_{n_c}}  \nonumber
\eea
for some coefficients $b_{f'_1,\ldots,f'_{n_c}}(k_1,\ldots,k_{n_c})$.
Likewise, the on-shell antibaryonic resolvent has form
\bea
\AB^{f_1,\ldots,f_{n_c}}(X_1,\ldots,X_{n_c}) &\eqcr&
[m^{-1}(X_1)]^{\!f_1}_{\,f'_1}\,
    \cdots\,[m^{-1}(X_{n_c})]^{\!f_{n_c}}_{\,f'_{n_c}}\,\times{}
\label{ABaryonFormula}\\
&&{}\times\sum_{k_1=0}^{p-1}\cdots\sum_{k_{n_c}=0}^{p-1}
    \tilde b^{f'_1,\ldots,f'_{n_c}}(k_1,\ldots,k_{n_c})\,
	X_1^{k_1}\cdots X_{n_c}^{k_{n_c}}  \nonumber
\eea
for some coefficients $\tilde b^{f'_1,\ldots,f'_{n_c}}(k_1,\ldots,k_{n_c})$.

The total antisymmetry of the baryonic and antibaryonic resolvents under
simultaneous permutations of the flavor indices and the variables
$X_1,\ldots,X_{n_c}$ translates into the total antisymmetry of the
$b_{f'_1,\ldots,f'_{n_c}}(k_1,\ldots,k_{n_c})$ and
$\tilde b^{f'_1,\ldots,f'_{n_c}}(k_1,\ldots,k_{n_c})$ coefficient
under simultaneous permutation of the flavors and the 
power indices $k_1,\ldots,k_{n_c}$.
But each power index $k_i$ has only $p$ allowed values $k_i=0,1,\ldots,(p-1)$
and hence each $(k_i,f_i)$ index combination can take only $p\times n_f$
distinct values, which means we cannot possibly antisymmetrize $n_c$ such
index pairs unless $n_c\le p\times n_f$.
Thus, {\it\blue the baryonic and the antibaryonic resolvents --- and hence
all the baryonic and the antibaryonic generators of the chiral ring ---
vanish on shell unless $p\times n_f\ge n_c$.}
In particular, in the (deformed) $\NN=2$ SQCD with linear quark masses
we have
\bea
\BB_{f_1,f_2\ldots,f_{n_c}}^{}(X_1,X_2,\ldots,X_{n_c}) &\eqcr&
\frac{ B_{f_1,f_2,\ldots,f_{n_c}}^{}}{(X_1^{}-\mu_{f_1^{}}^{})
	(X_2^{}-\mu_{f_2^{}}^{})\cdots(X_{n_c}^{}-\mu_{f_{n_c}^{}}^{})}\,,
	\label{BaryonF1}\\
\AB^{f_1,f_2\ldots,f_{n_c}}_{}(X_1,X_2,\ldots,X_{n_c}) &\eqcr&
\frac{\tilde  B^{f_1,f_2,\ldots,f_{n_c}}_{}}{(X_1^{}-\mu_{f_1^{}}^{})
	(X_2^{}-\mu_{f_2^{}}^{})\cdots(X_{n_c}^{}-\mu_{f_{n_c}^{}}^{})}\,,
	\label{ABaryonF1}
\eea
where $B_{f_1,f_2,\ldots,f_{n_c}}$ and $\tilde B^{f_1,f_2,\ldots,f_{n_c}}$
are the ordinary, $\Phi$--less baryon and antibaryon.
And therefore, all the
on-shell baryonic and antibaryonic generators vanish unless $n_f\ge n_c$.

Note that from the baryonic point of view, a theory with
non-renormalizable $\aq\Phi^p Q$ interactions of order $p\ge 2$ behaves
as an effective $\NN=2$ SQCD with $n_f^{\rm eff}=n_f\times p$ flavors.
Likewise, Cachazo {\it et al} showed \cite{CSW2} that
the Seiberg--Witten curve of a $p\ge 2$ theory ---
\be
Y^2\ -\ Y\times P(X)\ +\ \Lambda^{2n_c-pn_f}\,B(X)\ =\ 0
\label{SWsingle}
\ee
where $P(X)$ is a polynomial of degree $n_c$
 and $B(X)=\det(m(X))$ is a polynomial of degree $p\times n_f$
--- looks exactly like the $\NN=2$ SQCD curve for the same number $n_c$
of colors but $n_f^{\rm eff}=p\times n_f$ flavors.
Altogether, as far as the chiral ring is concerned,
a single quark flavor with $\aq\Phi^p Q$ interactions of order $p>1$
is physically equivalent to $p$ ordinary quark flavors.
And since many formul\ae\ look much more complicated for $p\ge 2$
we shall henceforth limit our presentation to the $p=1$ case.%
\footnote{%
	In the quiver theory we have a similar situation where
	a single quark flavor with a $p$-node hopping
	superpotential~(\ref{LongHop}) acts as $p$ ordinary flavors with
	nearest-neighbor hopping~(\ref{Whop}).
	Indeed, classically such a $p$-node-hopping flavor deconstructs
	up to $p$ light 5D flavors, and in the fully-quantum chiral ring of
	the quiver it is physically equivalent to $p$ ordinary flavors,
	although many formul\ae\ become more complicated.
	For example, \eq{XiProd} for the $B(X)$ becomes
	\be
	\alpha B(X)\ =\ \Lambda^{N(2n_c-pF)}\times
	\Det\left( X^{1/N}\delta_{\ell',\ell}\delta_{f'}^f\
		-\ (\Gamma_{\ell'-\ell})_{\!f'}^{\,f}\right)\quad
	\mbox{with respect to both}\ \ell',\ell\ \mbox{and}\
	f',f\ \mbox{indices}
	\ee
	(this is actually a polynomial of $X$ of degree $pF$ because of the
	translational $\ZZ_N$ invariance of the quiver),
	while formul\ae\ for the on-shell mesonic and baryonic resolvents
	would take pages simply to write down, never mind the derivations.
	Consequently, we have limited our presentation to the simpler case
	of the nearest-neighbor hopping.
	}

Classically, the baryonic branches of the moduli space have overdetermined
Coulomb moduli:
\be
B_{f_1,\ldots,f_{n_c}}\neq 0\ {\bf or}\ \tilde B^{f_1,\ldots,f_{n_c}}\neq 0\quad
{\bf requires}\quad (\varpi_1,\ldots,\varpi_{n_c})\
=\ (\mu^{}_{f_1^{}},\ldots,\mu^{}_{f_{n_c^{}}^{}}),
\label{ClassicalBB}
\ee
which in turn requires $\mu^{}_{f_1^{}}+\cdots+\mu^{}_{f_{n_c^{}}^{}}=0$
because $\varpi_1+\cdots+\varpi_{n_c}=\tr(\Phi)=0$.
In the quantum theory, a non-zero on-shell value of a baryonic or an
antibaryonic operator also over-determines the Coulomb moduli,
although their exact values receive quantum corrections.
To see how this works, consider a baryonic resolvent
$\BB_{f_1,\ldots,f_{n_c}}(X_1,\ldots,X_{n_c})$
at a point where the variables $X_1,\ldots,X_{n_c}$ are all equal to each other.
By construction (\ref{BaryonResolvent}),
\be
\BB_{f_1,\ldots,f_{n_c}}(X_1=\cdots=X_{n_c})\
=\ \epsilon_{j_1,\ldots,j_{n_c}}\prod_{i=1}^{n_c}
	\left(\frac{1}{X-\Phi}\,Q_{f_i^{}}^{}\right)^{\!j_i}\
=\ \det\left( \frac{1}{X-\Phi}\times Q_{f^{}_1,\ldots,f^{}_{n_c^{}}}\right)
\ee
where $X$ denotes $X_1=X_2=\cdots=X_{n_c}$ and
$Q_{f^{}_1,\ldots,f^{}_{n_c^{}}}$ is the $n_c\times n_c$ block of the
$\rm color\times flavor$ quark matrix corresponding to flavors
$f_1,\ldots,f_{n_c}$.
Note $\det(Q_{f^{}_1,\ldots,f^{}_{n_c^{}}})=B_{f^{}_1,\ldots,f^{}_{n_c^{}}}$
and therefore
\bea
\det\left( \frac{1}{X-\Phi}\times Q_{f^{}_1,\ldots,f^{}_{n_c^{}}}\right)
&=& \det\left(\frac{1}{X-\Phi}\right)\times B_{f^{}_1,\ldots,f^{}_{n_c^{}}}\,
    +\, \mbox{instantonic corrections}\hskip 45pt\\
&\eqcr& \langle\!\langle\mbox{quantum corrected}\rangle\!\rangle\
	\det\left(\frac{1}{X-\Phi}\right)\times
	B_{f^{}_1,\ldots,f^{}_{n_c^{}}}
\eea
where the second equality follows from the fact that any
instantonic term on the first line must involve a baryonic
operator with appropriate flavor indices
and all such operators are proportional to
the $B_{f^{}_1,\ldots,f^{}_{n_c^{}}}$.
Consequently,
\be
\BB_{f_1,\ldots,f_{n_c}}(X_1=\cdots=X_{n_c})\
\eqcr\ \langle\!\langle\mbox{quantum corrected}\rangle\!\rangle\
\det\left(\frac{1}{X-\Phi}\right)\times B_{f^{}_1,\ldots,f^{}_{n_c^{}}}\,,
\ee
and comparing this equation with the on-shell formula~(\ref{BaryonF1}) we
immediately see that a non-zero on-shell baryon
$B_{f^{}_1,\ldots,f^{}_{n_c^{}}}\neq 0$ requires
\be
\langle\!\langle\mbox{quantum corrected}\rangle\!\rangle\quad
\det\left(\frac{1}{X-\Phi}\right)\
=\ \prod_{i=1}^{n_c}\frac{1}{X-\mu^{}_{f_i^{}}}\,.
\label{BaryonReq}
\ee
Likewise, the antibaryonic resolvents satisfy
\be
\AB^{f_1,\ldots,f_{n_c}}(X_1=\cdots=X_{n_c})\
\eqcr\ \langle\!\langle\mbox{quantum corrected}\rangle\!\rangle\
\det\left(\frac{1}{X-\Phi}\right)\times\tilde B^{f_1,\ldots,f_{n_c}}\,,
\ee
and hence in light of \eq{ABaryonF1}, a non-zero on-shell antibaryon
$\tilde B^{f_1,\ldots,f_{n_c}}$ requires exactly the same determinant
condition~(\ref{BaryonReq}) as a non-zero on-shell baryon 
$B_{f_1,\ldots,f_{n_c}}$.
In other words, \eq{BaryonReq} is the quantum-corrected version of
the classical \eq{ClassicalBB} for the Coulomb moduli
of the baryonic branches.

The trouble with \eq{BaryonReq} is that in the quantum theory
a determinant of an operator-valued matrix has several
definitions yielding different results.
For example, the definition
$$
\refstepcounter{equation}\displaylines{
    (1)\hfill
    \det\nolimits_1\left(\frac{1}{X-\Phi}\right)\
    \eqdef\ \exp\left( \tr\left( \log\frac{1}{X-\Phi}\right)\right)
    \hfill \label{DetDef1} (\theequation)\cr }
$$
yields on-shell (for the $T(X)$ resolvent given by eqs.~(\ref{Tformula})
and (\ref{Xiformula}))
\be
\det\nolimits_1\left(\frac{1}{X-\Phi}\right)\
=\ \frac{1}{\Xi(X)}\,, \label{Det1}
\ee
but another definition
$$
\refstepcounter{equation}\displaylines{
    (2)\hfill \det\nolimits_2\left(\frac{1}{X-\Phi}\right)
    \eqdef\ {\cal D}_{n_c}\left(
	\tr\left(\frac{1}{X-\Phi}\right),\tr\left(\frac{1}{X-\Phi}\right)^2, 
	\ldots,\tr\left(\frac{1}{X-\Phi}\right)^{n_c}
	\right)\qquad\cr
    	\hfill \label{DetDef2} (\theequation)\cr
     }
$$
(where ${\cal D}_n(t_1,t_2,\ldots,t_n)$ is the Newton's polynomial
formula for the determinant of an ordinary $n\times n$ matrix $A$
in terms of the traces $t_1=\tr(A),\ t_2=\tr(A^2),\ldots,t_n=\tr(A^n)$)
yields
\be
\det\nolimits_2\left(\frac{1}{X-\Phi}\right)\
=\ \frac{1}{\Xi(X)}\times\frac{1}{n_c!}\,\frac{d^{n_c}\Xi}{dX^{n_c}}\,,
\label{Det2}
\ee
and yet another definition
$$
\refstepcounter{equation}
\tabskip 0pt
\halign to \displaywidth{%
    #\unskip\hfil\tabskip=0pt plus 1fil &
    \hfil $\displaystyle{#}$\tabskip=0pt &
    $\displaystyle{\enspace{}\eqdef{}\enspace #}$\hfil\tabskip=0pt plus 1fil &
    \hfil #\unskip\tabskip=0pt\cr
    (3) & \det\nolimits_3\left(\frac{1}{X-\Phi}\right) &
    \frac{1}{\det(X-\Phi)} & \label{DefDet3}(\theequation)\cr
    \vrule width 0pt height 30pt &&
    \frac{1}{{\cal D}_{n_c}\left(
	\tr(X-\Phi),\tr(X-\Phi)^2, \ldots,\tr(X-\Phi)^{n_c}
	\right)} &\cr
    }
$$
yields
\be
\det\nolimits_3\left(\frac{1}{X-\Phi}\right)\
=\ \frac{1}{\left[\Xi(X)\right]_+}\,. \label{Det3}
\ee
However, all such definitions yields exactly the same on-shell
determinant when $\Xi(X)$ happens to be a polynomial,
and in light of $1/\mbox{polynomial}(X)$ expression on the right hand
side of \eq{BaryonReq} it is clear that
\be
B_{f_1,\ldots,f_{n_c}}\neq 0\quad {\bf and/or}\quad
\tilde B^{f_1,\ldots,f_{n_c}}\neq 0\quad{\bf requires}\quad
\Xi(X)\ =\ \prod_{i=1}^{n_c}\left(X-\mu^{}_{f^{}_i}\right) .
\label{QuantumBB1}
\ee

Now consider the Seiberg--Witten curve (\ref{SWsingle}) of the
(deformed) $\NN=2$ SQCD.
In a classical baryonic vacuum, the $SU(n_c)$ is Higgsed down to nothing
by the quark and antiquark VEVs and there are no massless gluons or
photons whatsoever.
In the quantum theory we likewise have no massless photons at all,
which means a Seiberg--Witten curve of zero genus.
In other words,
\be
\Xi(X,Y)\ =\ Y\
=\ \half\left( P(X)\,+\,\sqrt{P^2(X)-4\Lambda^{2n_c-n_f}B(X)}\right)
\label{XiSingle}
\ee
should have at most one branch cut over the complex $X$ plane.
Furthermore, the spectral curves with $h\ge 1$ branch cuts belong
to the mesonic vacua with non-zero classical VEVs of $n_c-h$ quark flavors
(or to the non-classical pseudo-confining vacua),
but in the baryonic vacua all $n_c$ quark flavors develop
non-zero classical VEVs.
This suggests $h=0$ for the baryonic vacua, $i.\,e.$ no branch cuts
whatsoever;
in other words,
\be
P^2(X)\ -\ 4\Lambda^{2n_c-n_f}\,B(X)\ =\ K^2(X)
\ee
for some polynomial $K(X)$, and $\Xi(X)=\half(P(X)+K(X))$ is also
a polynomial function of~$X$.
Of course, this whole argument is rather heuristic, but it serves
as a qualitative explanation why \eq{QuantumBB1} imposes a polynomial
formula on the~$\Xi(X)$.

It remains to solve \eq{QuantumBB1} for the Coulomb moduli of a
baryonic vacuum.
Let us factorize
\be
B(X)\ \equiv\ \prod_{f=1}^{n_f}(X-\mu_f^{})\ =\ B_1(X)\times B_2(X)
\ee
where
\be
B_1(X)\ =\ \prod_{f=f_1,\ldots,f_{n_c}} (X-\mu_f^{})
\quad\mbox{and}\quad
B_2(X)\ =\ \prod_{f\neq f_1,\ldots,f_{n_c}} (X-\mu_f^{}) .
\ee
Eq.~(\ref{QuantumBB1}) amounts to $\Xi(X)=B_1(X)$ while \eq{XiSingle}
implies
\be
P^2(X)\ -\ 4\Lambda^{2n_c-n_f}\,B_1(X)B_2(X)\
=\ \left(2\Xi(X)\,-\,P(X)\right)^2\ =\ \left(2B_1(X)\,-\,P(X)\right)^2
\ee
and therefore
\be
P(X)\ =\ B_1(X)\ +\ \Lambda^{2n_c-n_f}\,B_2(X).
\label{BaryonicCorrections}
\ee
Thus, in the quantum theory, the Coulomb moduli of a baryonic vacuum
with $B_{f_1,\ldots,f_{n_c}}\neq 0$ and/or
$\tilde B^{f_1,\ldots,f_{n_c}}\neq 0$ are given by
\be
\prod_{i=1}^{n_c}(X-\varpi_i)\
=\ \prod_{f=f_1,\ldots,f_{n_c}} (X-\mu_f^{})\
+\ \Lambda^{2n_c-n_f}\,\prod_{f\neq f_1,\ldots,f_{n_c}} (X-\mu_f^{})
\label{QuantumBB}
\ee
where the first term on the right hand side is the classical
formula~(\ref{ClassicalBB}) while the second term is due to
single-instanton quantum effects.
Remarkably, there are no higher-order instantonic corrections.

Similar to its classical counterpart~(\ref{ClassicalBB}),
\eq{QuantumBB} over-determines the Coulomb moduli
$(\varpi_i,\ldots,\varpi_{n_c})$ because of the tracelessness constraint
\be
\tr(\Phi)\ = \oint\!\frac{dX}{2\pi i}\,XT(X)\
= \oint\!\frac{X\,d\,\Xi(X)}{2\pi i\>\Xi(X)}\ =\ 0.
\label{TraceFormula}
\ee
The Coulomb moduli~(\ref{QuantumBB}) provide for $\Xi(X)=B_1(X)$
and hence the integral~(\ref{TraceFormula}) evaluates
to simply $\sum_i\mu^{}_{f_i^{}}$.
In other words, the existence of a baryonic branch with
\be
B_{f_1,\ldots,f_{n_c}}\neq 0\quad {\bf or}\quad
\tilde B^{f_1,\ldots,f_{n_c}}\neq 0\quad{\bf requires}\quad
\mu^{}_{f^{}_1}\ +\ \mu^{}_{f^{}_2}\ +\ \cdots\ + \mu^{}_{f^{}_{n_c}}\ =\ 0
\ee
exactly as in the classical theory.

This completes our study of the baryonic aspects of the
single $SU(n_c)$ theory.
In the following section \S4.2 we shall see that the
baryons of the \suq\ quiver theory behave in a similar way.

%
\subsection{Baryonic Operators of the \suq\ Quiver}
At this point, we are ready to face the whole zoo of chiral
baryonic and antibaryonic operators of the \suq\ quiver,
so let us begin with the zoo's inventory.
Most generally, a chiral baryonic operator of the quiver
comprises $n_c$ quarks $Q_1,\ldots,Q_{n_c}$ located at arbitrary quiver
nodes~$\ell_1,\ldots,\ell_{n_c}$ connected by link chains to yet another
quiver node $\ell$ where the color indices of the $SU(n_c)_\ell$
are combined together into a gauge singlet.
Each of the $n_c$ quarks $Q_i$ has an independent flavor index
$f_i=1,\ldots,F$, and each chain connecting the $i^{\rm th\over{}}$ quark
node~$\ell_i$ to the combination node~$\ell$ can wrap any number of times
around the quiver in either direction independently on any other chain.
Altogether, this gives us a very large family of chiral baryonic operators
\be
\!B_{f_1^{},\ldots,f^{}_{n_c^{}}}^{}
    (\ell;\ell_1,\ldots,\ell_{n_c^{}};k_1,\ldots,k_{n_c^{}})\,
=\, \epsilon_{j_1^{},\ldots,j_{n_c^{}}^{}}^{} \prod_{i=1}^{n_c}\left(
    \left(\link_{\ell-1}\link_{\ell-2}\cdots\link_{\ell}\right)^{k_i^{}}
    \link_{\ell-1}\link_{\ell-2}\cdots\link_{\ell_i^{}}\,Q_{\ell_i^{},f_i^{}}
    \right)^{j_i^{}}\!
\label{Bzoo}
\ee
where  each $k_i$ runs from $-\infty$ to $+\infty$:
A positive $k_i$ means the link chain from $\ell_i$ to $\ell$
wraps $k_i$ times around the quiver in the forward direction
while a negative $k_i$ means the chain wraps backwards,
\be
\left(\link_{\ell-1}\link_{\ell-2}\cdots\link_{\ell}\right)^{(k_i^{}<0)}
\times\link_{\ell-1}\link_{\ell-2}\cdots\link_{\ell_i^{}}\
=\ \left(\link_\ell^{-1}\link_{\ell+1}^{-1}\cdots\link_{\ell-1}^{-1}
	\right)^{(-1-k_i^{}\ge0)}\times
\link_{\ell}^{-1}\link_{\ell+1}^{-1}\cdots\link_{\ell_i^{}-1}^{-1}\,.
\label{BackChain}
\ee

Note that the set of all possible baryonic operators (\ref{Bzoo}) is
redundant in several ways.
First, $B_{f_1^{},\ldots,f^{}_{n_c^{}}}^{}
    (\ell;\ell_1,\ldots,\ell_{n_c^{}};k_1,\ldots,k_{n_c^{}})$
is totally antisymmetric with respect to permutations
of the $(f_i^{},\ell_i^{},k_i^{})$ index triplets, $i.\,e.$
with respect to {\it simultaneous} permutations of the flavor,
quark-node and chain-wrap indices.
This follows from the Bose statistics of the squark and link operators
and from the totally antisymmetric contraction of the color indices
$j_1^{},\ldots,j_{n_c}^{}$.
Nevertheless, there is no antisymmetry with respect to permutations
of the flavor indices alone, apart from the $\ell_i$ and $k_i$ indices,
and therefore the {\it off-shell} chiral baryonic operators~(\ref{Bzoo})
exist for any non-zero flavor number~$F$.
However, similarly to the $\Phi$--baryons of the single $SU(n_c)$ theory,
all the baryonic operators~(\ref{Bzoo}) of the \suq\ quiver with $n_f<n_c$
vanish {\it on-shell\/}; we shall prove this statement later in this section.

The second redundancy is due to quark locations $\ell_i$ being defined
modulo $N$, hence in a baryonic operator, changing $\ell_i\to\ell_i\pm N$
is equivalent to changing the wrap number $k_i\to k_i\mp1$.
Finally, letting the $k_i$ run from $-\infty$ to $+\infty$ is redundant
because any backward-wrapping link chain~(\ref{BackChain}) is a linear
combination of all the forward-wrapping chains, and {\it vice verse}, any
forward-wrapping chain is a linear combination of the backward-wrapping
chains.
This follows from the fact that a single meromorphic resolvent
\bea
\frac{\link_{\ell-1}\cdots\link_{\ell_i^{}}}{X_i-\link\cdots\link} &
=&\sum_{k_i\ge 0}\left(\frac{1}{X_i}\right)^{k_i+1}\times
    \left(\link_{\ell-1}\link_{\ell-2}\cdots\link_{\ell}\right)^{k_i^{}}
    \link_{\ell-1}\link_{\ell-2}\cdots\link_{\ell_i^{}}\\
&=&-\sum_{k_i<0} X_i^{-1-k_i}\times\left(
	\link_\ell^{-1}\link_{\ell+1}^{-1}\cdots\link_{\ell-1}^{-1}
	\right)^{1-k_i^{}}
    \link_{\ell}^{-1}\link_{\ell+1}^{-1}\cdots\link_{\ell_i^{}-1}^{-1}\,.\qquad
\eea
summarizes both types of link chains but can be decomposed into a
convergent power series in terms of either only the forward-wrapping chains
or else only the backward-wrapping chains.

In light of the above redundancies, we can encode all the chiral
baryonic operators of the quiver in a family of baryonic resolvents
\be
\BB_{f_1^{},\ldots,f^{}_{n_c^{}}}^{}
    (\ell;\ell_1,\ldots,\ell_{n_c^{}};X_1,\ldots,X_{n_c^{}})\
=\ \epsilon_{j_1^{},\ldots,j_{n_c^{}}^{}}^{} \prod_{i=1}^{n_c}\left(
    \frac{\link_{\ell-1}\cdots\link_{\ell_i^{}}}{X_i-\link\cdots\link}\,
    Q_{\ell_i^{},f_i^{}}^{}\right)^{j_i^{}}
\label{Bresolvents}
\ee
where $X_1,\ldots,X_{n_c^{}}$ are independent complex numbers,
and the resolvents are totally antisymmetric with respect to
{\it simultaneous} permutations of the flavor indices $f_i$,
the quiver indices $\ell_i$, and the arguments $X_i$.
Similar to the mesonic resolvents~(\ref{Mresolvents}), the
baryonic resolvents~(\ref{Bresolvents}) of the quiver theory
trade the wrapping numbers $k_1,\ldots,k_{n_c^{}}$ for the complex
arguments $X_1,\ldots,X_{n_c^{}}$, but now we also have to contend with
the quiver indices $\ell_1,\ldots,\ell_{n_c^{}}$ and $\ell$.
The independent resolvents have all $\ell_i$ running from $\ell$
down to $\ell-N+1$; for the $\ell_i$ outside of this range,
there is a periodicity condition
\be
\eqalign{
\BB(\ell;\ldots,\ell_i=\ell-N,\ldots;X_1,\ldots,X_{n_c^{}})\ &
- X_i\times \BB(\ell;\ldots,\ell_i=\ell,\ldots;X_1,\ldots,X_{n_c^{}})\cr
&=\ -\epsilon_{j_1^{},\ldots,j_{n_c^{}}^{}}^{}\, Q_{\ell,f_i^{}}^{j_i^{}}
    \times\prod_{\imath'\neq i}\left(
    \frac{\link_{\ell-1}\cdots\link_{\ell_{\imath'}}}
	{X_{\imath'}-\link\cdots\link}\,
    Q_{\ell_{\imath'},f_{\imath'}}^{}\right)^{j_{\imath'}}\cr
&\omit\qquad which does not depend on the $X_i$.\hfil\cr
}\label{Bperiodicity}
\ee

The antibaryonic chiral operators of the quiver
theory comprise $n_c$ antiquarks located at independent quiver nodes
$\ell_i$ and connected by link chains to yet another quiver node $\ell$
where the color indices are combined into a gauge singlet.
The most general operator of this kind has form
\be
\!\widetilde{B}^{f_1^{},\ldots,f^{}_{n_c^{}}}_{}
    (\ell;\ell_1,\ldots,\ell_{n_c^{}};k_1,\ldots,k_{n_c^{}})\,
=\, \epsilon^{j_1^{},\ldots,j_{n_c^{}}^{}}_{} \prod_{i=1}^{n_c}\left(
    \aq_{\ell_i}^{f_i^{}}\,\link_{\ell_i-1}\link_{\ell_i-2}\cdots\link_\ell
    \left(\link_{\ell-1}\link_{\ell-2}\cdots\link_{\ell}\right)^{k_i^{}}
    \right)_{j_i^{}}\!
\label{Azoo}
\ee
for some flavor indices $f_i^{},\ldots,f_{n_c^{}}^{}$, quiver-node
indices $\ell_1,\ldots,\ell_{n_c^{}}$ and $\ell$, and wrapping
numbers $k_1.\ldots,k_{n_c^{}}$.
The antibaryonic operators~(\ref{Azoo}) are subject to the same
redundancy conditions as the baryonic operators~(\ref{Bzoo}),
thus we may encode them in a similar family of the antibaryonic
resolvents
\be
\AB^{f_1^{},\ldots,f^{}_{n_c^{}}}_{}
    (\ell;\ell_1,\ldots,\ell_{n_c^{}};X_1,\ldots,X_{n_c^{}})\
=\ \epsilon^{j_1^{},\ldots,j_{n_c^{}}^{}}_{} \prod_{i=1}^{n_c}\left(
    \aq_{\ell_i^{}}^{f_i^{}}\,
    \frac{\link_{\ell-1}\cdots\link_{\ell_i^{}}}{X_i-\link\cdots\link}
    \right)_{j_i^{}}
\label{Aresolvents}
\ee
Again, the $X_1,\ldots,X_{n_c^{}}$ are independent complex numbers,
and the resolvents are totally antisymmetric with respect to
{\it simultaneous} permutations of the flavor indices $f_i$,
the quiver indices $\ell_i$, and the arguments $X_i$.
The independent antibaryonic resolvents have all the $\ell_i$ running from
$\ell$ up to $\ell+N-1$; for the $\ell_i$ outside of this range,
there is a periodicity condition
\be
\eqalign{
\AB(\ell;\ldots,\ell_i=\ell+N,\ldots;X_1,\ldots,X_{n_c^{}})\ &
- X_i\times \AB(\ell;\ldots,\ell_i=\ell,\ldots;X_1,\ldots,X_{n_c^{}})\cr
\noalign{\vskip 5pt}
&=\ -\epsilon^{j_1^{},\ldots,j_{n_c^{}}^{}}_{}\, \aq_{\ell,j_i^{}}^{f_i^{}}
    \times\prod_{\imath'\neq i}\left(
    \aq_{\ell_{\imath'}}^{f_{\imath'}}
    \frac{\link_{\ell_{\imath'}-1}\cdots\link_\ell}
	{X_{\imath'}-\link\cdots\link}\,
    \right)_{j_{\imath'}}\cr
&\omit\qquad which does not depend on the $X_i$.\hfil\cr
}\label{Aperiodicity}
\ee

Thus far, we summarized the baryonic and antibaryonic generators
of the {\it off-shell} chiral ring of the \suq\ quiver.
Note that the resolvents (\ref{Bresolvents}) and (\ref{Aresolvents}) are
totally antisymmetric with respect to simultaneous permutations
of the $f_i$, the $\ell_i$ and the $X_i$, but there is no antisymmetry
with respect to flavor indices alone.
Consequently, there are non-trivial {\it off-shell} (anti) baryonic
generators for any flavor number $F\ge 1$.

The \underline{\it on-shell} baryonic operators are constrained by
equations of motions stemming from quark-dependent variations
of the antiquark fields.
Generalizing \eq{QDAV} from the single $SU(n_c)$ theory to the
\suq\ quiver, consider
\be
\delta\aq_{\ell_1,j}^{f}\
=\ \varepsilon\left(
	\frac{\link_{\ell-1}\cdots\link_{\ell_1}}{X_1-\link\cdots\link}
	\right)^{j_1^{}}_{\,\,j}\times
\epsilon_{j_1^{},\ldots,j_{n_c^{}}^{}}^{} \prod_{i=2}^{n_c}\left(
    \frac{\link_{\ell-1}\cdots\link_{\ell_i^{}}}{X_i-\link\cdots\link}\,
    Q_{\ell_i^{},f_i^{}}^{}\right)^{j_i^{}}
\label{QDAVquiver}
\ee
for an arbitrary combination of (anti) quark flavors $f$ and
$f_2^{},\ldots,f^{}_{n_c^{}}$.
This variation is anomaly-free since the $\delta\aq_{\ell_1,j}^{f}$ does not
depend on the antiquark field $\aq_{\ell_1,j}^{f}$ itself, hence
the resulting equation of motion is simply $\delta W_{\rm tree}\eqcr 0$.
Or not so simply,
\bea
\delta W_{\rm tree}\ 
&=& \gamma\,\delta\aq_{\ell_1,j}^{f}\times\left[
	\left(\link_{\ell_1-1}^{}Q_{\ell_1-1,f}^{}\right)^j\
	-\ \mu_f Q_{\ell_1,f}^j
	\right] \nonumber\\
&=& \gamma\varepsilon\times
    \epsilon_{j_1^{},j_2^{},\ldots,j_{n_c^{}}^{}}^{}\prod_{i=2}^{n_c}\left(
    	\frac{\link_{\ell-1}\cdots\link_{\ell_i^{}}}{X_i-\link\cdots\link}\,
    	Q_{\ell_i^{},f_i^{}}^{}\right)^{j_i^{}} \times{} \nonumber\\
&&\qquad\times\left[
    	\left(
	    \frac{\link_{\ell-1}\cdots\link_{\ell_1}}{X_1-\link\cdots\link}\,
	    \link_{\ell_1-1}Q_{\ell_1-1,f}
	    \right)^{j_1^{}}\
	-\ \mu_f\left(
	    \frac{\link_{\ell-1}\cdots\link_{\ell_1}}{X_1-\link\cdots\link}\,
	    Q_{\ell_1,f}\right)^{j_1^{}}
	\right] \nonumber\\
&=& \gamma\varepsilon\times\Bigl[ \BB_{f,f_2^{},\ldots,f_{n_c^{}}^{}}
	(\ell;\ell_1-1,\ell_2,\ldots,\ell_{n_c^{}};X_1,X_2,\ldots,X_{n_c^{}})
    \nonumber\\
&&\qquad\ -\ \mu_f\BB_{f,f_2^{},\ldots,f_{n_c^{}}^{}}
	(\ell;\ell_1,\ell_2,\ldots,\ell_{n_c^{}};X_1,X_2,\ldots,X_{n_c^{}})
    \Bigr] \nonumber\\
&\eqcr& 0.
\eea
In other words, in the on-shell chiral ring, decreasing the $\ell_1$
index of a baryonic resolvent $\BB_{f_1^{},\ldots,f_{n_c^{}}^{}}
(\ell;\ell_1,\ldots,\ell_{n_c^{}};X_1,\ldots,X_{n_c^{}})$ by 1
simply multiplies the resolvent by $\mu_{f_1^{}}^{}$, regardless
of the $X_1,\ldots,X_{n_c^{}}$ arguments.
And since all $n_c$ quarks of a baryon have equal status,
decreasing any of the $\ell_i$ indices by 1 multiplies the resolvent
by the $i{\rm th\over{}}$ quark mass, namely $\mu_{f^{}_i}^{}$.
Hence, by iteration
\be
\BB_{f_1^{},\ldots,f_{n_c^{}}^{}}
    (\ell;\ell_1,\ldots,\ell_{n_c^{}};X_1,\ldots,X_{n_c^{}})\
\eqcr\ \prod_{i=1}^{n_c}\mu_{f^{}_i}^{\ell-\ell_i}\times
\BB_{\,{\rm same}\,f_i^{}}(\mbox{all}\,\ell_i=\ell;\,\mbox{same}\,X_i)\quad
\ee
for any $\ell_i\le\ell$,
and consequently, in light of the periodicity conditions~(\ref{Bperiodicity}),
\be
\BB_{f_1^{},\ldots,f_{n_c^{}}^{}}
	(\ell;\ell_1,\ldots,\ell_{n_c^{}};X_1,\ldots,X_{n_c^{}})\
\eqcr\ \prod_{i=1}^{n_c}
	\frac{\mu_{f^{}_i}^{\ell-\ell_i}}{X_i-\mu_{f^{}_i}^N}
\times B_{f_1^{},\ldots,f_{n_c^{}}^{}}(\ell)
\label{BaryonQuiver}
\ee
where
\be
B_{f_1^{},\ldots,f_{n_c^{}}^{}}(\ell)\
=\ \epsilon_{j_1^{},\ldots,j_{n_c^{}}^{}}^{}
Q_{\ell,f^{}_1}^{j^{}_1}\cdots Q_{\ell,f^{}_{n^{}_c}}^{j^{}_{n^{}_c}}
\label{OrdB}
\ee
is the ordinary, local baryon at the quiver node~$\ell$.
Thus, {\blue\it on shell, all the baryonic resolvents~{\rm(\ref{Bzoo})}
of the quiver
--- and hence all the non-local baryon-like chiral operators ---
are proportional to the ordinary local baryons~{\rm(\ref{OrdB})}}.

Likewise, on-shell
\be
\AB^{f_1^{},\ldots,f_{n_c^{}}^{}}
	(\ell;\ell_1,\ldots,\ell_{n_c^{}};X_1,\ldots,X_{n_c^{}})\
\eqcr\ \prod_{i=1}^{n_c}
	\frac{\mu_{f^{}_i}^{\ell_i-\ell}}{X_i-\mu_{f^{}_i}^N}
\times \widetilde B^{f_1^{},\ldots,f_{n_c^{}}^{}}(\ell)
\label{ABaryonQuiver}
\ee
and all the antibaryonic resolvents~(\ref{Azoo}) are proportional to
the ordinary antibaryons.
And since the ordinary  baryons and antibaryons have antisymmetrized flavor
indices, it follows that similarly to the deformed $\NN=2$ SQCD theory,
all the  baryonic and antibaryonic generators of the quiver's chiral ring
vanish on shell unless $F\ge n_c$.
Furthermore, we shall see momentarily that baryonic branches of the
quiver theory do not involve massless quark flavors, so
{\it\blue it takes at least $n_f\ge n_c$ \underline{massive} quark flavors}
({\it cf.}\ eqs.~(\ref{Nflavors}) and (\ref{masses4D})
{\it\blue to get any anti/baryonic VEVs at all}.

Indeed, similarly to the single $SU(n_c)$ theory of the previous section,
baryonic branches of the \suq\ quiver have overdetermined Coulomb moduli.
To derive this result, we again focus on the anti/baryonic resolvents
with equal arguments $X_1=X_2=\cdots=X_{n_c^{}}$, but this time we also
use equal quiver indices $\ell_1=\ell_2=\cdots\ell_{n_c^{}}=\ell$.
Thus,
\be
\eqalign{
\BB_{f_1^{},\ldots,f_{n_c^{}}^{}}(\mbox{all}\,\ell_i=\ell;&
\,\mbox{all}\,X_i=X)\ ={}\cr
&=\ \epsilon_{j_1^{},\ldots,j_{n_c^{}}^{}} \prod_{i=1}^{n_c}
    \left(\frac{1}{X-\link\cdots\link}\,Q_{\ell,f_i^{}}\right)^{j_i^{}}\qquad
    \langle\!\langle\mbox{{\it cf.}\ \eq{Bresolvents}}\rangle\!\rangle\cr
&=\ \det\left(\frac{1}{X-\link\cdots\link}\times
	Q_{f_1^{},\ldots,f_{n_c^{}}^{}}(\ell)\right) \cr
\langle\!\langle\mbox{on shell}\rangle\!\rangle\cr
&\eqcr\ \det\left(\frac{1}{X-\link\cdots\link}\right)\times
    \left[ \det\left(Q_{f_1^{},\ldots,f_{n_c^{}}^{}}(\ell)\right)\,
	\equiv\,B_{f_1^{},\ldots,f_{n_c^{}}^{}}(\ell)\right]\cr
&\qquad\qquad+\ \mbox{instantonic corrections} \cr
&=\ B_{f_1^{},\ldots,f_{n_c^{}}^{}}(\ell)\times
    \langle\!\langle\mbox{quantum corrected}\rangle\!\rangle
    \det\left(\frac{1}{X-\link\cdots\link}\right) ,\cr
}\label{Bdeterminant}
\ee
and comparing this result with the on-shell formula (\ref{BaryonQuiver})
for the same baryonic resolvent, we immediately see that
a non-trivial baryonic VEV $\vev{B_{f_1^{},\ldots,f_{n_c^{}}^{}}(\ell)}\neq0$
{\bf requires}
\be
\langle\!\langle\mbox{quantum corrected}\rangle\!\rangle
\det\left(\frac{1}{X-\link\cdots\link}\right)\
=\ \prod_{i=1}^{n_c}\frac{1}{X-\mu^N_{f_i^{}}}\,.
\label{QuiverBaryonReq}
\ee
Likewise, for an antibaryonic resolvent with the same indices and arguments
we have
\be
\eqalign{
\AB^{f_1^{},\ldots,f_{n_c^{}}^{}}(\mbox{all}\,\ell_i=\ell;&
\,\mbox{all}\,X_i=X)\ ={}\cr
&=\ \epsilon^{j_1^{},\ldots,j_{n_c^{}}^{}} \prod_{i=1}^{n_c}
    \left(\aq_\ell^{f_i^{}}\,\frac{1}{X-\link\cdots\link}\right)_{j_i^{}}\qquad
    \langle\!\langle\mbox{{\it cf.}\ \eq{Bresolvents}}\rangle\!\rangle\cr
&=\ \det\left( \aq^{f_1^{},\ldots,f_{n_c^{}}^{}}(\ell)\times
	\frac{1}{X-\link\cdots\link}\right) \cr
\langle\!\langle\mbox{on shell}\rangle\!\rangle\cr
&\eqcr\ \left[ \det\left(\aq^{f_1^{},\ldots,f_{n_c^{}}^{}}(\ell)\right)\,
	\equiv\,\widetilde B^{f_1^{},\ldots,f_{n_c^{}}^{}}(\ell)\right]
    \times\det\left(\frac{1}{X-\link\cdots\link}\right)\cr
&\qquad\qquad+\ \mbox{instantonic corrections} \cr
&=\ \widetilde B^{f_1^{},\ldots,f_{n_c^{}}^{}}(\ell)\times
    \langle\!\langle\mbox{quantum corrected}\rangle\!\rangle
    \det\left(\frac{1}{X-\link\cdots\link}\right) ,\cr
}\label{Adeterminant}
\ee
and comparing this result with the on-shell formula (\ref{ABaryonQuiver})
we see that a non-trivial antibaryonic VEV
$\vev{\widetilde B^{f_1^{},\ldots,f_{n_c^{}}^{}}(\ell)}\neq0$
requires exactly the same determinant condition (\ref{QuiverBaryonReq})
as the baryonic VEV $\vev{B_{f_1^{},\ldots,f_{n_c^{}}^{}}(\ell)}\neq0$.

{}From the Coulomb moduli's point of view, the determinant constraint
(\ref{QuiverBaryonReq}) of the quiver theory is precisely analogous
to the determinant constraint (\ref{BaryonReq}) of the
single $SU(n_c)$ theory, and its solution is precisely analogous
to eqs.~(\ref{QuantumBB1}) and (\ref{QuantumBB}).
Specifically,
\be
B_{f_1,\ldots,f_{n_c}}\neq 0\quad {\bf and/or}\quad
\tilde B^{f_1,\ldots,f_{n_c}}\neq 0\quad{\bf requires}\quad
\Xi(X)\ =\ \prod_{i=1}^{n_c}\left(X-\mu^{}_{f^{}_i}\right) ,
\label{QuiverBB1}
\ee
which means the Seiberg--Witten curve (\ref{SeibergWitten}) of the
quiver should have no branch cuts at all, and the Coulomb moduli
$\varpi_i\equiv\ev_i^N$ are given by
\be
\prod_{i=1}^{n_c}(X-\varpi_i)\
=\ \prod_{f=f_1^{},\ldots,f_{n_c^{}}^{}} (X-\mu_f^N)\
+\ \alpha\,\prod_{f\neq f_1^{},\ldots,f_{n_c^{}}^{}} (X-\mu_f^N)
\label{QuiverBB}
\ee
where the first term on the right hand side describes the classical
location of the quiver's baryonic branch
($\ev_1^N=\mu_{f^{}_1}^N,\ldots,\ev_{n_c^{}}^N=\mu_{f^{}_{n_c^{}}}^N$)
while the second term is the instantonic correction.
Note that all $F$ quark flavors, massive or massless alike, must appear
in either product on the right hand side of \eq{QuiverBB}.
However, \eq{QuiverBB1} implies
\be
T(X)\ =\ \frac{1}{\Xi}\,\frac{d\Xi}{dX}\
=\,\sum_{i=1}^{n_c}\frac{1}{X-\mu_{f_i^{}}^N}
\ee
without any quantum corrections, and whereas the link resolvent
of the \suq~quiver must be regular at $X=0$ on the physical sheet,
it follows that {\it\blue the quark flavors
$f_1^{},\ldots,f_{n_c^{}}^{}$ involved in any anti/baryonic VEV must all be
massive}, hence the $n_f\ge n_c$ requirement rather than just $F\ge n_c$.

Eq.~(\ref{QuiverBB}) over-determines the Coulomb moduli of a baryonic branch
because only $n_c-1$ of these moduli are independent
while their product is constrained according to \eq{DetCon}.
Consequently, the baryonic branch involving quark flavors
$f_1^{},\ldots,f_{n_c^{}}^{}$ exists if and only if
\be
\prod_{\!f=f_1^{},\ldots,f_{n_c^{}}^{}\!} \mu_f^N\
+\ \alpha(-1)^F\!\!\prod_{\!f\neq f_1^{},\ldots,f_{n_c^{}}^{}\!} \mu_f^N\
=\ V^{Nn_c}\ \equiv\ V_1^{Nn_c}\ +\ V_2^{Nn_c}
\label{QBBCon}
\ee
where on the right hand side $V_1^{n_c}$ and $V_2^{n_c}$ are the two
roots of \eq{V1V2} as we shall explain in the following section~\S4.3.
Physically,  $V^{n_c}=v^{n_c}+{}$quantum corrections, and we shall see that
the corrections vanish if
the quiver has massless quarks: For $\Delta F>0$, $V_1^{n_c}=v^{n_c}$
while $V_2=0\ \Longrightarrow\ V^{n_c}=v^{n_c}$ exactly.
At the same time, any $\mu_f=0$ kills the second product on the left hand
side of \eq{QBBCon}, which leads us to the un-modified
classical condition
\be
\prod_{\!f=f_1^{},\ldots,f_{n_c^{}}^{}\!} \mu_f^N\ =\ v^{Nn_c} .
\label{CBBCon}
\ee
For $\Delta F=0$ (massive flavors only) the situation is more complicated:
Both products on the left hand side of \eq{QBBCon} have non-zero values,
and likewise both $V_1\neq 0$ and $V_2\neq0$ on the right hand side.
However, taking the second \eq{V1V2} to the $N{\rm th\over{}}$ power we have
\be
V_1^{Nn_c}\times V_2^{Nn_c}\ 
=\ \left( \Lambda^{2n_c-F}(-\gamma)^F\prod_{{\rm all}\,f}\mu_f^{}\right)^N\
=\ \left(\prod_{\!f=f_1^{},\ldots,f_{n_c^{}}^{}\!}\!\! \mu_f^N\right)\times
\left(\alpha(-1)^F\!\!\!\prod_{\!f\neq f_1^{},\ldots,f_{n_c^{}}^{}\!}\!\!
	\mu_f^N\right)
\ee
({\it cf.}\ \eq{Aformula} for the $\alpha$),
and comparing this equation with  \eq{QBBCon}
we immediately obtain a much simpler formula
\be
\prod_{\!f=f_1^{},\ldots,f_{n_c^{}}^{}\!} \mu_f^N\
=\ V_1^{Nn_c}\ {\bf or}\ V_2^{Nn_c}.
\label{QBBCon2}
\ee
In any case, {\it\blue the baryonic branch involving quark flavors
$f_1^{},\ldots,f_{n_c^{}}^{}$ exists if and only if the masses of these
flavors satisfy the classical product condition~{\rm(\ref{CBBCon})}
for $\Delta F>0$ or the quantum-corrected product
condition~{\rm(\ref{QBBCon2})} for $\Delta F=0$}.

This almost concludes our presentation of the baryonic aspects
of the \suq\ quiver theory, except for one minor point.
Classically, a baryonic branch of the quiver has $\ell$--independent VEVs
$\vev{B^{}_{f^{}_1,\ldots,f^{}_{n_c^{}}}}$ and
$\vev{\widetilde B_{}^{f^{}_1,\ldots,f^{}_{n_c^{}}}}$,
modulo an $e^{2\pi ik\ell/N}$ phase factor, but in the quantum theory
eqs.~(\ref{BaryonQuiver}) and~(\ref{ABaryonQuiver}) seem to allow for arbitrary
$\ell$--dependence of the ordinary baryons and antibaryons.
To plug this loophole, consider the baryonic resolvent with equal arguments
$X_1=X_2=\cdots=X_{n_c^{}}$ and quiver indices $\ell_1=\ell_2=\cdots
=\ell_{n_c^{}}=\ell-1$:
\be
\eqalign{
\BB_{f_1^{},\ldots,f_{n_c^{}}^{}}(\ell;\mbox{all}\,\ell_i=\ell-1;
	\,\mbox{all}\,X_i=X)\ &
=\ \epsilon_{j_1^{},\ldots,j_{n_c^{}}^{}} \prod_{i=1}^{n_c}\left(
	\frac{\link_{\ell-1}}{X-\link\cdots\link}\,
	Q_{\ell-1,f_i^{}}\right)^{j_i^{}}\cr
&=\ \det\left( \link_{\ell-1}\times\frac{1}{X-\link\cdots\link}
	\times Q_{f_1^{},\ldots,f_{n_c^{}}}^{}(\ell)\right) \cr
\langle\!\langle\mbox{on shell}\rangle\!\rangle \cr
&=\ \det(\link_{\ell-1})\times \det\left(
	\frac{1}{X-\link\cdots\link}
	\times Q_{f_1^{},\ldots,f_{n_c^{}}}^{}(\ell-1)
	\right) \cr
&\qquad\qquad+\ \mbox{instantonic corrections} \cr
&=\ \langle\!\langle\mbox{quantum corrected}\rangle\!\rangle
    \det(\link_{\ell-1})\times{} \cr
&\qquad\qquad\times\BB_{f_1^{},\ldots,f_{n_c^{}}^{}}
	(\ell-1;\mbox{all}\,\ell_i=\ell-1;\,\mbox{all}\,X_i=X), \cr
}\ee
and comparing this formula with \eq{BaryonQuiver}, we arrive at
\be
B_{f_1^{},\ldots,f_{n_c^{}}^{}}(\ell)\
=\ \frac{\mu_{f_1^{}}^{}\times\mu_{f_2^{}}^{}\times\cdots\times\mu_{f_{n_c}^{}}}
	{\langle\!\langle\mbox{quantum corrected}\rangle\!\rangle
		\det(\link_{\ell-1})}
\times B_{f_1^{},\ldots,f_{n_c^{}}^{}}(\ell-1).
\label{Bstep}
\ee
Likewise, for the antibaryons we have
\be
\eqalign{
\AB^{f_1^{},\ldots,f_{n_c^{}}^{}}(\ell;\mbox{all}\,\ell_i=\ell+1;
	\,\mbox{all}\,X_i=X)\ &
=\ \AB^{f_1^{},\ldots,f_{n_c^{}}^{}}(\ell+1;\mbox{all}\,\ell_i=\ell+1;
	\,\mbox{all}\,X_i=X)\times{}\cr
&\qquad\qquad\times \langle\!\langle\mbox{quantum corrected}\rangle\!\rangle
    \det(\link_{\ell}),\cr
}\qquad
\ee
and therefore
\be
\widetilde B^{f_1^{},\ldots,f_{n_c^{}}^{}}(\ell)\
=\ \frac{\mu_{f_1^{}}^{}\times\mu_{f_2^{}}^{}\times\cdots\times\mu_{f_{n_c}^{}}}
	{\langle\!\langle\mbox{quantum corrected}\rangle\!\rangle
		\det(\link_{\ell})}
\times \widetilde B^{f_1^{},\ldots,f_{n_c^{}}^{}}(\ell+1).
\label{Astep}
\ee
In both eqs.~(\ref{Bstep}) and~(\ref{Astep}) we have a quantum-corrected
link determinant, and while it may be hard to calculate the quantum
corrections here, they obviously do not depend on a particular link
$\link_\ell$.
Consequently,
\be
\forall\ell:\quad
B_{f_1^{},\ldots,f_{n_c^{}}^{}}(\ell)\ =\ C\times
B_{f_1^{},\ldots,f_{n_c^{}}^{}}(\ell-1)\quad \mbox{and}\quad
\widetilde B^{f_1^{},\ldots,f_{n_c^{}}^{}}(\ell)\ =\ C\times
\widetilde B^{f_1^{},\ldots,f_{n_c^{}}^{}}(\ell+1)\qquad
\label{Steps}
\ee
for some $\ell$-independent constant $C$,
and by periodicity of the quiver we must have $C^N=1\
\Longrightarrow\ C=e^{2\pi i k/N}$.
In other words, the anti/baryonic VEVs of the quantum quiver
do follow the classical rule of
\be
B_{f_1^{},\ldots,f_{n_c^{}}^{}}(\ell)\
=\ e^{+2\pi ik\ell/N}\,B^{}_{f_1^{},\ldots,f_{n_c^{}}^{}}\quad
{\rm and}\quad \widetilde B^{f_1^{},\ldots,f_{n_c^{}}^{}}(\ell)\
=\ e^{-2\pi ik\ell/N}\,\widetilde B_{}^{f_1^{},\ldots,f_{n_c^{}}^{}}.
\ee
And this does complete our presentation of the baryonic issues.

%
\subsection{Determinants of Link Chains.}
In this section we complete our study of the \suq~quiver's chiral
ring by calculating the quantum corrections to the determinants of 
link chains $\det(\link_{\ell_2}\cdots\link_{\ell_1})$,
especially the $\det(\link_N\cdots\link_1)$ which controls the
parameter $V^{n_c^{}}=v^{n_c^{}}+\cdots$ in eqs.~(\ref{DetCon})
and (\ref{QBBCon}).
For simplicity, we focus on the un-deformed deconstructive quivers.

In the quarkless case $F=0$, the corrections follow from `local' instantons
in the individual $SU(n_c)_\ell$ gauge factors rather than the `global'
instantons of the whole quiver.
Iterating the Seiberg formula \cite{seibergq}
\be
\det(\MM)\ =\ \BB\AB\ +\ \Lambda^{2n_c}
\label{SeibergNcNf}
\ee
for the ordinary SQCD with $n_f=n_c$, Rodr\'\i guez
\cite{rodriguez} and later Chang and Georgi \cite{georgi} found
for an open link chain
\be
\det(\link_{\ell_2}\cdots\link_{\ell_1})\
=\ \mathop{\rm Poly}\left[
    \prod_{\ell=\ell_1}^{\ell_2}\det(\link_\ell)\times
    \prod_{\ell=\ell_1+1}^{\ell_2}\left(1\,
	+\,\frac{\Lambda_\ell^{2n_c}}{\det(\link_\ell)\det(\link_{\ell-1})}
	\right)\right]
\label{OpenChain}
\ee
where $\mathop{\rm Poly}[\cdots]$ denotes the polynomial part of the
expression in the square brackets, that is, the terms without any net
$\det(\link_\ell)$ factors in the denominator.
For example,
\be
\eqalign{
    \det(\link_4\link_3\link_2\link_1)\ ={}\ &
    \det(\link_4)\det(\link_3)\det(\link_2)\det(\link_1)\
	+\ \det(\link_4)\det(\link_3)\Lambda_2^{2n_c}\cr
    &+\ \det(\link_4)\Lambda_3^{2n_c}\det(\link_1)\
	+\ \Lambda_4^{2n_c}\det(\link_2)\det(\link_1)\
 	+\ \Lambda_4^{2n_c}\Lambda_2^{2n_c}\cr
    &+\ \crossout{\frac{\det(\link_4)\Lambda_3^{2n_c}\Lambda_2^{2n_c}}
		{\det(\link_2)}}\
	+\ \crossout{\frac{\Lambda_4^{2n_c}\Lambda_3^{2n_c}\det(\link_1)}
		{\det(\link_3)}}
	+\ \crossout{\frac{\Lambda_4^{2n_c}\Lambda_3^{2n_c}\Lambda_2^{2n_c}}
		{\det(\link_3)\det(\link_2)}}\,. \cr
    }
\ee
Likewise, the closed link chain wrapped around the whole quiver has
determinant
\be
\det(\link_N\cdots\link_1)\
=\ \mathop{\rm Poly}\left[
    \prod_{\ell=1}^{N}\det(\link_\ell)\times
    \prod_{\ell=1}^{N}\left(1\,
	+\,\frac{\Lambda_\ell^{2n_c}}{\det(\link_\ell)\det(\link_{\ell-1})}
	\right)\right]\,;
\label{ClosedChain}
\ee
for the $\ell$--independent $\Lambda_\ell\equiv\Lambda$
and $\det(\link_\ell)\equiv v^{n_c}$, this expression evaluates to
\cite{IK1}
\be
\det(\link_N\cdots\link_1)\ =\ V_1^{Nn_c}\ +\ V_2^{Nn_c}
\label{VF0}
\ee
where $V_1^{n_c}$ and $V_2^{n_c}$ are the two
roots of the quadratic equation system
\be
V_1^{n_c}\ +\ V_2^{n_c}\ =\ v^{n_c},\qquad
V_1^{n_c}\times V_2^{n_c}\ =\ \Lambda^{2n_c}.
\label{V1V2F0}
\ee
In the deconstruction limit $N\to\infty$ the right hand side of \eq{VF0}
is dominated by the larger root, hence the quantum corrections may be
summarized as $v\to V=\max(V_1,V_2)$; as discussed in \cite{IK1}, this leads
to $(1/g_5^2)\ge 0$ in the deconstructed theory and prevents phase transitions.
However, for the present purposes we are interested in fixed--$N$ quivers
and exact holomorphic relations in the chiral ring, so both roots of
eqs.~(\ref{V1V2F0}) are equally important.

In the quiver theories with quarks, the link chain determinants are
affected by both local and global instantonic effects.
The local effects follow from integrating out the massive
(${\rm mass}=\gamma\mu_f$) quark-antiquark pairs from each individual
$SU(n_c)_\ell$ gauge group factor, thus
\be
\Lambda_\ell^{2n_c}\ \to\ \Lambda_\ell^{2n_c-F}\times
\prod_{f=1}^F(-\gamma\mu_f)\,,
\label{LambdaIO}
\ee
and then proceeding exactly as in the quarkless case, hence
\bea
\det(\link_{\ell_2}\cdots\link_{\ell_1}) &
=& \mathop{\rm Poly}\left[
    \prod_{\ell=\ell_1}^{\ell_2}\det(\link_\ell)\times
    \prod_{\ell=\ell_1+1}^{\ell_2}\left(
	1\,+\,\frac{\Lambda_\ell^{2n_c}\prod_f(-\gamma\mu_f)}
		{\det(\link_\ell)\det(\link_{\ell-1})}
	\right)\right]\
    +\ \cdots\,,\qquad \label{OpenChainF}\\
\det(\link_N\cdots\link_1) &
=& \mathop{\rm Poly}\left[
    \prod_{\ell=1}^{N}\det(\link_\ell)\times
    \prod_{\ell=1}^{N}\left(
	1\,+\,\frac{\Lambda_\ell^{2n_c}\prod_f(-\gamma\mu_f)}
		{\det(\link_\ell)\det(\link_{\ell-1})}
	\right)\right]\
    +\ \cdots \nonumber\\
&=& V_1^{Nn_c}\ +\ V_2^{Nn_c}\ +\ \cdots \label{ClosedChainF}
\eea
where
\be
V_1^{n_c}\ +\ V_2^{n_c}\ =\ v^{n_c},\qquad
V_1^{n_c}\times V_2^{n_c}\ =\ \Lambda^{2n_c-F}\prod_{f=1}^F(-\gamma\mu_f)
\label{V1V2}
\ee
and the `$\cdots$' stand for the non-local quantum corrections, if any.
Note that the products of quark masses in \eqrange{LambdaIO}{V1V2}
involves all $F$ quark flavors.
{\it When some of the flavors are exactly massless ($i.\,e.,\ \Delta F>0$)
the \underline{local} instanton corrections to the determinants
\eqrange{OpenChain}{ClosedChain} go away.}

The non-local effects arise from the quark and the antiquark
fields propagating between different quiver nodes according to the
hopping superpotential~(\ref{Whop}).
In a moment, we shall see that all such effects
must be completely global and involve all $N$ quiver nodes at once,
thus {\it eg.}
\be
\det(\link_N\cdots\link_1)\ =\ V_1^{Nn_c}\ +\ V_2^{Nn_c}\ +\ \dv
\quad\mbox{where}\quad \dv\
=\ O\Bigl(\bigl(\Lambda^{2n_c-F}\gamma^F\bigr)^N\Bigr)\ \equiv\ O(\alpha).
\label{DetGlobal}
\ee
Indeed, let us temporarily promote the gauge and
the superpotential couplings of the quiver theory to $\ell$--dependent
background fields and allow for generic flavor dependence of the quarks'
couplings and masses, thus
separate $\Lambda_\ell$ for each $SU(n_c)_\ell$ factor and
\begingroup
\postdisplaypenalty=10000
$$
\refstepcounter{equation}
W_{\rm tree}\ =\, \sum_{\ell=1}^N\Bigl(
    \tr(\aq_{\ell+1}\link_\ell Q_\ell\Gamma_\ell)\
    -\ \tr(m_\ell\aq_\ell Q_\ell)\
    +\ \beta s_\ell(\det\link_\ell\,-\,v_\ell^{n_c})
    \Bigr).\eqno (\theequation)\rlap{\strut\footnotemark}
$$\footnotetext{%
	Note the Yukawa coupling $\gamma$ being promoted to the matrices
	$[\Gamma_\ell]^f_{\,f'}$ and the quark masses $\gamma\mu_f$ to
	$[m_\ell]^f_{\,f'}$.
	}%
\par\noindent\endgroup
The promoted theory has a separate $[U(F)\times U(F)\times U(1)]_\ell$
flavor symmetry for each quiver node: For any $2N$ $U(F)$ matrices
$U_\ell$ and $\tilde U_\ell$ and any $N$ unimodular complex numbers $\eta_\ell$
we may transform
\be
Q'_\ell\ =\ Q_\ell U_\ell\,,\qquad
\aq'_\ell\ =\ \tilde U_\ell\aq_\ell\,,\qquad
\link'_\ell\ =\ \eta_\ell\link_\ell\,,\qquad
s'_\ell\ =\ \eta_\ell^{-n_c}s_\ell\,,
\label{FieldsFlavor}
\ee
\be
m'_\ell\ =\ \tilde U_\ell^{-1} m_\ell U_\ell^{-1}\,,\qquad
\Gamma'_\ell\ =\ \eta_\ell^{-1}\tilde U_{\ell+1}^{-1}\Gamma_\ell U_\ell^{-1}\,,
\qquad v'_\ell\ =\ \eta_\ell v_\ell\,,
\label{CouplingsFlavor}
\ee
\be
(\Lambda^{2n_c-F}_\ell)'\ =\ \Lambda_\ell^{2n_c-F}\times
\eta_\ell^{n_c}\eta_{\ell-1}^{n_c}\det(U_\ell)\det(\tilde U_\ell)\,.
\label{LambdaFlavor}
\ee
All quantum effects in the promoted theory must transform covariantly
under this exact $[U(F)\times U(F)\times U(1)]^N$ symmetry,
and this is a very strong constraint
on the holomorphic equations of the chiral ring.
Indeed, there are only $N+F$ independent holomorphic invariant combinations
of the background fields\footnote{%
	Generally,
	\be
	\#(\mbox{invariants})\ =\ \#(\mbox{fields})\ -\ \#(\mbox{symmetries})\
	+\ \#(\mbox{generically unbroken symmetries}).
	\ee
	The  $[U(F)\times U(F)\times U(1)]^N$ symmetry of the promoted theory
	is mostly spontaneously broken by the non-zero values of the
	background fields $m_\ell$, $\Gamma_\ell$, $v_\ell$ and $\Lambda_\ell$.
	Generically, only the $U(1)^F\subset U(F)_{\rm diag}=
	\diag\left([U(F)]^{2N}\right)$ flavor symmetry remains unbroken, hence
	\be
	\#(\mbox{invariants})\ =\ N(F^2+F^2+1+1)\ -\ N(F^2+F^2+1)\ +\ F\
	=\ N\ +\ F.
	\ee
	},
for example
\bea
y_\ell &
=& \frac{\Lambda_\ell^{2n_c-F}\det(m_\ell)}{v_\ell^{n_c} v_{\ell-1}^{n_c}}\,,
	\quad\ell=1,\ldots,N,\quad\mbox{and}\label{Yind} \\
C_k &=& \frac{\alpha b_k}{(v_1\cdots v_N)^{2n_c-k}}\,,
	\quad k=1,\ldots,F, \label{Cinv}
\eea
where
\be
\alpha\ =\ (-1)^F\,\prod_{\ell=1}^N\Lambda^{2n_c-F}\times
	\prod_{\ell=1}^N\det(-\Gamma_\ell)
\label{Agen}
\ee
and $b_k$ are coefficient of the characteristic polynomial
\be
B(X)\ \equiv\,\sum_{k=0}^F b_k X^k\ =\ \det\left(X\,
	-\,m_1\Gamma_1^{-1} m_2\Gamma_2^{-1} \cdots m_N\Gamma_N^{-1}\right) .
\label{Bgen}
\ee
Eventually, we will let the background fields take their usual values
$\Gamma_\ell\equiv\gamma\mbox{\large 1}_{F\times F}$ and
$m_\ell\equiv\gamma\mu$, and consequently \eq{Agen} would reduce to
the good old \eq{Aformula} while the polynomial~(\ref{Bgen}) would
become exactly as in \eq{XiProd}, thus
\be
b_F^{}\ =\ 1,\quad b_{F-1}^{}\ =\ -\sum_{f=1}^F\mu_f^N\,,\quad
b_{F-2}^{}\ =\ +\sum_{f<f'}\mu_f^N\mu_{f'}^N\,,\quad{\it etc.,\ etc.}
\label{Bspec}
\ee
For the moment, however, we need to keep the background fields completely
generic, and it is important to note that despite the appearance of
$\Gamma_\ell^{-1}$ in \eq{Bgen}, the products $\alpha b_k$ are actually
polynomial in all the $\Gamma_\ell$ fields as well as in the $m_\ell$ fields.

In general, the quantum corrections to link chain determinants may depend
on the Coulomb moduli $\varpi_j$ of the quiver theory.
The symmetry~(\ref{FieldsFlavor}) acts on the moduli according to
$\varpi'_j=\varpi_j\times(\eta_1\cdots\eta_N)$, hence most generally
\be
\det({\link_{\ell_2}\cdots\link_{\ell_1}})\
=\ v_{\ell_2}^{n_c}\cdots v_{\ell_1}^{n_c}\times\left[
	1\ +\ {\cal F}\left(y_\ell,C_k,
		\frac{\varpi_j}{(v_1v_2\cdots v_N)}\right)
	\right]
\label{GenericDetQ}
\ee
for some analytic function $\cal F$.
Actually, $\cal F$ has to be a polynomial of its arguments because of the
following considerations:
\begingroup
\parskip= 0pt plus 2pt
\begin{itemize}
\item
    In the un-deformed quiver theory there is no (pseudo) confinement
    or gaugino condensation, hence the determinants~(\ref{GenericDetQ}) should
    be single-valued functions of the
    ${\Lambda^{}_\ell}^{\mskip-9mu\red 2n_c-F}$
    and the ${v^{}_\ell}^{\mskip-7mu\red n_c}$
    as well as of the $\Gamma_\ell$ and $m_\ell$ matrices.
\item
    The determinants should not diverge 
    for any finite values of the background fields
    $m_\ell$, $\Gamma_\ell$, $v_\ell^{n_c}$ and $\Lambda_\ell^{2n_c-F}$ ---
    and this includes regular behavior for $m_\ell\to0$, $\Gamma_\ell\to0$,
    $\Lambda_\ell^{2n_c-F}\to0$ and especially $v_\ell^{n_c}\to 0$.
\item
    Likewise, the moduli dependence of the determinants~(\ref{GenericDetQ})
    should be regular throughout the Coulomb moduli space.
\end{itemize}
\par\noindent\endgroup
Furthermore, every term in $v_{\ell_2}^{n_c}\cdots v_{\ell_1}^{n_c}\times
\cal F$ must carry a non-negative integer power of every $v_\ell^{n_c}$
parameter.
Consequently, for open link chains of length $\ell_2-\ell_1+1<N$
\be
\det({\link_{\ell_2}\cdots\link_{\ell_1}})\
=\ v_{\ell_2}^{n_c}\cdots v_{\ell_1}^{n_c}\times\Bigl[
	1\ +\ {\cal F}\left(y_\ell\ \mbox{only}\right)\Bigr] ,
\label{ShortChain}
\ee
while for the closed chain of length $N$ (once around the quiver)
\be \det(\link_N\cdots\link_1)\
=\ (v_1\cdots v_N)^n_c\times\left[
    1\ +\ {\cal F}_1(y_\ell)\ +\ {\cal F}_2\left(
	\frac{\alpha b_k}{(v_1\cdots v_N)^{2n_c-k}},
	\frac{\varpi_j}{(v_1\cdots v_N)}
	\right) \right] .
\label{WholeQuiver}
\ee
Physically, the $O(y_\ell)$ effects  due to `local' instantons
in the individual $SU(n_c)_\ell$ factors, while the $O(\alpha)$ effects are
due to `globally coordinated' instanton effects in all the $SU(n_c)_\ell$
factors at once --- or equivalently, due to the `diagonal' instantons in the
$SU(n_c)_{\rm diag}$.
Thus, all quantum corrections to the determinant~(\ref{ShortChain}) are solely
due to the local instanton effects $\Longrightarrow$ the explicit part of
\eq{OpenChain} (without the `${}+\cdots$' part) is actually exact.
By comparison, the determinant~(\ref{WholeQuiver}) suffers from two completely
separate sets of quantum corrections, one purely local and the other
purely global.
The local corrections should be exactly as in \eq{ClosedChain}, hence
in terms of \eq{DetGlobal} the $V_1$ and $V_2$ are exactly as in \eq{V1V2}
the global correction is given by
\bea
\dv &
=& (v_1\cdots v_N)^{n_c}\times{\cal F}_2\left(
	\frac{\alpha b_k}{(v_1\cdots v_N)^{2n_c-k}},
	\frac{\varpi_j}{(v_1\cdots v_N)}
	\right) \nonumber\\
\noalign{\vskip 10pt}
&=& \alpha\sum_{k\ge n_c} b_k\times\pp^{(1)}_k(\varpi)\
    +\ \alpha^2\sum_{\mskip-20mu k_1+k_2\ge 3n_c\mskip-20mu }
	b_{k_1}b_{k_2}\times\pp^{(2)}_{k_1,k_2}(\varpi)
    \qquad\qquad \label{AlphaExpansion}\\
&&\quad{}+\ \alpha^3\sum_{\mskip-30mu k_1+k_2+k_3\ge 5n_c \mskip-30mu }
	b_{k_1}b_{k_2}b_{k_3}\times\pp^{(3)}_{k_1,k_2,k_3}(\varpi)\
    +\ \cdots \nonumber
\eea
where $\pp^{(1)}_k$, $\pp^{(2)}_{k_1,k_2}$,
$\pp^{(3)}_{k_1,k_2,k_3}$, {\it etc.}, are symmetric homogeneous
polynomials of the Coulomb moduli $(\varpi_1,\ldots,\varpi_{n_c})$
of respective degrees
$(k-n_c)$, $(k_1+k_2-3n_c)$, $(k_1+k_2+k_3-5n_c)$, {\it etc.}
Since the $b_k$ coefficients exist only for $k\le F$ ({\it cf.}\
\eqrange{Bgen}{Bspec}), the flavor number $F$ puts an upper limit on the
sums in \eq{AlphaExpansion}, which immediately gives us
several general rules:
\begingroup
\parskip=0pt plus 2pt
\renewcommand\labelenumi{(\Alph{enumi})}
\begin{enumerate}
\item  For $F<n_c$ all sums are empty and $\dv=0$, $i.\,e.$
    there is no global correction.
\item  For $F=n_c$ there is only one valid term
    \be
    \dv\ =\ \alpha\times b_F\times\pp^{(1)}_{F=n_c}\
    =\ \alpha\times1\times\mbox{a numeric constant},
    \label{Borderline}
    \ee
    hence the global correction exists but does not depend
    on the moduli~$\varpi_j$.
\item  For $F>n_c$ there are several valid terms and the global correction
    becomes moduli dependent.
\item  For $n_c\le F<\frac{3}{2}n_c$ only the first sum
    (in \eq{AlphaExpansion}) has valid terms, hence
    $\Delta V^{Nn_c}\propto\alpha^1\ \Longrightarrow{}$ the global
    corrections arises at the one diagonal instanton level only.
\item  For $\frac{3}{2}n_c\le F<2n_c$ several instanton levels contribute to
    the global correction $\dv$, up to the maximum of
    $\Bigl\lfloor\frac{n_c}{2n_c-F}\Bigr\rfloor$
    diagonal instantons.
\item  Finally, for $F=2n_c$ all instanton levels contribute to the
    global correction and \eq{AlphaExpansion} becomes an infinite power
    series rather than a finite polynomial.
\end{enumerate}
\endgroup

To understand the physical significance of these rules we need
to take a closer look at the closed link chain $\link_N\cdots\link_1$.
As a composite chiral field, the closed chain has adjoint-like
gauge quantum numbers, so the precise definition of its determinant
is somewhat ambiguous in the quantum theory.
To understand and resolve this ambiguity, consider the characteristic
``polynomial''
\be
\chi(X)\ =\ \det\Bigl(X\,-\,\link_N\cdots\link_1\Bigr),
\label{CharPol}
\ee
which may actually be a non-polynomial function of $X$, depending on the
specific definition of the determinant on the right hand side. 
For example, adapting the definitions (\ref{DetDef1}--\reftail{Det3})
to the present situation, we have
\bea
\det\nolimits_1(X-\link\cdots\link) &\eqdef&
\exp\Bigl[\tr\Bigl(\log(X-\link\cdots\link)\Bigr)\Bigr],\\
\det\nolimits_2(X-\link\cdots\link) &\eqdef&
{\cal D}_{n^{}_c}^{}\Bigl( \tr(X-\link\cdots\link),\ldots,
	\tr(X-\link\cdots\link)^{n_c}\Bigr),\\
\det\nolimits_3(X-\link\cdots\link) &\eqdef&
\left[\det\nolimits_2\left(\frac{1}{X-\link\cdots\link}\right)\right]^{-1} \\
&\eqdef& \left[ {\cal D}_{n^{}_c}^{}\left(
	\tr\frac{1}{X-\link\cdots\link}\,,\ldots,
	\tr\frac{1}{(X-\link\cdots\link)^{n_c}_{}}
	\right)\right]^{-1}, \qquad\qquad\nonumber
\eea
which respectively yield on shell
\bea
\mbox{the non-polynomial}\quad\det\nolimits_1(X-\link\cdots\link) &=&
\Xi(X)\,, \label{qDet1} \\
\mbox{the polynomial}\quad\det\nolimits_2(X-\link\cdots\link) &=&
\Bigl[\Xi(X)\Bigr]_+, \label{qDet2}\\
\mbox{and the non-polynomial}\quad\det\nolimits_3(X-\link\cdots\link) &=&
\frac{n_c!\,\Xi(X)}{(d/dX)^{n_c}\Xi(X)}\,.\qquad\qquad \label{qDet3}
\eea
However, expanding the three functions (\ref{qDet1}--\reftail{qDet3})
in power series of $X\to\infty$, we obtain exactly the same polynomial
parts for all three functions
\be
\Bigl[ \det\nolimits_1(X-\link\cdots\link) \Bigr]_+\
=\ \Bigl[ \det\nolimits_2(X-\link\cdots\link) \Bigr]_+\
=\ \Bigl[ \det\nolimits_3(X-\link\cdots\link) \Bigr]_+\
=\ \Bigl[\Xi(X)\Bigr]_+
\label{DetRule}
\ee
and only the negative-power parts are different.
Although its dangerous to generalize from just three data points,
we believe \eq{DetRule} should work for all sensible definitions
of the quantum determinant, and this gives us an unambiguous formula
for the polynomial part of the characteristic ``polynomial''~(\ref{CharPol}),
namely
\be
\Bigl[\chi(X)\Bigr]_+\ =\ \Bigl[\Xi(X)\Bigr]_+\
=\ P(X)\ -\,\sum_{d\ge1}\frac{(2d-2)!}{d!(d-1)!}\>\alpha^d
\left[\frac{[B(X)]^d}{[P(X)]^{2d-1}}\right]_+ \qquad\qquad
\label{CharSum}
\ee
where the second equality comes from expanding \eq{Xiformula} for
the $\Xi(X)$ in powers of $\alpha$.
Physically, the $P(X)$ term is the classical characteristic polynomial
of the quiver and the $\sum_d$ adds quantum corrections,
the $d{\rm th\over{}}$ term representing the effect of $d$
{\sl diagonal} instantons.
Note that for $F<2n_c$ the sum stops at a finite instanton level
$d_{\rm max}=\Bigl\lfloor\frac{n_c}{2n_c-F}\Bigr\rfloor$, and for
$F<n_c$ there are no quantum corrections at all and $\chi(X)=P(X)$.

Classically,
\be
(-1)^{n_c}\det(\link_N\cdots\link_1)\
=\ \mbox{free term of}\ \chi(X) \ \buildrel{\rm cl}\over=\ \chi(0)
\ee
but in the quantum theory we should reinterpret the free part of $\chi(X)$
as the free part of the polynomial part $[\chi(X)]_+$ because only the
polynomial part is unambiguous.
Or equivalently, we may identify the free term of $\chi(X)$ as the
coefficient of the $X^0$ term in the power series expansion around
$X\to\infty$, thus
\be
(-1)^{n_c}\det(\link_N\cdots\link_1)\
=\ \left.\Bigl[\chi(X)\Bigr]_+\right|_{X=0}\
=\ \Bigl[\chi(X)\Bigr]_0\
= \oint\!\frac{dX\,\chi(X)}{2\pi i\,X}
\qquad \label{DetXi}
\ee
where the integration contour is a very large circle on the physical
sheet of the Seiberg--Witten curve~(\ref{SeibergWitten}).
Hence, following the instanton expansion~(\ref{CharSum}), we write
\bea
(-1)^{n_c}\det(\link_N\cdots\link_1)\ =\ P(0) &-&
\alpha\!\oint\!\frac{dX\,B(X)}{2\pi i XP(X)}\
    -\ \alpha^2\!\oint\!\frac{dX\,B^2(X)}{2\pi i XP^3(X)}\nonumber\\
&-& 3\alpha^3\!\oint\!\frac{dX\,B^3(X)}{2\pi i XP^5(X)}\
    -\ \cdots \label{DetSum} \\
{}=\ P(0) &-&
\sum_{k\ge n_c} \alpha b_k\times\pph^{(1)}_k(\varpi)\nonumber\\
&-&\sum_{k_1+k_2\ge 3n_c}\alpha^2 b_{k_1}b_{k_2}\times
	\pph^{(2)}_{k_1,k_2}(\varpi)
    \qquad\qquad \label{DetExpansion}\\
&-&\sum_{k_1+k_2+k_3\ge 5n_c}\alpha^3 b_{k_1}b_{k_2}b_{k_3}\times
	\pph^{(3)}_{k_1,k_2,k_3}(\varpi)\
    -\ \cdots\qquad\qquad \nonumber
\eea
where
\be
\eqalign{
\pph^{(1)}_k(\varpi)\ &
= \oint\!\frac{dX\,X^{k-1}}{2\pi i\,P(X)}\,,\cr
\pph^{(2)}_{k_1,k_2}(\varpi)\ &
= \oint\!\frac{dX\,X^{k_1+k_2-1}}{2\pi i\,P^3(X)}\,,\cr
\pph^{(3)}_{k_1,k_2,k_3}(\varpi)\ &
=\ 3\!\oint\!\frac{dX\,X^{k_1+k_2+k_3-1}}{2\pi i\,P^5(X)}\,,\cr
\omit\span\omit\dotfill\cr
}\label{definepph}
\ee
are homogeneous symmetric polynomial of the quiver's moduli
$(\varpi_1,\ldots,\varpi_{n_c})$ of respective degrees $(k-n_c)$,
$(k_1+k_2-3n_c)$, $(k_1+k_2+k_3-5n_c)$, {\it etc.}
And because these degrees are exactly as for the $\pp^{(1)}_k(\varpi)$,
$\pp^{(2)}_{k_1,k_2}$, {\it etc.}, polynomials appearing in the expansion
(\ref{AlphaExpansion}), {\it the instanton expansion}~(\ref{DetExpansion})
{\it must satisfy exactly the same general rules \rm (A) \it through \rm (F).}
In particular,
\begingroup
\parskip=0pt plus 2pt
\renewcommand\labelenumi{(\Alph{enumi})}
\begin{enumerate}
\item For $F<n_c$
    \be
    (-1)^{n_c}\det(\link_N\cdots\link_1)\ =\ P(0)
    \ee
    without any instantonic corrections whatsoever, and therefore
    \be
    (-1)^{n_c}P(0)\ =\ V_1^{Nn_c}\ +\ V_2^{Nn_c}\,.
    \ee
    In other words, {\blue the quantum-corrected constraint on the
    $n_c$ redundant Coulomb moduli $\ev_i^N\equiv\varpi_i^{}$ of the quiver is}
    \be\blue
    \prod_{i=1}^{n_c}\varpi_i\ =\ V_1^{Nn_c}\ +\ V_2^{Nn_c}\,,\
    \rm exactly.\black
    \label{Pzero}
    \ee    
\item For $F=n_c$ there is a quantum correction at the one-diagonal-instanton
    level but this correction is moduli independent,
    \be
    (-1)^{n_c}\det(\link_N\cdots\link_1)\ =\ P(0)\ -\ \alpha.
    \ee
    This formula exactly parallels \eq{Borderline} up to an unknown
    numerical constant in the latter,
    and if that constant happens to be equal to
    $(-1)^{n_c-1}$ then \eq{Pzero} would remain valid for $F=n_c$
    as well as for $F<n_c$.
\item For $F>n_c$ the instantonic corrections become moduli dependent
    according to the polynomials~(\ref{definepph}).
    Comparing eqs.~(\ref{AlphaExpansion}) and (\ref{DetExpansion}) we find
    \bea
    (-1)^{n_c}P(0)\ =\ V_1^{Nn_c}\ +\ V_2^{Nn_c}
    &+&\alpha\sum_{k\ge n_c}b_k\left[
	\pp^{(1)}_k(\varpi)\,+\,(-1)^{n_c}\,\pph^{(1)}_k(\varpi)\right]
	\nonumber\\
    &+&\alpha^2\sum_{\mskip-20mu k_1+k_2\ge 3n_c \mskip-20mu }
	b_{k_1}b_{k_2}\left[ \pp^{(2)}_{k_1,k_2}(\varpi)\,
	+\,(-1)^{n_c}\,\pph^{(2)}_{k_1,k_2}(\varpi)\right] \nonumber\\
    &+&\cdots\,,\hbox{\vrule height 15pt width 0pt}
    \eea
   and if we are lucky and
   \be
   \mbox{all}\quad\pp^{(d)}_{k_1,\ldots,k_d}(\varpi)\
   \equiv\ -(-1)^{n_c}\pph^{(d)}_{k_1,\ldots,k_d}(\varpi)\
   =\ (-1)^{n_c-1}\frac{(2d-2)!}{d!(d-1)!} \oint\!\frac{dX}{2\pi i}\,
	\frac{X^{(k_1+\cdots+k_d-1)}}{[P(X)]^{(2d-1)}}
   \label{Conjecture}
   \ee
   then \eq{Pzero} continues to hold true for $F>n_c$.
\end{enumerate}
\endgroup

We wanted to conclude this section by proving that
the $\pp^{(d)}_{k_1,\ldots,k_d}(\varpi)$ polynomials
are indeed given by eqs.~(\ref{Conjecture}) and therefore
{\blue \eq{Pzero} does hold true for any $F<2n_c$} and maybe for $F=2n_c$
as well,\footnote{%
	To be precise, we believe that for $F=2n_c$ eqs.~(\ref{Conjecture})
	hold for the $\pp^{(1)}_k(\varpi)$ polynomials controlling the
	one-instanton-level corrections, but we are not at al sure about the
	higher instanton levels $d\ge 2$.
	The trouble with the $F=2n_c$ case is that the ultraviolet gauge
	couplings $\tau_\ell$ of the \suq\ quiver are asymptotically finite
	rather than asymptotically free, so we don't really know what happens
	to the theory beyond the weak coupling approximation.
	And even in the weak coupling limit, the sub-leading quantum
	corrections are liable to depend on the ultraviolet completion of
	the theory.
	}
but as of this writing our proof is only 95\% complete
\begin{pspicture}[0.2](-2,-2)(+2,+2)
    \pscircle[linewidth=0.4pt,fillstyle=solid,fillcolor=yellow](0,0){2}%
    \psarc[linecolor=red,linewidth=1.2pt](0,-2.1){1.5}{50}{130}%
    \pscircle*[linecolor=blue](-0.6,+0.8){0.3}%
    \pscircle*[linecolor=blue](+0.6,+0.8){0.3}%
\end{pspicture}.
It is also longer than it ought to be, so we present it in the Appendix
to this paper rather than here.

 
%
%
%
\section{Open Questions}
Having analyzed the chiral rings of deconstructive \suq\ quivers
in much detail, we would like to conclude this paper by
discussing the implications of the present research and the
open questions it raises.

The most immediate implication of our rings concerns the
quantum effects in deconstructed \sqcdv;
this will be discussed at length in ref.~\cite{dNK}, but we would like
to present a few highlights here.
From the deconstruction point of view, our most important results
are the Seiberg--Witten curve~(\ref{SeibergWitten}) and the
quantum-corrected constraint~(\ref{Pzero}) on the Coulomb moduli
of the quiver.
In the large quiver size limit $N\to\infty$ governed by the
$\rm 4D\leftrightarrow 5D$ map of moduli and parameters\footnote{%
    In the following \eqrange{ModuliMap}{CouplingMap}, the $\phi_i$
    denote the Coulomb moduli of the 5D SQCD, the $m_i$ are the
    5D quark masses, and the $g^2_{\rm 5D}$ is the 5D gauge coupling
    {\sl at the origin of the Coulomb moduli space}.
    }
\bea
\varpi_i &=& V^N\times\exp(Na\phi_i),\label{ModuliMap}\\
\mu_f^{} &=& V\times\exp(am_f^{}),\label{MassMap}\\
\alpha &=& \frac{V^{2n_cN}}{\prod_f\max(\mu_f^N,V^N_{})}\times
    \exp\left(-Na\,\frac{8\pi^2}{g_{\rm 5D}^2}\right)
    \label{CouplingMap}
\eea
the 4D abelian gauge couplings encoded in the Seiberg--Witten
curve~(\ref{SeibergWitten}) behave precisely as in a 5D gauge theory
compactified on a large circle of size $2\pi R=Na$.
The 5D abelian couplings implied by this procedure --- using dimensional
deconstruction as a UV completion of \sqcdv\ ---
are in perfect agreement with those of \sqcdv\ embedded in string or M theory,
which means they are intrinsic properties of the 5D theory.

Furthermore, thanks to \eq{Pzero},
the $V$ parameter in \eqrange{ModuliMap}{CouplingMap}
is the greater of $V_1,V_2$ roots of \eq{V1V2}.
Consequently, for $\Delta F=0$ there is a lower limit on the $V/\Lambda$ ratio
and hence according to \eq{CouplingMap} the inverse 5D gauge coupling
also has a finite lower limit;
on the other hand, for $\Delta F>0$ there are no limits
and the $g_{\rm 5D}^{-2}$ ranges all the way from $+\infty$ to $-\infty$.
In 5D terms this means that \sqcdv\ theories with maximal Chern--Simons
levels $\kcs=n_c-\half n_f$ have only the positive-coupling phase
(see ref.~\cite{IK1} for the $n_f=0$, $\kcs=n_c$ case), but for
lower Chern--Simons levels there are both positive-coupling and
negative-coupling phases.

We would like to extend our techniques from the deconstructed \sqcdv\
to other deconstructed 5D gauge theories, but this remains an open question.
Naively, the simplest extension is promoting the $SU(n_c)$ gauge theory
to the $U(n_c)$:
Classically, all one has to do in 4D is to promote each $[SU(n_c)]_\ell$
factor to a $[U(n_c)]_\ell$, dispense with the $s_\ell$ singlet fields
and the $W_\Sigma$ part of the superpotential ({\it cf.}\ \eq{Wsigma}),
and expand the theory around a vacuum with non-zero link VEVs
$\vev{\link_\ell}\propto\mbox{\large\bf 1}_{n_c\times n_c}\,$.
Unfortunately, the abelian gauge fields of such a $[U(n_c)]^N$ quiver
suffer from triangular anomalies
\be
\psset{linewidth=0.5pt,unit=0.2cm}
\ovalnode[linecolor=blue]{common}{
	SU(n_c)_\ell^{}\ \mbox{or}\ U(1)_\ell^{}}
\pspicture[0.47](-9,-8)(+23,+8)
\psset{linewidth=1pt,arrowscale=1.5}
\pscircle*(0,-8){0.5}
\pscircle*(0,+8){0.5}
\pscircle*(+14,0){0.5}
\psline{->}(0,-7.5)(0,+7.5)
\psline{->}(0.44,+7.75)(+13.56,+0.25)
\psline{->}(+13.56,-0.25)(+0.44,-7.75)
\rput[l](+14.0,0){\psCoil[coilaspect=0,coilwidth=1]{120}{2970}}%
\rput[l]{180}(-0.1,+8){\psCoil[coilaspect=0,coilwidth=1]{120}{2970}}%
\rput[l]{180}(+0.4,-8){\psCoil[coilaspect=0,coilwidth=1]{300}{3150}}%
\rput[l](+2.75,0){$\link$ fields}
\pnode(-9,+8){upper}
\pnode(-9,-8){lower}
\endpspicture
\nccurve[angleA=90,angleB=180,linecolor=blue]{->}{common}{upper}
\nccurve[angleA=270,angleB=180,linecolor=blue]{->}{common}{lower}
\ovalnode[linecolor=blue]{third}{
	U(1)_{\ell+1}^{}\ \mbox{or}\ U(1)_{\ell-1}^{}}
\label{U1anomalies}
\ee
and the quantum theory does not work.
To cancel the anomalies we need additional chiral superfields
({\it cf.}\ \cite{Dudas,Falkowski} for the $[U(1)]^N$ quiver)
with non-trivial $SU(n_c)_\ell^{}$, $U(1)_\ell^{}$ and $U(1)_{\ell\pm1}^{}$
quantum numbers, for example
\be
\left.\eqalign{
A_\ell\ &
=\ \Bigl({\bf\square}_{\,\ell\,},({\bf 1}^{-n_c})_{\ell+1}\Bigr)\cr
B_\ell\ &
=\ \Bigl(\overline{\bf\square}_{\,\ell\,},({\bf 1}^{+n_c})_{\ell-1}\Bigr)\cr
C_\ell\ &
=\ \Bigl(({\bf 1}^{-n_c})_{\ell\,},({\bf 1}^{+n_c})_{\ell+1}\Bigr)\cr
}\right\}\ \mbox{for all}\ \ell=1,\ldots,N.
\label{cancellers}
\ee
Then we need to endow all these fields with suitable superpotential
couplings and find a vacuum state of the theory where all the light particles
correspond to Kaluza--Klein modes of the 5D $U(n_c)$ theory compactified
on a circle, {\it and nothing else}.
There will be of course all kinds of particles with $O(V)$ masses, and
we need to find their effects on the Chern--Simons interactions of
the 5D theory and perhaps adjust the number $\Delta F$ of
massless quark flavors.
At this point we have a complicated 4D $[U(n_c)]^N$ theory which is
no longer described by a simple quiver diagram~(\ref{quiver}), and now
the real work begins: Evaluating the chiral ring
of the theory and  its implications for the quantum deconstruction.

A bigger open question concerns
quantum deconstruction of other types of 5D gauge theories,
for example $SO(n)$ or $Sp(n)$.
Again, one must first deconstruct the 5D theory at the classical level
and verify the quantum consistency of the resulting 4D quiver theory,
and then one must study the chiral ring in all its glorious details.
But the really big challenge comes from deconstructing 2 extra dimensions
at once:  Start with a 6D SYM theory with 16 supercharges, discretize
the $x^4$ and $x^5$ coordinates into a 2D lattice, and interpret the
result as a 4D, $\NN=1$ gauge theories with a complicated quiver.
Classically, the procedure is well known \cite{deconstructing2D};
for example the $SU(n)$ SYM theory in 6D deconstructs to the
$[SU(n)]^{N^2}$ theory in 6D with a quiver diagram forming a 2D
triangular lattice
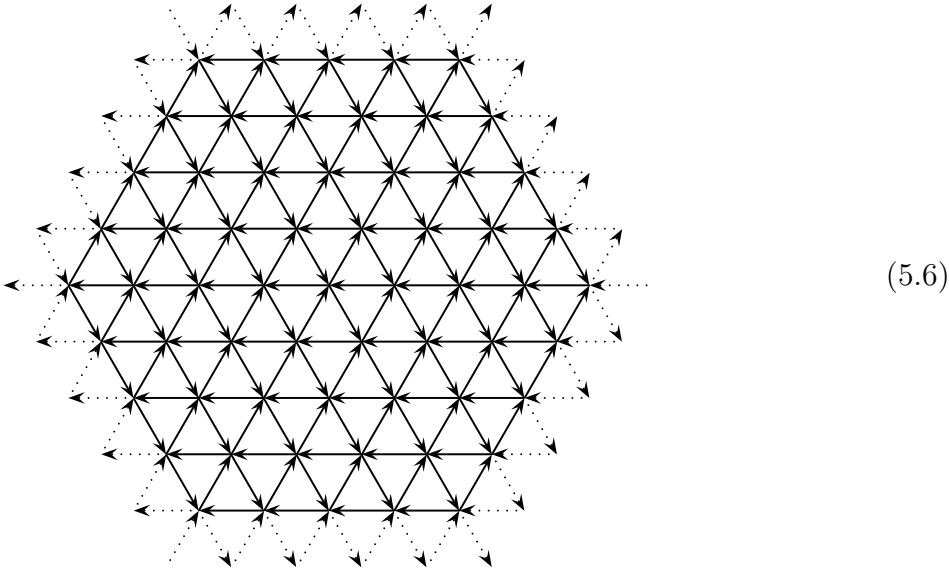
\be
\psset{linewidth=0.75pt,xunit=0.433cm,yunit=0.75cm,arrowscale=1.5}
\def\seg(#1)#2{%
    \count255=#2 \multiply\count255 by 2
    \edef\jj{(\number\count255,0)}
    \rput(#1){%
	\psline[linestyle=dotted]{<-}(-2,0)(0,0)
	\expandafter\rput\jj{\psline[linestyle=dotted]{<-}(0,0)(2,0)}
	}
    \multirput(#1)(+2,0){#2}{\psline{<-}(0,0)(+2,0)}
    }
\def\dir#1{%
    \rput{#1}(0,0){%
	\seg(-4,-4){4}
	\seg(-5,-3){5}
	\seg(-6,-2){6}
	\seg(-7,-1){7}
	\seg(-8,0){8}
	\seg(-7,+1){7}
	\seg(-6,+2){6}
	\seg(-5,+3){5}
	\seg(-4,+4){4}
	}
    }
\pspicture[](-9,-5)(+9,+5)
\dir{0}
\dir{120}
\dir{240}
\endpspicture
\label{trianglattice}
\ee
But at the quantum level, the chiral ring of this quiver
poses a formidable challenge because every closed loop
on the lattice gives rise to an independent generator of the ring,
{\it eg.} $\Tr(\link_9\link_8\cdots\link_2\link_1)$
for the loop of 9 links on the following picture:
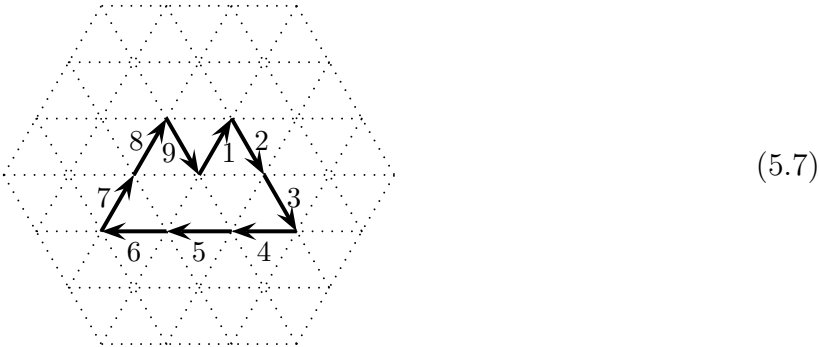
\be
\psset{xunit=0.433cm,yunit=0.75cm}
\pspicture[](-6,-3)(+6,+3)
\psset{linewidth=0.75pt,linestyle=dotted}
\def\lines#1{%
    \rput{#1}(0,0){
	\psline(-3,-3)(+3,-3)
	\psline(-4,-2)(+4,-2)
	\psline(-5,-1)(+5,-1)
	\psline(-6,0)(+6,0)
	\psline(-5,+1)(+5,+1)
	\psline(-4,+2)(+4,+2)
	\psline(-3,+3)(+3,+3)
	}
    }
\lines{0}
\lines{120}
\lines{240}
\psset{linewidth=1.5pt,linestyle=solid,arrows=->,arrowscale=1.5}
\psline(0,0)(+1,+1) \rput[l](+0.7,+0.4){\small 1}
\psline(+1,+1)(+2,0) \rput[l](+1.7,+0.6){\small 2}
\psline(+2,0)(+3,-1) \rput[l](+2.7,-0.4){\small 3}
\psline(+3,-1)(+1,-1) \rput[t](+2,-1.2){\small 4}
\psline(+1,-1)(-1,-1) \rput[t](0,-1.2){\small 5}
\psline(-1,-1)(-3,-1) \rput[t](-2,-1.2){\small 6}
\psline(-3,-1)(-2,0) \rput[r](-2.7,-0.4){\small 7}
\psline(-2,0)(-1,+1) \rput[r](-1.7,+0.6){\small 8}
\psline(-1,+1)(0,0) \rput[r](-0.7,+0.4){\small 9}
\endpspicture
\ee

Another open question is finding a {\sl simple}
string model of the deconstructed
\sqcdv\ and investigating the string origin of the quantum effects
discussed in this paper at the 4D field theory level.
Ironically, while some complicated quiver theories do have simple string
models --- for example the lattice quiver~(\ref{trianglattice})
deconstructing the 6D SYM obtains via $n_c\times N^2$ fractional
D3 branes at a $\CC^3/\ZZ_N\times\ZZ_N$ orbifold point \cite{deconstructing2D}
--- the simplest known string model of the \suq\ quiver~(\ref{quiver})
involves {\sl brane webs} on an orbifold \cite{LPT}.
The issue is not 5D versus 6D, and other, more complicated  deconstructed
5D gauge theories do have web-less string models.
For example, $2N\times n$ fractional D3 branes at the
$\CC^3/\ZZ_{2N}[1,1,-2]$ orbifold point
give rise to the 4D theory with the
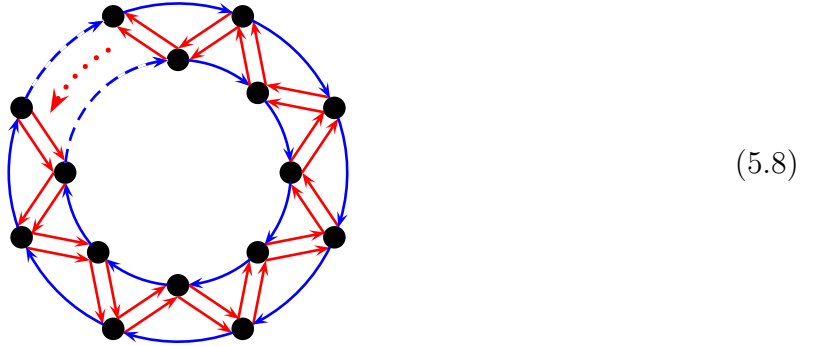
\be
\psset{unit=1.5cm,linewidth=1pt,arrowscale=1.2}
\def\segment#1{\rput{#1}(0,0){%
	\pscircle*(+1,0){0.1}
	\pscircle*(+1.386,+0.574){0.1}
	\psarc[linecolor=blue]{<-}(0,0){1}{5}{40}
	\psarc[linecolor=blue]{<-}(0,0){1.5}{26}{64}
	\psline[linecolor=red]{->}(+1,+0.1)(+1.3,+0.546)
	\psline[linecolor=red]{->}(+1.1,0)(+1.424,+0.482)
	\psline[linecolor=red]{<-}(+1.1,0)(+1.424,-0.482)
	\psline[linecolor=red]{<-}(+1,-0.1)(+1.3,-0.546)
	}}
\pspicture[](-1.6,-1.6)(+1.6,+1.6)
\segment{90}
\segment{45}
\segment{0}
\segment{315}
\segment{270}
\segment{225}
\segment{180}
\pscircle*(-1.386,+0.574){0.1}
\psarc[linecolor=blue]{<-}(0,0){1.5}{161}{199}
\psarc[linecolor=white](0,0){1.5}{116}{154}
\psarc[linecolor=white](0,0){1}{95}{175}
\psarc[linecolor=blue,linestyle=dashed]{<-}(0,0){1.5}{116}{154}
\psarc[linecolor=blue,linestyle=dashed]{<-}(0,0){1}{95}{175}
\psarc[linecolor=red,linestyle=dotted,linewidth=2pt]{->}(0,0){1.25}{115}{155}
\endpspicture
\label{TwistedBand}
\ee
quiver diagram which deconstructs an $SU(n)\times SU(n)$ theory in 5D.
But this theory has a very different chiral ring than the rings discussed
in \S\S3--4 of this paper, and it needs to be worked out in detail before
we can analyze its implications for the string theory.

Finally, there is an open question concerning Dijkgraaf--Vafa matrix models
of the chiral quiver theories.
For the deconstructive \suq\ quiver theories discussed in this article,
we found a matrix model where the bi-fundamental link fields $\link_\ell$
correspond to unitary complex matrices integrated over
the $SU(n_c)$ group manifolds instead of the usual $\CC^{n^2}$ linear spaces.
The details of this model will be presented in a forthcoming article
\cite{dNKmatrix}.
At the same opportunity, we shall also derive the effective superpotential
for the gaugino condensates~(\ref{CondensateIntegrals}) of the quiver theory.

\filbreak\noindent \underline{Acknowledgments:}
We would like to thank Ofer Aharoni, Yaron Oz, Jan Louis and Stefan Theisen
for interesting conversations on the subject.
V.~K.\ thanks the hospitality of the  Tel Aviv University's High--Energy
dept.\ during his multiple visits there.

This article is based on research supported by
the US National Science Foundation (grant PHY--0071512),
the Robert~A.\ Welsh foundation,
the Israeli Science Foundation, and
the German--Israeli Foundation for Scientific Research (GIF).
However, none of these funding agencies should be held responsible
for any errors or deficiencies of this paper.
All the blame --- or hopefully the praise --- belongs to the authors
and nobody else.

 
%
%
%
\appendix
\section{Appendix}
In this appendix we (almost) prove that
the $\pp^{(d)}_{k_1,\ldots,k_d}(\varpi)$ polynomials in \eq{AlphaExpansion}
are indeed given by eqs.~(\ref{Conjecture}) and therefore
\eq{Pzero} does hold true for any $F<2n_c$ and maybe for $F=2n_c$.
Specifically we shall prove  that
(1) the polynomials $\pp^{(d)}_{k_1,\ldots,k_d}
(n_c;\varpi_1,\ldots,\varpi_{n_c})$ have the same
form for all quivers with the same color number~$n_c$ regardless of the
quiver's size $N$ or the flavor number $F$
(as long as $k_1,\ldots,k_d\le F\le 2n_c$ and $N\ge2$), and
(2) they are recursively related to the $\pp^{(d)}_{k'_1,\ldots,k'_d}
(n'_c;\varpi_1,\ldots,\varpi_{n'_c})$ polynomials for smaller values
of the $k'_1,\ldots,k'_d$ indices and/or color numbers $n'_c$.
These relations are consistent with eqs.~(\ref{Conjecture}) and could be used
to prove them all by mathematical induction from a few special cases.
Alas, verifying those special cases remains a loophole; we hope to close
it soon, but we have not done it yet.

We begin by proving the $N$ independence of the $\pp^{(d)}_{k_1,\ldots,k_d}$
polynomials.
Consider what happens when one of the $v_\ell$ parameters of the
promoted quiver theory becomes very large, say $v_1\gg{}$any other mass
scale of the theory.
Physically, in the $v_1\to\infty$ limit we have a high-energy threshold
due to the large semiclassical VEV $\vev{\link_1}=v_1\times
\mbox{\large 1}_{n_c\times n_c}$ (modulo gauge transforms):
The $SU(n_c)_1\times SU(n_c)_2$ gauge symmetry is
Higgsed down to a single $SU(n_c)_{1+2}$ factor, the $\link_1$ chiral field
is eaten up, and the $Q_1$ quarks and $\aq_2$ antiquarks field become massive.
Integrating out these heavy fields we see that the effective
low-energy theory  has the same quiver structure as the high-energy theory
--- except for the $N^{\rm low}=N^{\rm high}-1$ --- and the same coupling
parameters except for the
$$
\refstepcounter{equation}
\Lambda_{1+2}^{2n_c-F}\ =\ \Lambda_1^{2n_c-F}\Lambda_2^{2n_c-F}\times
\frac{\det(v_1\Gamma_1)}{v_1^{2n_c}}\quad\mbox{and}\quad
m_{1+2}\ =\ m_1\times\frac{1}{v_1\Gamma_1}\times m_2\,.
\eqno(\theequation)\rlap{\strut\footnotemark}
$$
\footnotetext{%
	The quarks $Q_{1+2,f}\approx Q_{2,f}$ and the antiquarks
	$\aq_{1+2}^{f'}\approx \aq_1^{f'}$ of the low energy theory acquire
	small masses $[m_{1+2}]^{\!f'}_{\,f}$ via the see-saw mechanism when
	the heavy fields $Q_{1,f}$ and $\aq_2^{f'}$ are integrated out.
	}%
In terms of the $\alpha$ and $b_k$ parameters this means
\be
\alpha^{\rm low}\ =\ \frac{\alpha^{\rm high}}{v_1^{2n_c-F}}\quad
\mbox{and}\quad b_k^{\rm low}\ =\ \frac{b_k^{\rm high}}{v_1^{F-k}}
\label{ABlowN}
\ee
while the Coulomb moduli of the low-energy and the high-energy theories
are related according to
\be
\varpi_j^{\rm high}\ =\ v_1\times\varpi_j^{\rm low} .
\ee
Consequently, \eq{AlphaExpansion} of the low-energy theory becomes
\bea
\left[\dv\right]^{\rm low} &
=& \sum_d (\alpha^{\rm low})^d \sum_{k_1,\ldots,k_d}
	b^{\rm low}_{k_1}\cdots b^{\rm low}_{k_d}\times
	\pp^{(d)\,\rm low}_{k_1,\ldots,k_d}(\varpi_j^{\rm low}) \nonumber\\
&=& \sum_d (\alpha^{\rm high})^d \sum_{k_1,\ldots,k_d}
	b^{\rm high}_{k_1}\cdots b^{\rm high}_{k_d}\times v_1^{\sum k-2dn_c}
	\times\pp^{(d)\,\bf low}_{k_1,\ldots,k_d}
		(v_1^{-1}\,\varpi_j^{\rm high}) \qquad\nonumber\\
&=& \frac{1}{v_1^{n_c}}\times
    \sum_d (\alpha^{\rm high})^d \sum_{k_1,\ldots,k_d}
	b^{\rm high}_{k_1}\cdots b^{\rm high}_{k_d}\times
	\pp^{(d)\,\bf low}_{k_1,\ldots,k_d}(\varpi_j^{\rm high})
    \label{DVlowN}
\eea
where the last equality follows from the $\pp^{(d)}_{k_1,\ldots,k_d}$
being homogeneous polynomials of respective degrees $\sum k-(2d-1)n_c$.
By comparison, the high energy theory has
\be
\left[\dv\right]^{\rm high}\
=\,\sum_d (\alpha^{\rm high})^d \sum_{k_1,\ldots,k_d}
b^{\rm high}_{k_1}\cdots b^{\rm high}_{k_d}\times
\pp^{(d)\,\bf high}_{k_1,\ldots,k_d}(\varpi_j^{\rm high})
\label{DVhighN}
\ee
regardless of the $v_1$ parameter being large or small because the global
correction is completely independent of any of the $v_\ell$.
On the other hand, in the $v_1\to\infty$ limit quantum effects associated
with the $\link_1$ link field become small because of asymptotic freedom
(assuming $F<2n_c$), hence
\be
\det(\link_N\cdots\link_2\link_1)\
\to\  \det(\link_N\cdots\link_2)\times v_1^{n_c}
\ee
and therefore the global corrections to these determinants should
obey the same scaling law
\be
\left[\dv\right]^{\rm high}\ =\ \left[\dv\right]^{\rm low}\times v_1^{n_c}\,.
\ee
Consequently, substituting eqs.~(\ref{DVlowN}) and (\ref{DVhighN}) into
this formula and comparing the moduli-dependent coefficients of similar
$(\alpha^{\rm high})^d b^{\rm high}_{k_1}\cdots b^{\rm high}_{k_d}$
we see that we must have
\be
\pp^{(d)\,\rm high}_{k_1,\ldots,k_d}(\varpi_j)\
=\ \pp^{(d)\,\rm low}_{k_1,\ldots,k_d}(\varpi_j)
\quad\forall k_1+\cdots+k_d\ge (2d-1)n_c\,.
\ee
And therefore by induction, the polynomials
$\pp^{(d)}_{k_1,\ldots,k_d}(\varpi)$ have exactly the same form for all
quiver sizes $N\ge2$. \ $\cal Q.~E.~D.$

Next, let us show that the flavor number $F$ also does not affect
the form of the $\pp^{(d)}_{k_1,\ldots,k_d}(\varpi)$ polynomials
(as long as $F\ge k_1,\ldots,k_d$).
For this purpose we no longer need the $\ell$--dependent matrices
of quark masses and Yukawa couplings, so let
$[\Gamma_\ell]^{\!f'}_{\,f}\equiv\gamma\delta^{f'}_f$ and
$[m_\ell]^{\!f'}_{\,f}\equiv\gamma\mu_f\delta^{f'}_f$, and consider the
limit in which one of the quark flavors becomes very heavy, say
$\mu_1\to\infty$.
Again, we integrate out the heavy fields and derive the low-energy effective
theory which has $F^{\rm low}=F^{\rm high}-1$ and
\be
\left[\Lambda_\ell^{2n_c-F}\right]^{\rm low}\
=\ \left[\Lambda_\ell^{2n_c-F}\right]^{\rm high}\times (-\gamma\mu_1)\
\Longrightarrow\ \alpha^{\rm low}\ =\ -\mu_1^N\times\alpha^{\rm high} .
\label{FscalingA}
\ee
Let us keep the low-energy physics fixed while $\mu_1\to\infty$,
thus fixed $\alpha^{\rm low}$ and fixed $\mu_2,\ldots,\mu_F$
${}\Longrightarrow{}$ fixed $B^{\rm low}(X)$.
In terms of the high-energy theory, this means
\be
\left[\alpha B(X)\right]^{\rm high}\
=\ \frac{X-\mu_1^N}{-\mu_1^N}\times\left[\alpha B(X)\right]^{\rm low}\
\Longrightarrow\ [\alpha b_k]^{\rm high}\
\becomes{\mu_1\to\infty}\ [\alpha b_k]^{\rm low}
\ee
and therefore
\be
\left[\dv\right]^{\rm high}\ \becomes{\mu_1\to\infty}\,
\sum_d (\alpha^{\rm low})^d \sum_{k_1,\ldots,k_d}
b^{\rm low}_{k_1}\cdots b^{\rm low}_{k_d}\times
\pp^{(d)\,\bf high}_{k_1,\ldots,k_d}(\varpi_j^{\rm high}) .
\label{DVhighF}
\ee
On the other hand, decoupling of the heavy quark flavor implies
\bea
\left[\dv\right]^{\rm high} & \becomes{\mu_1\to\infty} &
\left[\dv\right]^{\rm low}\quad\mbox{for}\
	\varpi_j^{\rm low}\equiv\varpi_j^{\rm high}
	\label{DVlowF}\\
&=& \sum_d (\alpha^{\rm low})^d \sum_{k_1,\ldots,k_d}
	b^{\rm low}_{k_1}\cdots b^{\rm low}_{k_d}\times
	\pp^{(d)\,\bf low}_{k_1,\ldots,k_d}(\varpi_j^{\rm high}) .
	\nonumber
\eea
Comparing \eqrange{DVhighF}{DVlowF} we immediately see that we should have
\be
\pp^{(d)\,\rm high}_{k_1,\ldots,k_d}(\varpi_j)\
=\ \pp^{(d)\,\rm low}_{k_1,\ldots,k_d}(\varpi_j)
\quad\forall F\ge k_1,\ldots,k_d
\ee
and therefore by induction, the polynomials
$\pp^{(d)}_{k_1,\ldots,k_d}(\varpi)$ have exactly the same form for all
sufficiently large flavor numbers. \ $\cal Q.~E.~D.$

Now let us relate quiver theories with different color numbers.
To integrate out a color we need each link field $\link_\ell$
to have one very large eigenvalue $\ev_1\to\infty$, or in gauge
invariant terms we need a very large Coulomb modulus
$\varpi_1=\ev_1^N\to\infty$.
To make sure this modulus stays large, we trap it on a mesonic branch
of the moduli space where $\varpi_1=\mu_1^N=\mu_2^N$ and then take
the degenerate mass to infinity.
Integrating out all fields which become superheavy in this limit,
we arrive at the effective low-energy theory which now has
\be
F^{\rm low}\ =\ F^{\rm high}\,-\,2,\qquad
n_c^{\rm low}\ =\ n_c^{\rm high}\,-\,1,\qquad\mbox{and}\quad
\alpha^{\rm low}\ =\ \alpha^{\rm high}.
\ee
Classically
\be
\det(\link_N\cdots\link_1)^{\rm high}\ \buildrel{\rm cl}\over=\
\ev_1^N\times \det(\link_N\cdots\link_1)^{\rm low} ,
\ee
but the quantum corrections may also have sub-leading contributions, thus
we look for
\be
\left[\dv\right]^{\rm high}\
=\ \varpi_1\times \left[\dv\right]^{\rm low}\ +\ \cdots
\label{DVhighlowC}
\ee
where the `$\cdots$' denote terms which do not grow in the
$\varpi_1=\mu_1^N=\mu_2^N\to\infty$ limit.
To be precise, \eq{DVhighlowC} holds when we identify
$(\varpi_2,\varpi_2,\ldots,\varpi_{n_c})^{\rm high}=
(\varpi_1,\varpi^{\rm low})$ and
\be
B^{\rm high}(X)\ =\ (X-\varpi_1)^2\times B^{\rm low}(X)\
\Longrightarrow\ b_k^{\rm high}\
=\ b_{k-2}^{\rm low}\,-\,2\varpi_1 b_{k-1}^{\rm low}\,
	+\,\varpi^2 b_k^{\rm low}. 
\ee
Let us plug
these identifications and the respective eqs.~(\ref{AlphaExpansion})
for both the high-energy and the low energy theories into \eq{DVhighlowC}.
The result looks rather messy, but grinding through the algebra and matching
similar powers of $\alpha^{\rm high}=\alpha^{\rm low}$ and similar products
$b_{k_1}^{\rm low}\cdots b_{k_d}^{\rm low}$ on both sides we arrive at
\begingroup
\def\hh_#1{\pp_{#1}^{(1)\,\rm high}(\varpi_1,\varpi^{\rm low})}
\interdisplaylinepenalty=10000
\bea
\hh_k\ -\ 2\varpi_1 \hh_{k-1} &+&
\varpi_1^2 \hh_{k-2} \label{polyrec1}\\
\noalign{\vskip 7pt}
&=& \varpi_1\times \pp^{(1)\,\rm low}_{k-2}(\varpi^{\rm low})\ +\ \cdots
    \qquad\nonumber
\eea
for the one-diagonal-instanton level, and more generally
\bea
\sum_{q_1,\ldots,q_d=0,1,2} {2\choose q_1}\cdots{2\choose q_d}\,
    (-\varpi_1)^{q_1+\cdots+q_d} &\times&
\pp^{(d)\,\rm high}_{k_1-q_1,\ldots,k_d-q_d}(\varpi_1,\varpi^{\rm low})
    \label{polyrecG}\\
&=& \varpi_1\times \pp^{(d)\,\rm low}_{k_1-2,\ldots,k_d-2}(\varpi^{\rm low})\
    +\ \cdots \qquad\nonumber
\eea
\endgroup
where the `$\cdots$' denotes terms independent of the $\varpi_1$ modulus.

Eqs.~(\ref{polyrecG}) give us recursive relations between the $\pp$ polynomials
of quiver theories with different color numbers.
It is easy to see that these relations are consistent with
eqs.~(\ref{Conjecture}):
Since the $\pph^{(d)}_{k_1,\ldots,k_d}$ polynomials do not depend on how
the index sum $K=k_1+\cdots+k_d$ is partitioned into individual
$k_1,\ldots,k_d$ indices, the left hand side of \eq{polyrecG} becomes
\be
\eqalign{
\sum_{Q=0}^{2d}{2d\choose Q}(-\varpi_1)^Q &
\times\left[ \pp^{(d)\,\rm high}_{K-Q}(\varpi_1,\varpi^{\rm low})\,
	=\,(-1)^{n_c-1}c_d\oint\!\frac{dX}{2\pi i}\,
	\frac{X^{K-Q-1}}{[P^{\rm high}(X)]^{(2d-1)}}\right] \cr
&=\ (-1)^{n_c-1}c_d\oint\!\frac{dX}{2\pi i}\,
    \frac{X^{K-2d-1}\times(X-\varpi_1)^{2d}}{[P^{\rm high}(X)]^{(2d-1)}} \cr
&=\ (-1)^{n_c-2}c_d\oint\!\frac{dX}{2\pi i}\,
    \frac{(\varpi_` X^{K-2d-1}-X^{k-21})}{[P^{\rm low}(X)]^{(2d-1)}} \cr
&=\ \varpi_1\times\pp^{(d)\,\rm low}_{K-2d}(\varpi^{\rm low})\
	-\ \pp^{(d)\,\rm low}_{K-2d+1}(\varpi^{\rm low})\cr
}\label{VerifyRec}
\ee
where $c_d=\frac{(2d-2)!}{d!(d-1)!}$, the third line follows from
$P^{\rm high}(X)=(X-\varpi_1)\times P^{\rm low}(X)$, and
the second term on the last line does not depend on the $\varpi_1$ modulus
in perfect agreement with the right hand side of \eq{polyrecG}.

Working in the other direction, the recursive formul\ae~(\ref{polyrecG})
allow us to completely determine all of the
$\pp^{(d)}_{k_1,\ldots,k_d}(n_c;\varpi)$ polynomials for all color
numbers $n_c$ {\it\blue provided we already know a few of these polynomials}.
Indeed, consider the one-instanton level and suppose we already know
the $\pp^{(1)}_k(n_c-1,\varpi)$ polynomials for the $n_c-1$ colors.
Then for the $n_c$ colors, eqs.~(\ref{polyrec1}) consecutively
determine all but one of the $\pp^{(1)}_k(n_c;\varpi)$ polynomials
according to
\begingroup
\interdisplaylinepenalty=10000
\bea
\pp^{(1)}_{n_c+1}(n_c;\varpi_1,\varpi_2,\ldots,\varpi_{n_c}) &
=& 2\varpi_1\times\pp^{(1)}_{n_c}(n_c;
		\crossout{\varpi_1},\crossout{\varpi_2},
		\ldots,\crossout{\varpi_{n_c}})
	\label{recurex1}\\
&&{}+\ \varpi_1\times\pp^{(1)}_{n_c-1}(n_c-1;\crossout{\varpi_2},
		\ldots,\crossout{\varpi_{n_c}})\
    +\ \varpi_1\mbox{--independent},\nonumber\\
\noalign{\penalty 300}
\pp^{(1)}_{n_c+2}(n_c;\varpi_1,\varpi_2,\ldots,\varpi_{n_c}) &
=& 2\varpi_1\times\pp^{(1)}_{n_c+1}(n_c;
		\varpi_1,\varpi_2,\ldots,\varpi_{n_c})
	\label{recurex2}\\
&&{}-\ \varpi_1^2\times\pp^{(1)}_{n_c}(n_c-1;
		\crossout{\varpi_1},\crossout{\varpi_2},
		\ldots,\crossout{\varpi_{n_c}})\nonumber\\
&&{}+\ \varpi_1\times\pp^{(1)}_{n_c}(n_c-1;\varpi_2,\ldots,\varpi_{n_c})\
    +\ \varpi_1\mbox{--indep},\nonumber\\
\noalign{\penalty 300}
\pp^{(1)}_{n_c+3}(n_c;\varpi_1,\varpi_2,\ldots,\varpi_{n_c}) &
=& 2\varpi_1\times\pp^{(1)}_{n_c+2}(n_c;
		\varpi_1,\varpi_2,\ldots,\varpi_{n_c})
	\label{recurex3}\\
&&{}-\ \varpi_1^2\times\pp^{(1)}_{n_c+1}(n_c-1;
		\varpi_1,\varpi_2,\ldots,\varpi_{n_c})\nonumber\\
&&{}+\ \varpi_1\times\pp^{(1)}_{n_c+1}(n_c-1;
		\varpi_2,\ldots,\varpi_{n_c})\
    +\ \varpi_1\mbox{--indep},\nonumber\\
\noalign{\penalty 300}
\omit\span\omit\span\omit\dotfill\qquad \cr
\noalign{\penalty 500}
\pp^{(1)}_{2n_c}(n_c;\varpi_1,\varpi_2,\ldots,\varpi_{n_c}) &
=& 2\varpi_1\times\pp^{(1)}_{2n_c-1}(n_c;
		\varpi_1,\varpi_2,\ldots,\varpi_{n_c})
	\label{recurex4}\\
&&{}-\ \varpi_1^2\times\pp^{(1)}_{2n_c-2}(n_c-1;
		\varpi_1,\varpi_2,\ldots,\varpi_{n_c})\nonumber\\
&&{}+\ \varpi_1\times\pp^{(1)}_{2n_c-2}(n_c-1;
		\varpi_2,\ldots,\varpi_{n_c})\
    +\ \varpi_1\mbox{--indep},\qquad\quad\nonumber
\eea
\endgroup
where the $\varpi_1$--independent term on the right hand side
of each equation is uniquely determined by the requirement that the
polynomial on the left hand side is homogeneous and totally symmetric
in all of the Coulomb moduli $(\varpi_1,\varpi_2,\ldots,\varpi_{n_c})$.
For example, suppose we already know that
\be
\pp^{(1)}_3(n_c=2)\,=\,-(\varpi_2+\varpi_3),\quad
\pp^{(1)}_3(n_c=3)\,=\,+1,\quad
\pp^{(1)}_4(n_c=3)\,=\,+(\varpi_1+\varpi_2+\varpi_3);
\ee
then \eq{recurex2} tells us
\be
\pp^{(1)}_5(n_c=3)\ =\ \varpi_1^2\ +\ \varpi_2(\varpi_2+\varpi_3)\
+\ \varpi_1\mbox{--independent}
\ee
and the only homogeneous symmetric polynomial of this form is
\be
\pp^{(1)}_5(n_c=3)\ =\ (\varpi_1^2+\varpi_2^2+\varpi_4^2)\
+\ (\varpi_1\varpi_2+\varpi_1\varpi_3+\varpi_2\varpi_3).
\ee

The one exception to this method is the zero-degree case of $k=n_c$
where \eq{polyrec1} reduces to a triviality
\be
\pp^{(1)}_{n_c}(n_c;\crossout{\varpi_1},\crossout{\varpi_2},
	\ldots,\crossout{\varpi_{n_c}})\
=\ \varpi_1\mbox{--independent}
\ee
and there is no symmetry argument to determine the
numerical constant $\pp^{(1)}_{n_c}$.
To plug this hole we need a separate recursive relation
\be
\pp^{(1)}_{n_c}(n_c)\ =\ -\pp^{(1)}_{n_c-1}(n_c-1),
\label{recurex0}
\ee
so let us cook up yet another integrating-out scheme.
Consider the quiver theory with $F=n_c$ in the limit of
$\varpi_1=\mu_1^N\to\infty$ while the remaining moduli and quark masses
remain finite.\footnote{%
	We may force $\varpi_1\equiv\mu_1^N$ while $\mu_1\to\infty$
	by working with discrete Higgs vacua~(\ref{HiggsVac}) of the
	theory with a slightly deformed superpotential.
	We should allow for generic roots of the deformation polynomial
	$\WW(X)$ in order to keep the remaining Coulomb moduli
	$\varpi_2,\ldots,\varpi_{n_c}$ generic, but the overall magnitude
	of the deformation should be kept infinitesimal to assure that
	the quantum corrections to the link chain determinants remains
	as in the undeformed theory.
	}
Integrating out the fields which become superheavy in this limit
we arrive at the low-energy theory with
\be
F^{\rm low}\ =\ n_c^{\rm low}\ =\ F^{\rm high}-1\ =\ n_c^{\rm high}-1\qquad
\mbox{and}\quad \alpha^{\rm low}\ =\ \frac{-1}{\mu_1^N}\times\alpha^{\rm high}.
\label{Iout1C1F}
\ee
Consequently, according to \eq{Borderline} for the low energy theory
\be
\left[\dv\right]^{\rm low}\
=\ \alpha^{\rm low}\times\left[\pp^{(1)}_{n_c}(n_c)\right]^{\rm low}
\ee
and therefore
\be
\left[\dv\right]^{\rm high}\
=\ \varpi_1\times \left[\dv\right]^{\rm low}\ +\ O(1)\
=\ -\alpha^{\bf high}\times\left[\pp^{(1)}_{n_c}(n_c)\right]^{\rm low}\
+\ O(\varpi_1^{-1}).
\label{DVlowhigh1CF}
\ee
On the other hand, we may apply \eq{Borderline} to the high-energy theory
itself since it also has $F=n_c$, thus
\be
\left[\dv\right]^{\rm high}\
=\ \alpha^{\rm high}\times\left[\pp^{(1)}_{n_c}(n_c)\right]^{\rm high},
\ee
and comparing this formula to \eq{DVlowhigh1CF} we immediately see that the
$\pp^{(1)}_{n_c}$ constants of the two theories must be related
according to \eq{recurex0}.

Altogether, eqs.~(\ref{recurex0}) and (\ref{recurex1}--\reftail{recurex4})
allow us to derive all of the $\pp^{(1)}_k(\varpi)$ polynomials for 
$n_c$ colors from the similar polynomials for $n_c-1$ colors.
Hence by induction in $n_c$, {\it once we verify the induction base}
\be
\pp^{(1)}_2(n_c=2)\,=\,-1,\quad
\pp^{(1)}_3(n_c=2)\,=\,-(\varpi_1+\varpi_2),\quad
\pp^{(1)}_4(n_c=2)\,=\,-(\varpi_1^2+\varpi_2^2+\varpi_1\varpi_2)
\label{Ibase1}
\ee
{\it then all of the $\pp^{(1)}_k$ polynomials for all color numbers must be
exactly as in eqs.}~(\ref{Conjecture}).

At the higher instanton levels $d\ge2$ we have a similar situation:
Once we know the $\pp^{(d)}_{k_1\,\ldots,k_d}(n_c-1,\varpi)$ polynomials
for $n_c-1$ colors, eqs.~(\ref{polyrecG}) let us sequentially construct most
of the $\pp^{(d)}_{k_1\,\ldots,k_d}(n_c,\varpi)$ polynomials for $n_c$ colors
according to formul\ae\ similar to \eqrange{recurex1}{recurex4}.
Again, the $\varpi_1$--independent terms can be uniquely determined by the
total symmetry of the  polynomials in all $n_c$ Coulomb moduli
$(\varpi_1,\varpi_2,\ldots,\varpi_{n_c})$.
And once again, in the zero-degree cases of $k_1+\cdots+k_d=(2d-1)n_c$
the symmetry argument does not work and we need an additional recursive
formula like (\ref{recurex0}) to obtain the numerical constants
\be
\pp^{(d)}_{k_1,\ldots,k_d}(n_c;\crossout{\varpi_1},
	\crossout{\varpi_2},\ldots,\crossout{\varpi_{n_c}})\,
=\, (-1)^{n_c-1}\,\frac{(2d-2)!}{d!(d-1)!}\quad
\forall\ k_1+\cdots+k_d=(2d-1)n_c\,.
\label{ZeroDegree}
\ee
{\it Once we verify these constants we may use induction in $n_c$, and
given the induction base}
\be
\eqalign{
\pp^{(d)}_{3,4,\ldots,4}(n_c=2)\ &
=\ -\frac{(2d-1)!}{d!(d-1)!}\,(\varpi_1+\varpi_2),\cr
\pp^{(d)}_{4,4,\ldots,4}(n_c=2)\ &
=\ -\frac{(2d-1)!}{d!(d-1)!}\Bigl(
	d(\varpi_1+\varpi_2)^2\,-\,\varpi_1\varpi_2\Bigr),\cr
}\label{IbaseG}
\ee
{\it all of the $\pp^{(d)}_{k_1\,\ldots,k_d}$ for all $n_c$ must be
exactly as in eqs.~{\rm(\ref{Conjecture})} and \blue eq.~{\rm(\ref{Pzero})}
must hold true for all numbers of colors and flavors.}

This completes our argument.
To complete the \underline{proof} we would need to close the remaining
loopholes, namely
(A) verify the zero-degree formul\ae~(\ref{ZeroDegree}) for $d\ge2$ instantons,
and (B) verify the induction base eqs.~(\ref{Ibase1}) and (\ref{IbaseG}).
This work is in progress, but we have not yet finished it.

 
%
%

\end{document}